\documentclass[twocolumn]{aastex701}
\newcommand{\Msun}{M$_{\odot}$}
\newcommand{\Mbh}{$M_{\rm BH}$}
\newcommand{\Mnsc}{$M_{\rm NSC}$}

\newcommand{\Mstar}{$M_{\star}$}

\newcommand{\Lsun}{L$_\odot$}

\newcommand{\ml}{\emph{M/L}}

\newcommand{\hst}{HST}

\newcommand{\kms}{km~s$^{-1}$}

\newcommand{\go}{G$_{\rm outer}$}

\newcommand{\jamcyl}{\textsc{JAM}$_{\rm cyl}$}
\newcommand{\jamsph}{\textsc{JAM}$_{\rm sph}$}
\newcommand{\vrms}{$V_{\rm rms}$}
\newcommand{\jam}{\textsc{JAM}}
\newcommand{\mge}{\textsc{MGE}}
\newcommand{\ppxf}{\textsc{PPXF}}
\newcommand{\adamet}{\textsc{AdaMet}}
\newcommand{\osiris}{\textsc{OSIRIS}}
\newcommand{\powerbin}{\textsc{PowerBin}}
\newcommand{\mlhst}{$M/L_{\rm F814W}$}

\newcommand{\Vrms}{$\sqrt{V^2+\sigma^2}$}

\hypersetup{colorlinks, linkcolor=blue, citecolor=blue, urlcolor=blue}

\usepackage{multirow}
\usepackage{footnote}
\usepackage{newtxtext,newtxmath}
\usepackage{graphicx}	% Including figure files
\usepackage{amsmath}	% Advanced maths commands
\usepackage{caption}
\usepackage{xcolor}
\graphicspath{{./}{figures/}}
\received{October 6, 2026}
\submitjournal{ApJ}

\shorttitle{IMBH Measurement in IC~342 with Keck/OSIRIS}
\shortauthors{Tuan N. Le et al.}
\begin{document}
\title{A $10^5\,M_\odot$ Black Hole in the Bulgeless Spiral Galaxy IC~342: First Stellar-Dynamical \Mbh\ Measurement with Keck/\osiris\ LGS-AO}

%%%%%%%%%%%%%%%%%%%%%%%%%%%%%%%
\correspondingauthor{Dieu D.\ Nguyen} 
\email{dieun@umich.edu}
%%%%%
\author[0009-0009-0015-1208]{Tuan N. Le}
\affiliation{Faculty of Physics—Engineering Physics, University of Science, VNU-HCM, Ho Chi Minh City, Vietnam}
\affiliation{Viet Nam National University Ho Chi Minh City, Vietnam}
\email{tuan.le.nutshell@gmail.com}
%%%%%
\author[0000-0002-5678-1008]{Dieu D.\ Nguyen}
\email{dieun@umich.edu}
\affiliation{Department of Astronomy, University of Michigan, 1085 South University Avenue, Ann Arbor, MI 48109, USA}
%%%%%
\author[0000-0002-2126-3905]{Mark Durr\'e}
\email{mdurre@swin.edu.au}
\affiliation{Centre for Astrophysics and Supercomputing, Swinburne University of Technology, Hawthorn, VIC 3122, Australia}
%%%%%
\author[0000-0002-6496-9414]{Alister W.\ Graham}
\email{alister.w.graham@gmail.com}
\affiliation{Centre for Astrophysics and Supercomputing, Swinburne University of Technology, Hawthorn, VIC 3122, Australia}
%%%%%
\author[0000-0001-5802-6041]{Elena Gallo}
\email{egallo@umich.edu}
\affiliation{Department of Astronomy, University of Michigan, 1085 South University Avenue, Ann Arbor, MI 48109, USA}
%%%%%
\author[0009-0006-5852-4538]{Hai N. Ngo}
\affiliation{Department of Astronomy, University of Michigan, 1085 South University Avenue, Ann Arbor, MI 48109, USA}
\email{hai10hoalk@gmail.com}
%%%%%
\author[0009-0005-8845-9725]{Tien H.\ T.\ Ho}
\email{htien2808@gmail.com}
\affiliation{Faculty of Physics—Engineering Physics, University of Science, VNU-HCM, Ho Chi Minh City, Vietnam}
\affiliation{Viet Nam National University Ho Chi Minh City, Vietnam}
%%%%%
\author[0009-0004-3689-8577]{Tinh Q.\ T.\ Le}
\email{lethongquoctinh01@gmail.com}
\affiliation{Department of Physics, International University, VNU-HCM, Ho Chi Minh City, Vietnam}
\affiliation{Viet Nam National University Ho Chi Minh City, Vietnam}
%%%%%
\author[0000-0003-2017-0982]{Thiem Hoang}
\email{thiemhoang@kasi.re.kr}
\affiliation{Korea Astronomy and Space Science Institute, Daejeon 34055, Republic of Korea}
\affiliation{Department of Astronomy and Space Science, UST, 217 Gajeong-ro, Yuseong-gu, Daejeon 34113, Republic of Korea} 
%%%%%
\author[0000-0002-7924-3253]{Igor V.\ Chilingarian}
\email{igor.chilingarian@cfa.harvard.edu}
\affiliation{Center for Astrophysics – Harvard and Smithsonian, 60 Garden St. Cambridge, MA, 02138 USA}
\affiliation{Observatoire de Paris, LUX, CNRS, PSL University, 75014, Paris, France}
%%%%%
\author[0000-0002-4622-796X]{Roberto Soria}
\email{roberto.soria@inaf.it}
\affiliation{INAF-Osservatorio Astrofisico di Torino, Strada Osservatorio 20, I-10025 Pino Torinese, Italy}
\affiliation{Sydney Institute for Astronomy, School of Physics A28, The University of Sydney, Sydney, NSW 2006, Australia}
%%%%%
\author[0000-0002-4436-6923]{Fan Zou}
\email{fanzou@umich.edu}
\affiliation{Department of Astronomy, University of Michigan, 1085 South University Avenue, Ann Arbor, MI 48109, USA}
%%%%%
\author[0009-0008-6050-5736]{Huy G.\ Tong}
\email{tonggiahuy191203@gmail.com}
\affiliation{Institute of Astronomy, National Tsing Hua University, Kuang Fu Rd. Sec. 2, Hsinchu, 30013, Taiwan}
%%%%%
\author[0000-0001-9649-2449]{Que T.\ Le}
\email{ltque@hcmiu.edu.vn}
\affiliation{Department of Physics, International University, VNU-HCM, Ho Chi Minh City, Vietnam}
\affiliation{Viet Nam National University Ho Chi Minh City, Vietnam}
%%%%%
\author[0009-0005-1314-9992]{Huy M.\ B.\ Tran}
\email{tran.ba@ufl.edu}
\affiliation{Department of Astronomy, University of Florida, P.O. Box 112055, Gainesville, FL32611, USA}

%%%%%%%%%%%% ABSTRACT OF PAPER %%%%%%%%%%%%%
\begin{abstract}

We present the first direct stellar-dynamical measurement of an intermediate-mass black hole (IMBH) in IC~342 (the ``Hidden Galaxy''), a nearby nearly face-on ($i\approx31^\circ$) bulgeless spiral at $D=3.45\pm0.13$~Mpc, using near-infrared integral-field spectroscopy with Keck/\osiris\ Laser Guide Star Adaptive Optics. The heavy foreground extinction ($A_{\mathrm{F814W}}\approx0.74$~mag) renders the nucleus inaccessible to optical spectroscopy, making $K$-band observations at $\sim$2.3~$\micron$ the only viable path to high-resolution nuclear kinematics. Two-dimensional stellar kinematics extracted from the CO bandhead absorption features via \ppxf\ at $0\farcs062$ resolution reveal a central velocity dispersion rising from $\sim$42~\kms\ at $r\gtrsim0\farcs2$ to a peak of $\sim$49~\kms, providing kinematic evidence for a central dark mass embedded within the prominent nuclear star cluster (NSC; $M_\mathrm{NSC}\approx1.17\times10^7$~\Msun). Modeling these kinematics with an ensemble of four \jam\ models, spanning cylindrical and spherical velocity-ellipsoid alignments and two PSF representations, combined with an \mge\ mass model constructed from a three-S\'ersic components and an exponential disk decomposition of archival \hst/WFC3 F814W imaging and incorporating a physically motivated stellar mass-to-light ratio gradient, driven by the young central NSC population, yields $M_\mathrm{BH}=(1.95\pm0.61)\times10^5$~\Msun. Crucially, accounting for this stellar population gradient is essential for the detection: standard constant-$M/L$ models systematically overestimate the central stellar mass, effectively hiding the black hole signature and yielding only an upper limit. The stellar orbits are near-isotropic ($\beta\approx0$), typical of fast-rotating late-type spirals. This measurement places IC~342 among the rare bulgeless spirals with a direct dynamical \Mbh\ at the IMBH mass scale, where both the NSC and BH are directly characterized, providing a new anchor for compact massive object co-evolution models at the low-mass end of the BH mass spectrum.

\end{abstract}

%%% You can use the \uat command to link your UAT concepts back its source.
%%% https://astrothesaurus.org/concept-select/

%%%%%%%%%%%%%% BODY OF PAPER %%%%%%%%%%%%%%%

\section{Introduction}\label{sec:intro}

The nuclei of late-type galaxies (LTGs) are among the most information-rich yet observationally challenging environments in the local Universe. Unlike the smooth cores of elliptical galaxies, LTG nuclei are shaped by a complex interplay of secular processes: bar-driven gas inflows, episodic starbursts, and the gradual accumulation of mass at the dynamical center. Two distinct central massive objects (CMOs) can inhabit these nuclei simultaneously: nuclear star clusters \citep[NSCs; e.g.,][]{Balcells2003, Boker2004, Seth2006} and massive black holes \citep[BHs; \Mbh~$\sim 10^4$--$10^6$~\Msun; e.g.,][]{Nguyen17, Nguyen19}, yet the physical relationship between them \citep{Graham2016} remains one of the most pressing open questions in galaxy evolution \citep[e.g.,][]{Neumayer2020, Nguyen22, Nguyen2025b}.

NSCs are compact ($r_e \sim~$2--5 pc), dense stellar systems common at the centers of both LTGs and dwarf early-type galaxies \citep[ETGs: e.g.,][]{1985AJ.....90.1759S, Boker2002, Graham2003, Cote2006}, with typical masses \Mnsc~$\sim 10^5$--$10^8$~\Msun \citep{Scott2013}. Rather than being mutually exclusive, NSCs and BHs are now known to coexist across a broad range of galaxy types, from the Milky Way \citep[e.g.,][]{1985ApJ...293..445S, 1987ARA&A..25..377G, Graham2009, Schodel2014} to nearby spirals \citep[e.g.,][]{2008AJ....135..747G, Seth2008, Nguyen14} and dwarf ellipticals \citep[e.g.,][and references therein]{Seth2010, Nguyen18, Thater23, Graham2025}, implying either independent formation channels or a physical coupling through shared gas supply and feedback.

The origin of this co-existence is deeply contested, with two competing pictures for how the two CMOs influence one another. In the first, the NSC comes first and \emph{feeds} the BH: NSC formation, through the infall and coalescence of globular clusters or through secular gas inflows \citep{Tremaine1975, Antonini2013, Arca-Sedda2014}, assembles a dense central stellar system whose high stellar densities and runaway collisions provide both the seed and the fuel for subsequent BH growth \citep{Debattista2006, NeumaerSeth2011, 2023MNRAS.526.4908C}. In the second, the BH ultimately \emph{limits} the NSC: once sufficiently massive, the central BH suppresses further NSC growth through accretion feedback that heats and expels the nuclear gas reservoir \citep{Antonini2015ApJ...812...72A}. In merger-built galaxies, binary-BH scouring can additionally erode the cluster dynamically. Together, these processes produce an anti-correlation between \Mbh\ and \Mnsc\ at 
intermediate masses \citep{Antonini2015ApJ...806L...8A, Neumayer2020}. The observed demographics reflect this competition: low-mass galaxies are NSC-dominated, high-mass galaxies are BH-dominated, and the handover occurs near a bulge mass of \Mstar~$\sim 10^{10}$~\Msun\ (or $\sigma_e \sim 65$~\kms), where the two CMOs overlap in mass and frequently co-exist \citep{NeumaerSeth2011, GrahamSoria2019}. The most powerful diagnostic of this competition is a \emph{direct, dynamical} \Mbh\ measurement in a galaxy whose NSC mass, age, and structural parameters are already well-characterized, precisely the transition regime that IC~342 occupies.

Additionally, calibrating the \Mbh--host scaling relations \citep[][]{1998AJ....115.2285M, Ferrarese00, Gebhardt00, Graham2001, Haring04, Kormendy13, 2016ASSL..418..263G} across the full range of galaxy morphologies is essential for understanding BH--galaxy co-evolution \citep{2023MNRAS.522.3588G} and speciation \citep{Graham2026MNRAS.549ag782G}. Yet of the $\sim$200 direct, dynamical $M_\mathrm{BH}$ measurements published to date, the majority reside in elliptical galaxies and lenticular galaxies with classical bulges \citep{Saglia16, Sahu19a, Sahu19b}. Fewer than $\sim$30 measurements exist in late-type spiral galaxies (type $\geq$ Sbc), and the sample thins sharply toward bulgeless or pseudobulge-dominated Scd--Sd systems \citep{2018ApJ...869..113D}. This morphological census bias is not accidental: for LTGs with $\sigma_e \lesssim 65$ km\,s$^{-1}$, scaling relations predict \Mbh~$\lesssim 10^6$~\Msun, placing the gravitational sphere of influence (SOI\footnote{Defined as $R_\mathrm{SOI} = GM_\mathrm{BH}/\sigma_e^2$, the radius within which the BH dominates the stellar gravitational potential, where the enclosed stellar mass equals approximately \Mbh.}) at only $\sim$0\farcs1 even for the nearest targets \citep[$D \sim 3$~Mpc;][]{Ferrarese2005, Nguyen23}. Resolving such small SOIs demands sub-0\farcs1 angular resolution, achievable in the optical with \textit{Hubble Space Telescope} (\hst) and the Multi Unit Spectroscopic Explorer (MUSE) on the Very Large Telescope (VLT), which is rendered ineffective by the heavy internal and/or foreground dust extinction prevalent in LTG nuclei.

The advent of near-infrared (NIR) adaptive optics (AO) assisted integral-field spectroscopy (IFS) on 8--10\,m telescopes has changed this picture. Observations targeting the CO bandhead absorption features at $\sim$2.3~\micron\ achieve angular resolutions of $\lesssim$0$\farcs$1 while simultaneously penetrating dust extinction that would render optical counterparts unusable. When combined with Jeans Anisotropic Modeling (\jam\footnote{v7.2.4: \url{https://pypi.org/project/jampy/}}; \citealt{Cappellari08, Cappellari2020}) and Multi-Gaussian Expansion (\mge\footnote{v6.0.4: \url{https://pypi.org/project/mgefit/}}; \citealt{Emsellem94, Cappellari2002}) decompositions of archival \hst\ photometry, this approach has yielded robust SMBH mass determinations \citep[\Mbh~$\approx 10^6$--$10^{10}$~\Msun;][]{Nguyen2025a, Nguyen_2026_highz}; specially, in LTGs \citep{Nguyen_2026_4258, Nguyen_2026_M81} and down to the intermediate-mass black hole (IMBH) mass (\Mbh~$\approx 10^2$--$10^5$~\Msun) frontier \citep{Ngo2025c, Ngo2025b, Nguyen_2025_NSC, LeTinh2026}.

\subsection{IC~342 as the Ideal Laboratory}\label{sec:target}

IC~342 (the ``Hidden Galaxy'') is a low-mass \citep[$\log_{10}(M_\star/M_\odot)=10.37$;][]{Pilyugin2023}, nearby, face-on \citep[$i \approx 31^\circ$;][]{Crosthwaite2000} LTG (SABcd) with a Tip-of-the-Red-Giant-Branch (TRGB) distance of $D = 3.45\pm0.13$ Mpc \citep{Wu2014}, corresponding to a physical scale of $\sim$16~pc~arcsec$^{-1}$, when adopting a flat $\Lambda$CDM cosmology with $H_0 = 70$~\kms~Mpc$^{-1}$, $\Omega_{\Lambda} = 0.7$, and $\Omega_{\rm m} = 0.3$. It is the nearest massive bulgeless galaxy, and the dominant member of the IC~342/Maffei group \citep{Karachentsev2003}.  Despite its intrinsic brightness, its proximity to the Galactic plane ($b \approx +10.6^\circ$) subjects it to severe foreground extinction of $A_{\mathrm{F814W}} \approx 0.74$ mag \citep{Schlafly2011}, rendering the nucleus effectively inaccessible to optical spectroscopy. NIR observations are therefore not merely advantageous, they are \emph{the only viable path} to high-resolution nuclear kinematics in this galaxy.

The nuclear region of IC~342 has been extensively characterized across wavelengths, making it the best-studied bulgeless spiral galaxy nucleus in the local Universe. Photometric study shows that its central region is occupied by a prominent and young ($\sim$60--630 Myr) NSC with an effective radius of $r_e \approx$ few pc and a stellar velocity dispersion of $\sigma_{\rm NSC} = 33 \pm 3$~\kms\ \citep{Boker1999}. It is the most luminous among known NSCs in nearby LTGs and it displays a bar-like isophotal elongation on scales of $\sim$110~pc linked to a weak central stellar bar \citep{Boker1997}.  The nucleus is embedded within a starburst ring of diameter $\sim$70 pc delineated by molecular gas, hosting two bright HII-region complexes of age $\sim$5 Myr \citep{Boker1997, Schinnerer2003}. High-resolution CO(2--1) interferometry (synthesized beam $\approx$1\farcs2) reveals ongoing non-circular inflow of molecular gas into the central few parsec at a rate of 0.003--0.14~\Msun~yr$^{-1}$ \citep{Schinnerer2003}, suggesting that conditions for both continued NSC growth and BH feeding are actively maintained. At larger scales, \textit{Chandra} and \textit{XMM-Newton} imaging reveals an extended nuclear X-ray source spatially coincident with the NSC, consistent with either starburst emission or an embedded low-luminosity active galactic nucleus \citep[LLAGN;][]{Bauer2003, Mak2008}.

However, IC~342 lacks a direct, dynamical \Mbh\ measurement. The only existing constraint comes from \citet{Boker1999}, who applied one-dimensional (1D) spherical Jeans models to aperture-integrated CO-bandhead spectroscopy of the NSC ($\sigma_\mathrm{NSC} = 33 \pm 3$~\kms) and derived a dynamical NSC mass of $\approx 6 \times 10^6$~\Msun\ together with an upper limit of \Mbh~$< 5 \times 10^5$~\Msun\ at their adopted distance of $D = 1.8$~Mpc. Since both quantities scale linearly with distance, they become \Mnsc~$\approx 1.17 \times 10^7$~\Msun\ and \Mbh~$< 9.6 \times 10^5$~\Msun\ at our adopted TRGB distance of $D = 3.45$~Mpc \citep{Wu2014}.

\subsection{This Work}\label{sec:thiswork}

We present the first stellar-dynamical \Mbh\ measurement in IC~342 using NIR IFS with Keck/\osiris\ observation assisted by the laser-guide-star AO (LGS-AO) system. With a point spread function (PSF) full width at half maximum (FWHM) of $\approx$0\farcs053--0\farcs062 (Section~\ref{sec:osirsK_psf}), our data resolve the predicted SOI down to \Mbh~$\sim 10^5$~\Msun\ at the adopted distance, enabling a direct, marginally resolved dynamical \Mbh\ in this system. Combined with the well-characterized NSC properties \citep{Boker1997, Boker1999}, this measurement directly probes the NSC--BH coexistence regime, tests whether the low-mass end of the \Mbh--$\sigma_e$ relation extends to bulgeless Scd-type spirals, and provides a critical anchor for both the direct \Mbh\ census in LTGs and theoretical models of CMO co-evolution in the secular-evolution regime.

This paper is structured as follows. Section~\ref{sec:osiris_observations} details the Keck/\osiris\ observations, data reduction, and its PSF derivation. We extracted  the nucleus stellar kinematics of IC~342 in Section~\ref{sec:losvd} and parametrize the galaxy mass model in Section~\ref{sec:photomodel}. Section~\ref{sec:dynamical-models} presents the \Mbh\ measurement via \jam\ dynamical modeling, and Section~\ref{sec:discussion} discusses its implications. We summarize our findings in Section~\ref{sec:conclusion}.

\section{\osiris\ Observations and Data Reduction} \label{sec:osiris_observations}

\subsection{IFS Observations and Data Reduction }\label{sec:obs}

We observed the nucleus of IC~342 with the OH Suppressing InfraRed Imaging Spectrograph  \citep[\osiris;][]{Larkin2006} with the Keck I LGS-AO system \citep{vanDam2006, Wizinowich2006}. \osiris{} is a lenslet array IFU spectrograph with a $2048 \times 2048$ Hawaii-2 detector and spectral resolution in the \textit{K} band of R $\sim$ 3600. The observations were conducted on 15 Jan 2025, and consisted of two 900-second object frames, with an intervening 900-second sky frame (i.e., 30 min on object), in the standard object-sky-object sequence, at an airmass of 1.54. The plate scale was 0\farcs02, which at the distance of IC~342 (3.45 Mpc) equates to 0.32 pc pixel$^{-1}$. We observed with the $Kbb$ filter (1965-2381 nm), which is ideally suited for CO bandhead observations. 

Each object frame was reduced using the \osiris{} data reduction pipeline (DRP) version 6.0, using the sky frames for background subtraction. The DRP uses ``Rectification Matrices" which incorporate all the flat fielding, bias and dark frames in one, along with mapping each pixel to a lenslet/spectral element. These matrices are retrieved from the Keck repository; for our data, these were dated Feb 2025. The resulting data cubes have fields of view of 0\farcs38 $\times$ 1\farcs28 (19 $\times$ 64 pixels). 

The data cubes were centered and combined, then trimmed for display purposes, resulting in images with about 1040 spectra. The resulting data cube values were cleaned of spikes and geometric artifacts; these are mainly a product of the DRP. Telluric correction was performed with observation of the A0 star, HIP~12906 (mag$_K = 7.77$). The stellar spectrum is interpolated over the hydrogen absorption features and a representative blackbody is assumed. This is divided through the data cube spectra to remove the telluric absorption features. Flux calibration is performed by integrating the stellar spectrum over a clean region (away from telluric and stellar features), then matched to a reference Vega spectrum.

\subsection{\osiris\ Point Spread Functions}\label{sec:osirsK_psf}

\subsubsection{Synthetic diffraction-limited PSF}\label{sec:synthetic}

The Keck/\osiris\ PSF at $2.3~\micron$ was modeled using the diffraction-limited routine \texttt{osirispsf.pro} from the \osiris\ Data Reduction Pipeline \citep[DRP;][]{Lockhart2019}, which computes the PSF as the squared modulus of the Fourier transform of the complex telescope pupil wavefront. The pupil is constructed via the companion routine \texttt{osirispupil.pro}, encoding the Keck aperture geometry (primary mirror, secondary obscuration, and spider supports) at a physical plate scale $du = \lambda / (N_{\rm pix}\,\delta\theta)$, where $\lambda$ is the effective wavelength, $N_{\rm pix}$ the padded pupil array size, and $\delta\theta$ the detector plate scale in radians per pixel. Zero-padding enforces two conditions: $du < 0.10$~m~pixel$^{-1}$ (adequate pupil sampling) and $N_{\rm pix}\,du > 12$~m (full coverage of the 10~m aperture). A phase ramp is then applied for sub-pixel PSF centroid positioning before the wavefront is Fourier-transformed and rebinned to the native detector scale. For IC~342, we adopted $\delta\theta = 20$~mas~pixel$^{-1}$ and $\lambda = 2.29~\micron$ (blue edge of the CO bandhead fitting window). The resulting PSF image was normalized to unit flux and parameterized in \mge\ form using the \texttt{mge\_fit\_sectors} routine from the \textsc{MgeFit} package \citep{Cappellari2002}; the best-fit parameters are listed in Table~\ref{tab:mge_psf}.

Although this theoretical PSF omits residual wavefront errors and the extended AO halo, the Keck LGS-AO system delivers typical $K$-band Strehl ratios of 30--40\% under normal seeing conditions \citep{vanDam2006}, ensuring a core-dominated PSF that is adequate for dynamical modeling purposes.

\subsubsection{Empirical PSF Constrained from the Standard Star HIP~12906}\label{sec:hip12906}

The empirical \osiris\ PSF was constructed from a dedicated standard star observation obtained immediately after the IC~342 science exposures, with the AO loops closed and identical instrument settings ($20\,\mathrm{mas\,pixel^{-1}}$, \textit{Kbb} filter). We observed HIP~12906 with a short exposure time to avoid saturation; the PSF image was formed by collapsing the normalized, median-combined data cube over 2.050--2.360~\micron, spanning the spectral range used for stellar kinematic extraction and fully contained within the \textit{Kbb} bandpass. 

\begin{figure}
    \centering
    \includegraphics[width=0.9\linewidth]{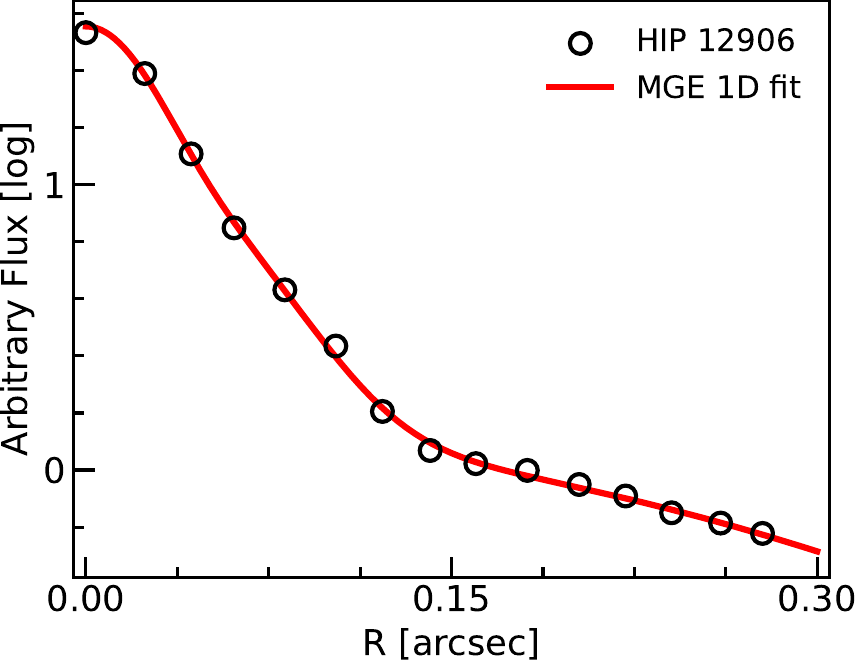}
    \caption{Circularly-averaged radial profile of the empirical Keck/\osiris\ Kbb PSF (open circles), constructed from the standard star HIP~12906 over 2.050--2.360~$\micron$, overlaid with its best-fit MGE model (red; Table~\ref{tab:mge_psf}). The flux is normalized and shown on a logarithmic scale.}
    \label{fig:empirical-PSF}
\end{figure}

The PSF image was parameterized using the \texttt{mge\_fit\_sectors} routine of \textsc{MgeFit}\footnote{v5.0.14: \url{https://pypi.org/project/mgefit/}} \citep{Cappellari2002}; the resulting \mge\ model is reported in Table~\ref{tab:mge_psf} and convolved with all \jam\ model predictions prior to comparison with the data (Section~\ref{sec:dynamical-models}). The circularly-averaged radial profile of this empirical 
\osiris\ PSF and its best-fit MGE model are shown in 
Figure~\ref{fig:empirical-PSF}.

\begin{table}
\caption{Keck/\osiris\ $Kbb$ PSF MGE models (pixel size = 0\farcs02)  \label{tab:mge_psf}}
\begin{tabular}{cccc}
\hline\hline
$j$&(Light fraction)$_j$&$\sigma_j~({\rm arcsec})$&FWHM$^{\rm tot}_{\rm PSF}~({\rm arcsec})$ \\
(1) & (2) & (3) & (4) \\
\hline
\multicolumn{4}{c}{Empirical PSF (standard star HIP~12906)}\\
\hline
1 & 0.148 & 0.018 & \multirow{2}{*}{0.062} \\
2 & 0.852 & 0.036 & \\
\hline
\multicolumn{4}{c}{Synthetic diffraction-limited PSF}\\
\hline
1 & 0.985 & 0.023 & \multirow{3}{*}{0.053} \\
2 & 0.007 & 0.128 & \\
3 & 0.008 & 1.405 & \\
\hline
\end{tabular}
\tablecomments{Columns list: (1) Gaussian component index; (2) fractional contribution to the total light; (3) Gaussian dispersion along the major axis; (4) intensity-weighted mean FWHM of the PSF.}
\end{table}

\begin{table*}
\caption{Optimal XSL Template Stars for IC~342}
\centering
\begin{tabular}{ccccccc}
\hline\hline
No.&Star Name & XSL ID & Sp.~Type & $T_{\rm eff}$~(K) & $\log g$ & OSIRIS \\
(1) & (2) & (3) & (4) & (5) & (6) & (7)\\
\hline
1  &BMB 106                & X0169 & M6         & 3270  & 0.40 & 0.50 \\
2  &HD~111631              & X0903 & M0V        & 3888  & 4.56 & 0.11 \\
3  &{}[W65]~c2             & X0591 & C          & --    & --   & 0.10 \\
4  &IRAS~09484$-$6242      & X0860 & R          & --    & --   & 0.05 \\
5  &{}[W65]~c2             & X0609 & C          & --    & --   & 0.05 \\
6  &HIP~75423              & X0684 & M2V        & 3490  & 4.78 & 0.04 \\
7  &SHV~0518161$-$683543   & X0530 & C          & 4078  & 0.20 & 0.04 \\
8  &SHV~0606101$-$724012   & X0121 & M7         & 3337  & 0.12 & 0.04 \\
9  &HD~57060               & X0116 & O7\,Iafpvar& 31840 & 3.41 & 0.02 \\
10 &HD~46375B              & MD014 & M2.0V      & --    & --   & 0.02 \\
11 &BMB 289                & X0155 & M9         & --    & --   & 0.01 \\
12 &ISO-MCMS~J005332.4$-$730501 & X0031 & --    & 4391  & 1.05 & 0.01 \\
13 &SHV~0542111$-$683837   & X0810 & C          & 4080  & 0.47 & 0.01 \\
\hline
\end{tabular}
\tablecomments{(1) Order number; (2) Star name; (3) XSL DR3 identifier \citep{Verro2022}; (4) Spectral type; (5) Effective temperature; (6) Surface gravity; (7) Fractional contribution to the total light from the optimal template fit to the global OSIRIS spectrum.}
\label{tab:xsl-templates}
\end{table*}

\section{Stellar Kinematics Measurements}\label{sec:losvd}

\subsection{Defining the Galaxy Center}\label{sec:center}

Defining the photometric center of IC~342 requires caution because of the multi-layered dust environment along the line-of-sight (LOS). Optical \hst\ images are unreliable for centroiding by the heavy foreground extinction \citep[$A_{\mathrm{F814W}} \approx 0.74$~mag;][]{Schlafly2011} combined with patchy internal obscuration from the nuclear starburst ring. Archival \hst/WFC3 IR imaging exists in several narrow- and medium-band filters, but each is unsuitable: F128N, F164N, and F187N capture Pa$\beta$, [\ion{Fe}{2}]~$\lambda1.644~\micron$, and Pa$\alpha$ emission, respectively, whose peaks are dominated by the two circumnuclear HII complexes located $\sim$4\arcsec\ east and west of the NSC \citep{Boker1997}, systematically offset from the stellar mass center. F127M and F153M are designed for planetary atmosphere studies, do not reliably trace galactic stellar continuum. 

We therefore adopt the surface-brightness peak of the \osiris\ $Kbb$ white-light image as the coordinate origin, assuming the central BH coincides with the NIR continuum peak. At $2.3~\micron$ \citep[$A_{2.3~\micron} \approx 0.15$~mag;][]{Schlafly2011}, this image is effectively immune to both foreground and internal dust, achieves the highest angular resolution available, and is spatially registered to the science data by construction.

The nuclear centroid was determined as the intensity-weighted  barycenter of the 120 brightest pixels within a $0\farcs2 \times 0\farcs4$ aperture of the OSIRIS $K_{bb}$ white-light image. Reassuringly, this photometric peak coincides to within a single spaxel with both the central peak of the velocity-dispersion map and the zero-crossing of the stellar rotation field, so that the photometric, dispersion, and kinematic centers are mutually consistent. We adopt this common location as the dynamical center and the coordinate origin for all subsequent LOSVD maps, mass models, and figures. We anchor on the OSIRIS continuum rather than the \hst\ imaging because at $2.3~\micron$ the NIR peak is largely immune to the dust-induced centroid shifts affecting the obscured optical and near-IR \hst\ images.

\subsection{\ppxf\ Setup $\&$ The Optimal Stellar Template}\label{sec:optimal_template}

Stellar kinematics of the IC~342 nucleus were extracted from the Keck/\osiris\ \textit{Kbb}-band data cube via the Penalized PiXel-Fitting method (\ppxf\footnote{v8.2.1: \url{https://pypi.org/project/ppxf/}}; \citealt{Cappellari2004, Cappellari2017, Cappellari23}), fitting the full \textit{Kbb} spectral range of 2.050--2.360~\micron\ and thereby including the CO bandhead absorption features at the red end of the bandpass.  Following the results of \citet{Nguyen2025a, Nguyen_2026_4258, Nguyen_2026_M81}, who found that LOSVDs derived from the empirical X-shooter Spectral Library \citep[XSL\footnote{DR3 with 830 spectra from 683 stars:\url{http://xsl.u-strasbg.fr}};][]{Verro2022} agree within 3\% with those from the higher-resolution PHOENIX synthetic library \citep{Husser2013}, we adopt XSL as our primary template source.   The XSL provides broad spectral coverage (3000--25,000~\AA) at $R\sim10,000$, encompassing a diverse range of O--M and evolved stars.

\begin{table}[!t]
\centering
\caption{Keck/\osiris\ Kinematic Data of the IC~342 Nucleus}
\vspace{-3mm}
\begin{tabular}{cccccc}
\hline \hline
$\Delta$R.A.&$\Delta$Decl.&$V$&$\Delta V$&$\sigma$&$\Delta\sigma$\\
(arcsec) & (arcsec) & (\kms) & (\kms) & (\kms) & (\kms) \\
\hline
$-$0.018 &  0.012  &  1.11  & 1.21 & 49.49 & 1.26 \\
 0.002   &$-$0.008 &  0.51  & 1.11 & 49.04 & 1.15 \\
 0.022   & 0.032   &  2.31  & 1.11 & 48.88 & 1.15 \\
 0.022   &$-$0.008 &  0.14  & 1.08 & 48.47 & 1.12 \\
 0.022   & 0.012   &  1.65  & 1.16 & 48.41 & 1.20 \\
 0.002   & 0.012   &  1.31  & 1.07 & 48.29 & 1.11 \\
 0.002   & 0.032   &  2.77  & 1.08 & 48.17 & 1.12 \\
...      &    ...  &    ...  &  ... &  ...   & ...  \\[1mm]
\hline
\end{tabular}
\noindent\parbox{\linewidth}{(This table is available in its entirety in machine-readable form)}
\label{tab:kin}
\end{table}

\begin{figure*}[!th]
 \centering
   	\includegraphics[width=\linewidth]{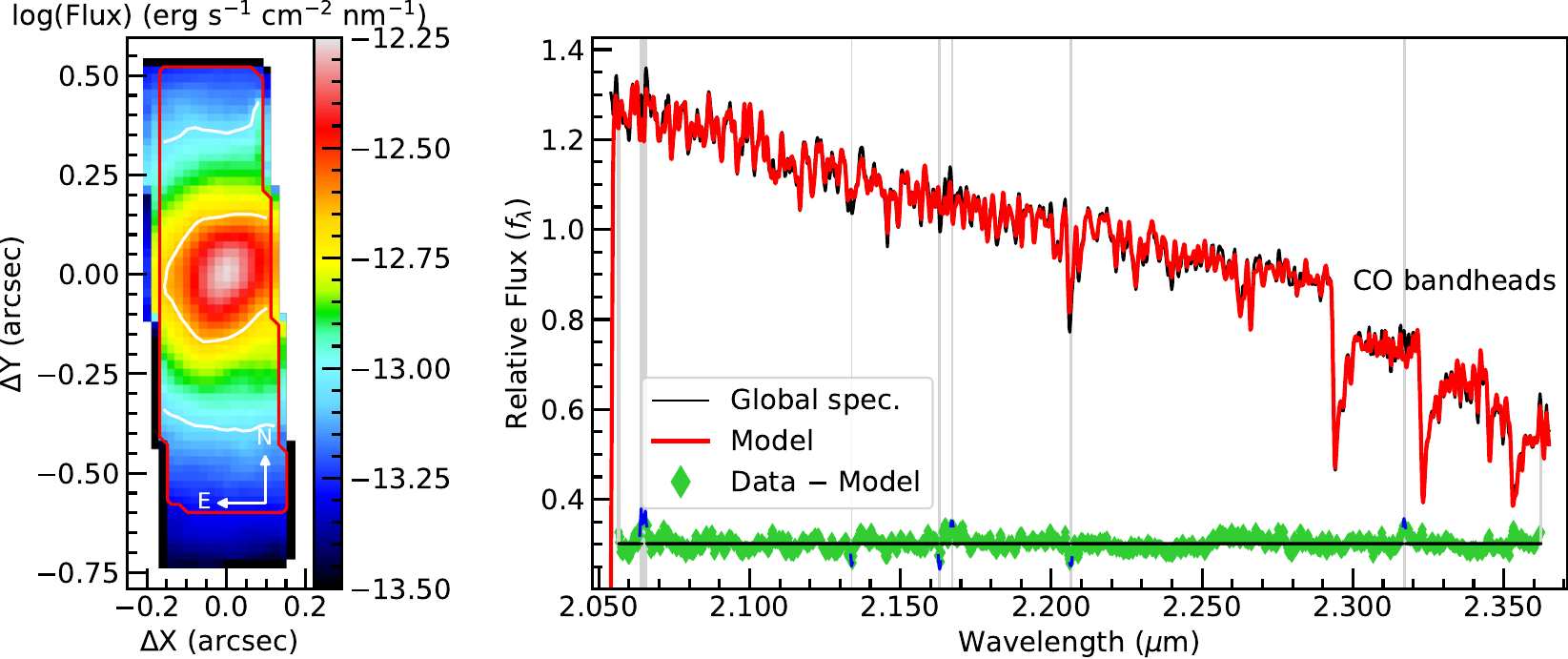} 
	\caption{\textit{Left:} Keck/\osiris\ $Kbb$-band white-light image of the IC~342 nucleus, constructed by collapsing the data cube along the spectral axis (logarithmic stretch); white contours trace the stellar continuum in steps of $1\,\mathrm{mag\,arcsec}^{-2}$. \textit{Right:} Global \ppxf\ fit to the nuclear spectrum obtained by co-adding all spaxels within the red aperture (left panel). The observed spectrum (black) and best-fit XSL template convolved with the optimal LOSVD (red) are shown over the spectral range of 2.050--2.360~\micron; fit residuals are plotted as green dots.}
	\label{fig:Intensity_maps}
\end{figure*}   

\begin{figure*}[!th]
    \centering\includegraphics[width=\textwidth]{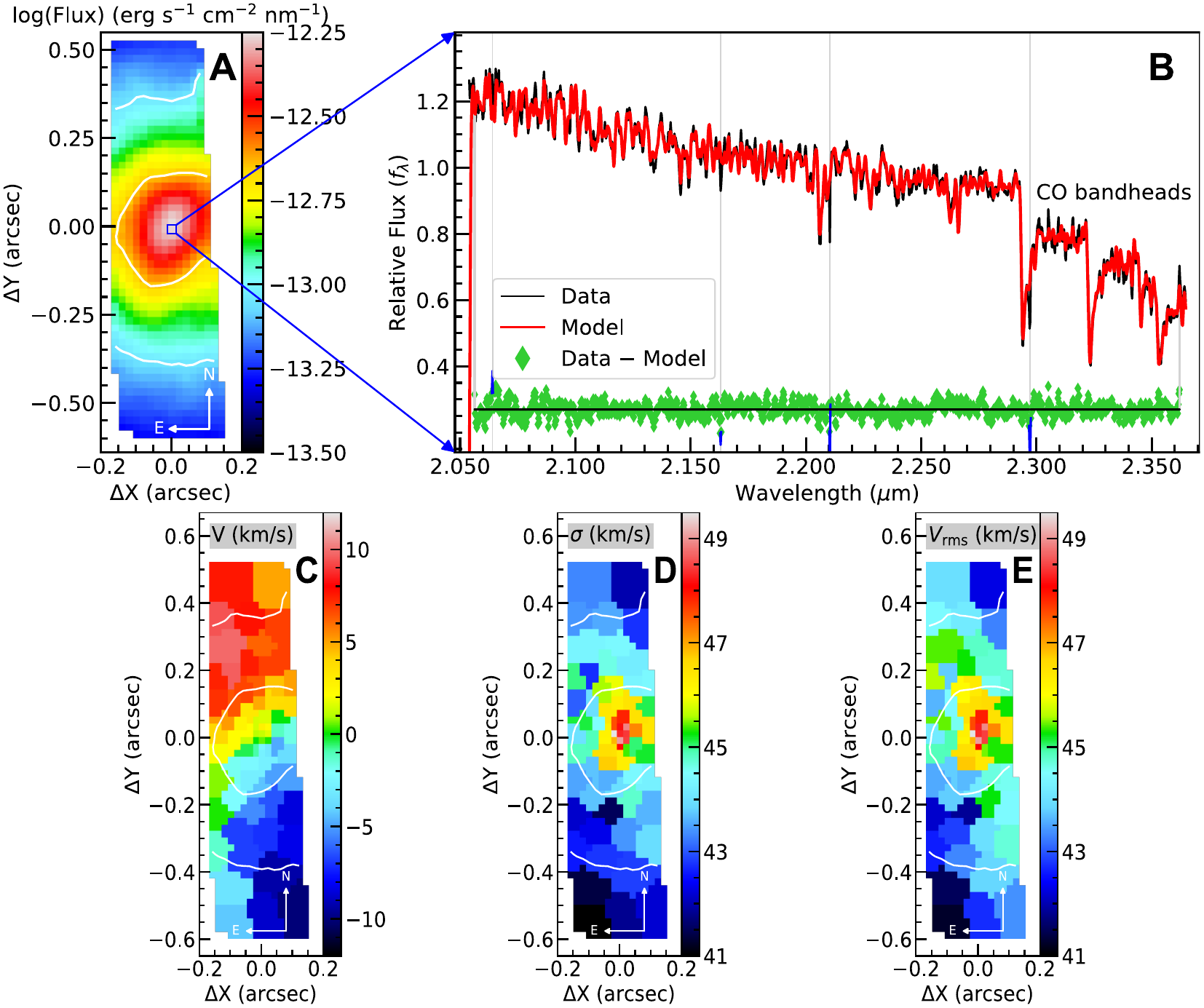}
    \caption{Keck/\osiris\ $Kbb$-band view of the IC~342 nucleus from stellar kinematic extraction. All panels are restricted to the spatial region enclosed by the red aperture in Figure~\ref{fig:Intensity_maps}, where spectra are free from artificial detector artifacts.  \textit{Panel~A:} Nuclear white-light image obtained by spectrally collapsing the \osiris\ data cube (logarithmic intensity scale, same as left panel of Figure~\ref{fig:Intensity_maps}). \textit{Panel~B:} Representative \ppxf\ spectral fit at the spaxel marked in Panel~A. The \osiris\ spectrum (black) is shown across the the spectral range of 2.050--2.360~$\micron$, with the best-fit fixed optimal XSL template overplotted in red; green points show fit residuals, and vertical gray lines indicate spectral regions excluded from the fit. \textit{Panels~C--E:} 2D stellar kinematic maps of the LOS velocity $V$, velocity dispersion $\sigma$, and $V_\mathrm{rms} \equiv \sqrt{V^2 + \sigma^2}$, with white contours overlaid at $1~\mathrm{mag~arcsec^{-2}}$ intervals tracing the stellar continuum morphology.}
    \label{fig:kin-map}
\end{figure*}

To account for the mixed stellar populations in the IC~342 nucleus and mitigate template-mismatch bias \citep{Marconi2000, Silge2003}, we derived a fixed optimal XSL template from the full DR3 library of 830 spectra from 683 stars and from a large spatial region's integrated spectrum, namely the  global spectrum. The global spectrum was formed by co-adding all spaxels within the red aperture in the left panel of Figure~\ref{fig:Intensity_maps}, achieving $\mathrm{S/N} \sim 240$ per spectral pixel. It was then logarithmically rebinned to a constant velocity scale of 32~\kms~$\mathrm{pixel}^{-1}$ following Eq.~8 of \citet{Cappellari2017}, and the XSL templates were broadened to the instrumental resolution by convolution with a Gaussian kernel whose dispersion was computed via Eq.~5 of \citet{Cappellari2017}. The fixed optimal XSL template was subsequently derived by fitting these broadened templates to the global spectrum with \ppxf\ under a Gaussian LOSVD (\texttt{moments} $= 2$) to obtain the rotational velocity $V$ (w.r.t. the systemic velocity $V_{\rm sys}$ of the nucleus) and velocity dispersion $\sigma$ as our primary kinematic quantities. 

Since additive Legendre polynomials can alter the depth of the CO bandhead absorption features and bias the recovered LOSVDs in the \textit{Kbb}-band \osiris\ data, we disabled them in all \ppxf\ fits (\texttt{degree} $= -1$). Multiplicative polynomials (\texttt{mdegree}) of orders 0--4 were tested to correct for residual continuum shape, sky subtraction, and calibration systematics; the recovered kinematics stabilize and become insensitive to polynomial order at \texttt{mdegree} $\geq 2$ across the full fitting window ($2.050$--$2.360~\micron$), so we adopt \texttt{mdegree} $= 2$ as our fiducial choice to ensure stable LOSVD recovery while minimizing overfitting.

This fitting process produced a fixed optimal stellar template, defined as a linear combination of 13 giant-star spectra that faithfully reproduce the full wavelength range of the \osiris\ spectra (Figure~\ref{fig:Intensity_maps}, right panel). The fixed optimal template is dominated by four giant stars: BMB~106 (contributing 50\% of the flux), HD~111631 (11\%), [W65]c2 (10\%), and IRAS 09484-6242 (5\%). These four stars alone provide a relatively good fit to the global spectrum; however, the remaining nine stars contribute an additional 34\% of the flux. The properties and fractional flux contributions of all 13 XSL template stars are listed in Table~\ref{tab:xsl-templates}. The global best-fit kinematics are $V = 34 \pm 2$~\kms\ (heliocentric) and $\sigma = 41 \pm 2$~\kms; uncertainties are the $1\sigma$ scatter from 200 Monte Carlo realizations \citep{Cappellari2004}.  While this \osiris\ heliocentric velocity is consistent with the NED\footnote{\url{https://ned.ipac.caltech.edu}} value, the central velocity dispersion measured here exceeds that of \citet{Boker1999} by a factor of $\sim$1.2. This offset is expected: \citet{Boker1999} reported an aperture-integrated $\sigma$ over the NSC using seeing-limited long-slit spectroscopy on the NASA Infrared Telescope Facility (IRTF), whereas our \osiris\ measurements resolve the central few parsec individually, where the gravitational influence of any central dark mass elevates the local velocity dispersion above the NSC-integrated value. The elevated central $\sigma$ is therefore consistent with (and indeed motivates) our dynamical search for a central dark mass in IC~342.

\subsection{Two-Dimensional Stellar Kinematic Maps}\label{sec:kinemap}

Two-dimensional stellar LOSVDs across the IC~342 nucleus were extracted by first adaptively binning spaxels with \powerbin\footnote{v1.1.11: \url{https://pypi.org/project/powerbin/}} \citep{Cappellari2025} to a target S/N of 150 per spectral pixel. Spaxels within $r_\mathrm{unbin} = 0\farcs05$ (close to one PSF FWHM of OSIRIS element) were left unbinned to preserve the spatial resolution critical for kinematic measurements in the innermost nuclear region, where individual spaxel S/N exceeds 90 per spectral pixel; spaxels at $r > r_\mathrm{unbin}$ were grouped adaptively, yielding $N = 70$ bins in total. Each resulting spectrum was logarithmically rebinned in wavelength and fitted with \ppxf\ \citep{Cappellari2017}, using the fixed optimal XSL template broadened to the instrumental resolution with second-degree multiplicative Legendre polynomials to correct for residual continuum shape differences, sky subtraction, and calibration systematics (following Section~\ref{sec:optimal_template}; disable additive Legendre polynomials and set \texttt{moments} $= 2$), recovering the stellar LOS velocity $V$, dispersion $\sigma$, and their associated uncertainties.

The resulting kinematic maps are presented in Figure~\ref{fig:kin-map}: the logarithmic integrated intensity (panel A), a representative \ppxf\ fit to a central spaxel (panel B), and 2D maps of $V$ (panel C), $\sigma$ (panel D), and root-mean-squared velocity \Vrms\ (panel E). The systemic heliocentric velocity, derived from the \osiris\ velocity map, is $v_{\rm sys} = 35 \pm 2$~\kms, which is 4~\kms\ larger than the NED value. After subtracting $v_{\rm sys}$, the nucleus exhibits a rotation reaching $|V| \approx 10 \pm 2$~\kms\ at the edge of the \osiris\ FOV.

We measure a global kinematic position angle of PA$_{\rm kin} = 38.4^\circ \pm 1^\circ$, consistent with the kinematic position angle of $\mathrm{PA} = 37^{\circ}$ derived from the \ion{H}{1} velocity field by \citet{Crosthwaite_2000}. Both $v_{\rm sys}$ and PA$_{\rm kin}$ were derived from the velocity map using the PAFIT\footnote{v2.0.8: \url{https://pypi.org/project/pafit/}} package \citep{Krajnovic2006}. The $\sigma$ profile, which dominates over $V$, rises from $\sim$$42 \pm 1.3$~\kms\ at $r \ge 0\farcs2$, to a central peak of $\sim$$49.0 \pm 1.3$~\kms. This central increase in both $\sigma$ and \vrms\ provides strong evidence for the presence of a central compact dark mass. We present these stellar-kinematic measurements in Table~\ref{tab:kin}.

\subsection{No Evidence of AGN in the Keck/OSIRIS Kbb IFU}\label{sec:no-AGN}

Following recent VLT/SINFONI \citep{Cappellari2009}, VLT/MUSE \citep{Simon2024}, and JWST/NIRSpec \citep{Nguyen_2026_4258, Nguyen_2026_M81} studies, we checked for AGN continuum dilution of the CO bandhead features, which can bias the recovered $\sigma$ and mimic spurious stellar population or $M/L$ variations in the nucleus. We quantified this using the stellar line-strength parameter $\gamma \equiv \mathrm{stars}/(\mathrm{stars + AGN})$ \citep{vanderMarel1993, Cappellari2009}, which approaches unity where AGN contamination is negligible and drops below the unity toward the center if an AGN continuum is present. We find $\gamma \approx 1$ across the entire \osiris\ FOV, indicating no detectable AGN contribution in the nucleus of IC~342 in the NIR wavelength; consequently, unlike \citet{Nguyen_2026_4258, Nguyen_2026_M81, Simon2024}, no AGN subtraction is applied to the galaxy photometric model.

This conclusion is further supported by multiwavelength observations of the IC~342 nucleus. The central X-ray source detected by \textit{Chandra} and \textit{XMM-Newton} has a luminosity of $L_X \lesssim 10^{39}$~erg~s$^{-1}$ \citep{Mak2008, Zhang2024}, well within the regime expected for normal star-forming galaxies and consistent with high-mass X-ray binary (HMXB) populations associated with the nuclear starburst. The source is also spatially extended rather than point-like, arguing against a well-defined AGN \citep{Mak2008}. Similarly, the extended radio emission from the IC~342 nucleus is fully consistent with starburst-driven synchrotron emission rather than AGN-powered jets \citep{Balser2017}. We therefore conclude that any AGN contribution to the nuclear emission of IC~342 is negligible, and that the $\gamma \approx 1$ recovered from our \osiris\ data reflects genuine stellar-continuum dominance rather than AGN dilution.

\begin{figure}[h]
    \centering
    \includegraphics[width=0.45\textwidth]{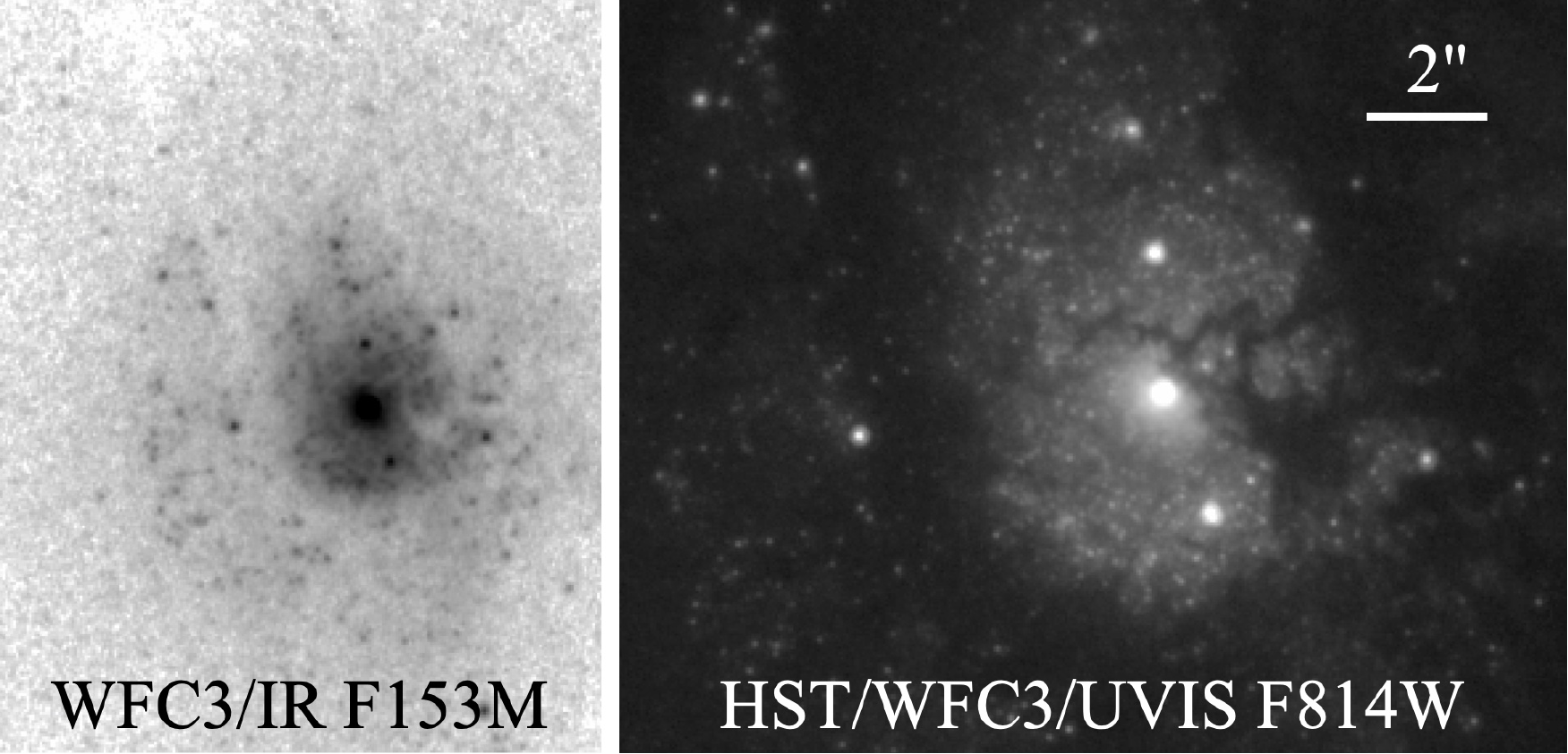}
    \caption{Nucleus of IC~342. The central dense nuclear star cluster is surrounded by a somewhat ill-formed and knotty $\sim$3--4$\arcsec$ ($\sim$48--64~pc) nuclear `disk'. (Images from HST Proposal ID 12163. PI: Barth.)}
    \label{fig:IC342_nucleus-2}
\end{figure}

\begin{figure}[h]
    \centering\includegraphics[width=0.48\textwidth]{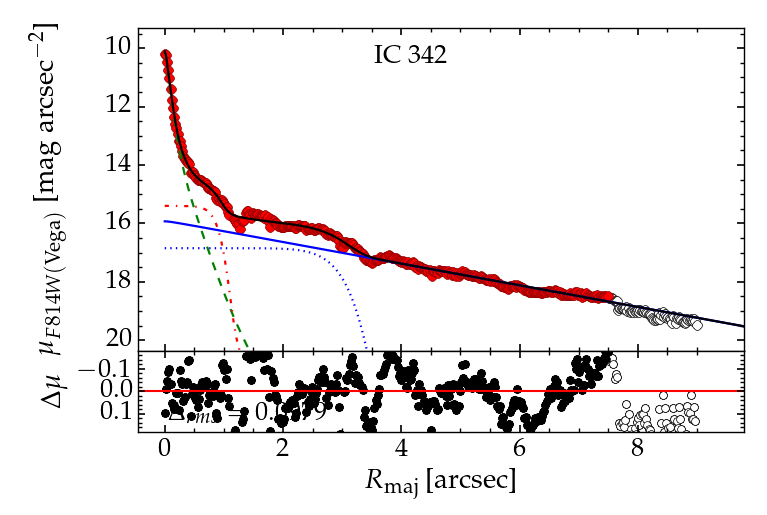}
    \caption{Four-component (three S\'ersic + exponential disk) surface brightness decomposition of the IC~342 nucleus from \hst/WFC3 UVIS F814W imaging. \textit{Upper panel:} The observed major-axis profile (filled circles) is overlaid with the best-fit total model (solid line), decomposed into an inner S\'{e}rsic component tracing the NSC (dashed line), an outer S\'{e}rsic component describing the central stellar distribution (dash-dotted line), and an exponential disk (dotted line). \textit{Lower panel:} Fit residuals (\texttt{data$-$model}) in mag~arcsec$^{-2}$.}
    \label{fig:ic342_f814w_SB}
\end{figure}

\section{Galaxy Photometric Model}\label{sec:photomodel}

The \hst/WFC3/UVIS F814W surface brightness profile of IC~342 was modeled using archival imaging (Proposal ID 12163; PI: Barth) with a pixel scale of $0\farcs04$ pixel$^{-1}$. This image was adopted for the structural analysis because it provides the highest available spatial resolution (PSF FWHM $=0\farcs069$) while being less affected by the galaxy's patchy internal dust than shorter-wavelength optical images. Figure~\ref{fig:IC342_nucleus-2} shows the galaxy light at 0.814 and 1.53~$\mu$m.  Dust lanes and heavily obscured regions were masked using \textsc{SAOImage ds9} prior to the isophote analysis, which used the {\sc mskregions} task in \textsc{IRAF}.

We use a 1D profile rather than a full 2D MGE fit to average the light distribution, thereby smoothing out the uneven, non-axisymmetric features from the patchy dust. The isophotes were fit using the \textsc{IRAF} tasks {\sc Isofit} and {\sc Cmodel} \citep{Ciambur2015}, while allowing the ellipticity and position angle to vary with radius but keeping the photometric centre fixed at the kinematic centre (Section~\ref{sec:center}). This constraint was imposed because patchy dust extinction produces apparent shifts in the isophotal centroids with radius that would otherwise bias a fully free-geometry fit. The resulting F814W surface brightness profile, shown in Figure~\ref{fig:ic342_f814w_SB}, was calibrated to the Vega magnitude system using a WFC3/UVIS F814W photometric zero-point of 24.747~mag.

We performed a decomposition of the geometric-mean and major-axis surface brightness profile using a Python~3 version of the \textsc{Profiler} code \citep{Ciambur2016}\footnote{Updated to Python~3 by A.~W.~Graham.}. The light was modeled with four S\'ersic  components. During the fitting process, each intrinsic S\'ersic profile was first convolved with the image PSF, and the convolved profiles were then summed to reproduce the observed (PSF-blurred) light profile.

The best-fitting major-axis model comprises: an inner S\'ersic component tracing the NSC with $\mu_{e,1} = 12.31$~mag~arcsec$^{-2}$, $r_{e,1} = 0\farcs14$ ($\approx$2.34~pc), and $n_1 = 2.37$; a second S\'ersic component with $\mu_{e,2} = 15.44$~mag~arcsec$^{-2}$, $r_{e,2} = 0\farcs63$, and $n_2 = 0.12$ tracing the puffy stars and star cluster around the NSC; an immediate S\'ersic component with $\mu_{e,3} = 16.87$~mag~arcsec$^{-2}$, $r_{e,3} = 2\farcs05$, and $n_3 = 0.10$ that captures what may likely be considered a nuclear disk in less well-resolved galaxies located at greater distances \citep{Balcells2007}. The large-scale disk is described by the outer S\'ersic component, which is consistent with an exponential component with $\mu_0 = 15.90$~mag~arcsec$^{-2}$ and $h = 2\farcs93$ ($\approx$49~pc). Fitting the geometric-mean axis yields magnitudes of 13.92 and 14.66 mag for the inner and middle components, respectively, and thus the inner component makes up 66.4 percent of the combined F814W flux of these two components.

Our four-component decomposition yields NSC parameters that differ substantially from the two-component {\sc GALFIT} fit (i.e., S\'{e}rsic + exponential) of \citet{Carson15}, who report only for the central excess (NSC), $n = 3.79^{+0.48}_{-0.40}$ and $r_e = 1.38^{+0.19}_{-0.13}$~pc from the same F814W image but fitting just the inner $1\farcs3$, While the effective radii agree by $\sim$70\% (2.34 vs. 1.38 pc), the S\'{e}rsic indices differ by $\sim$60\% (2.37 vs. 3.79). Discrepancy is expected given \citeauthor[][their Figure~6]{Carson15} adopted a single S\'{e}rsic plus exponential disk model, so their model's inner component absorbs flux from the extended inner stellar envelope ($r \sim 0\farcs4$--$0\farcs8$), that our fit separates as an independent component from the dense NSC, inflating both $n$ and the apparent magnitude relative to the innermost component of our model. Indeed, the summed magnitude ($m_{\rm F814W} \approx 13.47$~mag) of our inner two S\'{e}rsic components, coming from the fit to the geometric-mean radial profile, and thus taking the ellipticity profile into account, agrees to within 0.04~mag with the \citeauthor{Carson15} value ($m_{\rm F814W} = 13.43$~mag).
% confirming that the two decompositions are photometrically consistent.  
% 
% We caution, however, that the second and third S\'{e}rsic components, extending beyond $r \sim 1\arcsec$, do not represent physically distinct structures but instead reflect the complex morphology of the IC~342 nucleus at optical wavelengths. Their parameters should, therefore, not be over-interpreted;
However, not separating these two features, evident in the image (Figure~\ref{fig:IC342_nucleus-2}), forces the model with a single inner S\'ersic function to have a high-$n$ tail in an effort to capture the excess flux at intermediate radii ($r\sim 0\farcs4$--$0\farcs8$), in turn demanding a high central concentration/spike at radii well below the image resolution. Not only does it yield a stronger amplitude in the snake-like residual light profile, but it impacts the optimal extraction of the BH mass.

We caution that while our second and third S\'{e}rsic components broadly capture the complex morphology seen at optical wavelengths, their parameters should not be over-interpreted.  Rather, their inclusion enabled us to better isolate and quantify the innermost component, required for the \jam\ dynamical modeling. Beyond $\sim$3$\arcsec$, the decline in the light profile shown in Figure~\ref{fig:ic342_f814w_SB} is somewhat due to dust (see Figure~\ref{fig:IC342_nucleus-2}).

Our best-fit S\'{e}rsic functions were converted into an \mge\ model using the \texttt{mge\_fit\_1d} routine \citep{Cappellari2002}, applied to the circularized 1D major-axis profile sampled on a logarithmic radial grid. The observed axial ratio $q'_j$ of each Gaussian component was assigned from the ellipticity measured at the corresponding photometric scale, as required by the \jam\ dynamical models (Section~\ref{sec:dynamical-models}). The resulting \mge\ model is listed in Table~\ref{table:mge-table}.

\begin{table}
\centering
\caption{\hst/WFC3 UVIS F814W MGE model}    
\begin{tabular}{cccc} 
\hline\hline
$j$ &$\Sigma_{\star,j}$ (\Lsun\ ${\rm pc^{-2}})$ &$\sigma_j$ (arcsec) &$q'_j=b_j/a_j$\\
(1) & (2) & (3) & (4)\\ 
\hline
1  &2986143 &0.025 &0.94\\
2  &1408677 &0.057 &0.94\\
3  &469108  &0.126 &0.94\\
4  &80265   &0.428 &0.94\\
5  &17084   &1.765 &0.94\\
6  &3825    &4.879 &0.94\\
\hline
\end{tabular}
\parbox[t]{0.47\textwidth}{\small{\bf Notes:} Columns list: (1) Gaussian component index; (2) Luminosity surface density; (3) Gaussian dispersion along the major axis; (4) Observed axial ratio} 
\label{table:mge-table}
\end{table}

\section{Dynamical Modelling}\label{sec:dynamical-models}

\subsection{Jeans Anisotropic Models}\label{sec:jam}

The observed root-mean-squared velocity (\Vrms) maps of IC~342 were modeled with both the cylindrically aligned (\jamcyl; \citealt{Cappellari08}) and spherically aligned (\jamsph; \citealt{Cappellari2020}) implementations of \jam, each under the assumption of an axisymmetric gravitational potential. The two implementations bracket the range of physically plausible velocity-ellipsoid orientations and thereby quantify how this modeling choice affects the inferred \Mbh.

In \jamcyl, the velocity ellipsoid is tied to cylindrical coordinates $(R,\,\phi,\,z)$, with anisotropy $\beta_z \equiv 1 - (\sigma_z/\sigma_R)^2$ ($\sigma_z \neq \sigma_R = \sigma_\phi$); in \jamsph, it is tied to spherical coordinates $(r,\,\theta,\,\phi)$, with anisotropy $\beta_r \equiv 1 - (\sigma_\theta/\sigma_r)^2$ ($\sigma_r \neq \sigma_\theta = \sigma_\phi$). Both are computed through the \texttt{jam\_axi\_proj} routine of \textsc{JamPy} \citep{Cappellari2020} via \texttt{align=`cyl'} or \texttt{align=`sph'}, producing the projected second-moment map $\langle V^2_\mathrm{los}\rangle^{1/2} \equiv$~\Vrms. For \jamcyl, the intrinsic second moments $\langle\sigma_z^2\rangle$ and $\langle\sigma_\phi^2\rangle$ are approximated numerically to accelerate model evaluation; \jamsph\ instead computes the BH gravitational potential exactly without approximation.

The four free parameters common to both models are: \Mbh, sampled logarithmically; and three linear parameters: \mlhst, the inclination $i$ (converted internally to the minimum intrinsic axial ratio $q_\mathrm{min}$), and the orbital anisotropy (reported as dispersion ratios $\sigma_z/\sigma_R$ and $\sigma_\theta/\sigma_r$ for \jamcyl\ and \jamsph, respectively). In both \jamcyl\ and \jamsph, \mlhst\ is parameterized to vary spatially as a linear function of radius (Section~\ref{sec:default_mlvary}) rather than being held constant.  For the orbital anisotropy, \jamcyl\ instead assumes a spatially uniform value, whereas \jamsph\ adopts a logistic function of radius (Section~\ref{sec:logistic_anisotropy}). Search ranges and initial guesses for all free parameters are given in Table~\ref{tab:adamet}.

In total, we perform four \jam\ model runs to bracket the main sources of systematic uncertainty in our dynamical analysis. These span two choices of velocity-ellipsoid alignment (e.g., \jamcyl\ and \jamsph) and two representations of the \osiris\ PSFs in $Kbb$ band (e.g., empirical or diffraction-limited) as characterized in Section~\ref{sec:osirsK_psf} and listed in Table~\ref{tab:mge_psf}.

\subsection{Inclination Prior and MGE Deprojection  Degeneracy}\label{sec:inclination}

The photometric deprojection from a 2D surface brightness distribution into a 3D intrinsic mass density is formally non-unique for any galaxy not viewed precisely edge-on \citep{Rybicki1987, Magorrian1999}. In the MGE formalism, one particular deprojection is obtained analytically for each assumed inclination $i$ \citep{Cappellari2002}, but the resulting intrinsic density can vary substantially across the family of solutions consistent with the observed surface brightness when $i$ is small \citep{Lablanche2012}. This degeneracy is particularly severe for nearly face-on systems: \citet{Smith2019} and \citet{Dominiak2026MNRAS.548ag546D} demonstrated for NGC~524 ($i \lesssim 40^\circ$) and NGC~1387 ($i \lesssim 13.^\circ6$), respectively, that low inclinations introduce strongly asymmetric posteriors and inflate the uncertainties on both \Mbh\ and $M/L$, while \citet{Lablanche2012} showed explicitly that JAM models of a simulated galaxy at $i = 25^\circ$, can produce intrinsic density distributions that are visually indistinguishable in projection yet physically very different in 3D.

IC~342 is nearly face-on at $i \approx 31^\circ$ \citep{Crosthwaite2000}, placing it squarely in this problematic regime. Treating $i$ as a free parameter in the JAM fit would therefore not constrain the inclination from the stellar kinematics alone, but would instead allow the MCMC sampler to explore a wide range of degenerate 3D mass distributions, inflating the uncertainties on \Mbh\ without providing additional physical insight. We therefore fixed $i = 31^\circ$ throughout all JAM model variants, adopting the kinematic inclination measured from the HI velocity field by \citet{Crosthwaite2000}, which is well constrained independently of the stellar dynamics and represents the most reliable available inclination estimate for this galaxy.

\subsubsection{Linearly Varying \mlhst\ Profile}\label{sec:default_mlvary}

While the heavy foreground extinction toward IC~342 ($A_{\mathrm{F814W}} \approx 0.74$~mag) precludes verifying any radial stellar population variation with optical data, our JAM models consistently disfavor a spatially constant \mlhst\ when fitting the \osiris\ stellar kinematics. We therefore adopt a radially varying \mlhst\ profile that increases linearly from a central value $M/L_{\rm F814W}^{0\arcsec}$ to an outer value $M/L_{\rm F814W}^{0\farcs15}$ at $r = 0\farcs15$, beyond which \mlhst\ is held fixed at $M/L_{\rm F814W}^{0\farcs15}$. The transition radius of $0\farcs15$ corresponds to the shorter dimension of the \osiris\ FOV, within which the 2D stellar kinematics provide the tightest spatial constraints on the \mlhst\ gradient. See Section~\ref{sec:mlconst} for more details

In practice, this profile is implemented in \jam\ by assigning a distinct (\mlhst)$_j$ to each Gaussian component of dispersion $\sigma_j$ in the MGE model (Table~\ref{table:mge-table}):
% \begin{equation}
% \left(\frac{M}{L_{\rm F814W}}\right)_j = 
% \begin{cases}
%    M/L_{\rm F814W}^{0\arcsec} + 
%    \dfrac{M/L_{\rm F814W}^{0\farcs15} - 
%    M/L_{\rm F814W}^{0\arcsec}}{0\farcs15}\,\sigma_j, 
%    & \sigma_j < 0\farcs15\\[6pt]
%    M/L_{\rm F814W}^{0\farcs15}, & \sigma_j \ge 0\farcs15
% \end{cases}
% \end{equation}
% 
\begin{equation}
\Upsilon_j = 
\begin{cases}
    \Upsilon^{0\arcsec} + 
    \dfrac{\Upsilon^{0\farcs15} - \Upsilon^{0\arcsec}}{0\farcs15}\,\sigma_j, 
    & \sigma_j < 0\farcs15\\[10pt]
    \Upsilon^{0\farcs15}, & \sigma_j \ge 0\farcs15
\end{cases}
\end{equation}
with $\Upsilon = M/L_{\rm F814W}$.

\begin{figure*}[]
\centering
    \includegraphics[width=0.85\textwidth]{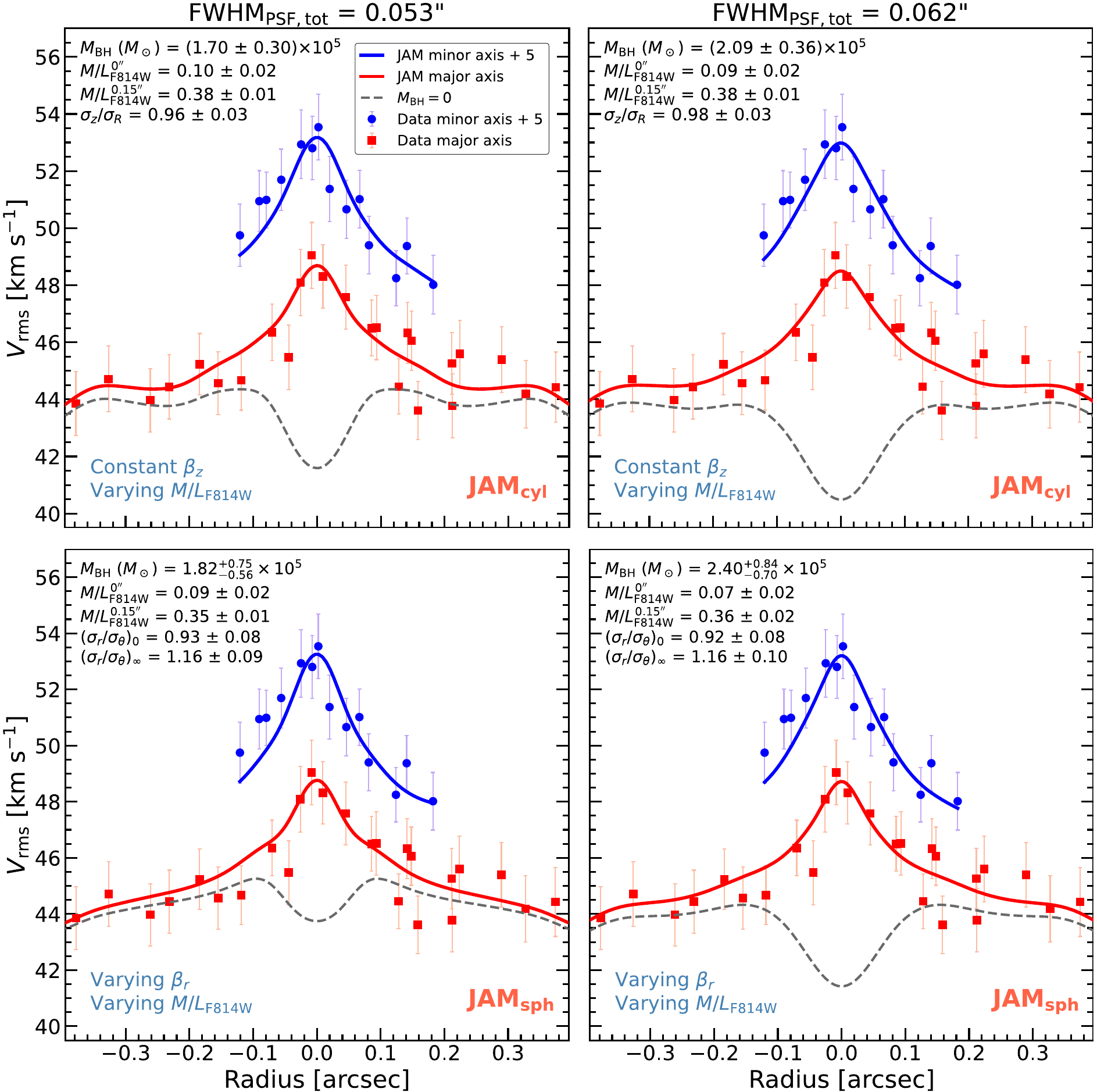} 
    \caption{1D stellar kinematics extracted from the OSIRIS $Kbb$  data cubes of IC~342, shown along the major axis (red, offset by $+5$~\kms\ for clarity) and the minor axis (blue), overlaid with the best-fit JAM$_{\rm cyl}$ (top) and JAM$_{\rm sph}$ (bottom) models, for the two PSF models ($\mathrm{FWHM}=0\farcs053$, left; $0\farcs062$, right), and with the $M_\mathrm{BH}=0$ model shown as the grey dashed curve.}
\label{fig:1D-profile}
\end{figure*}

\begin{figure*}[!t]
     \includegraphics[width=\textwidth]{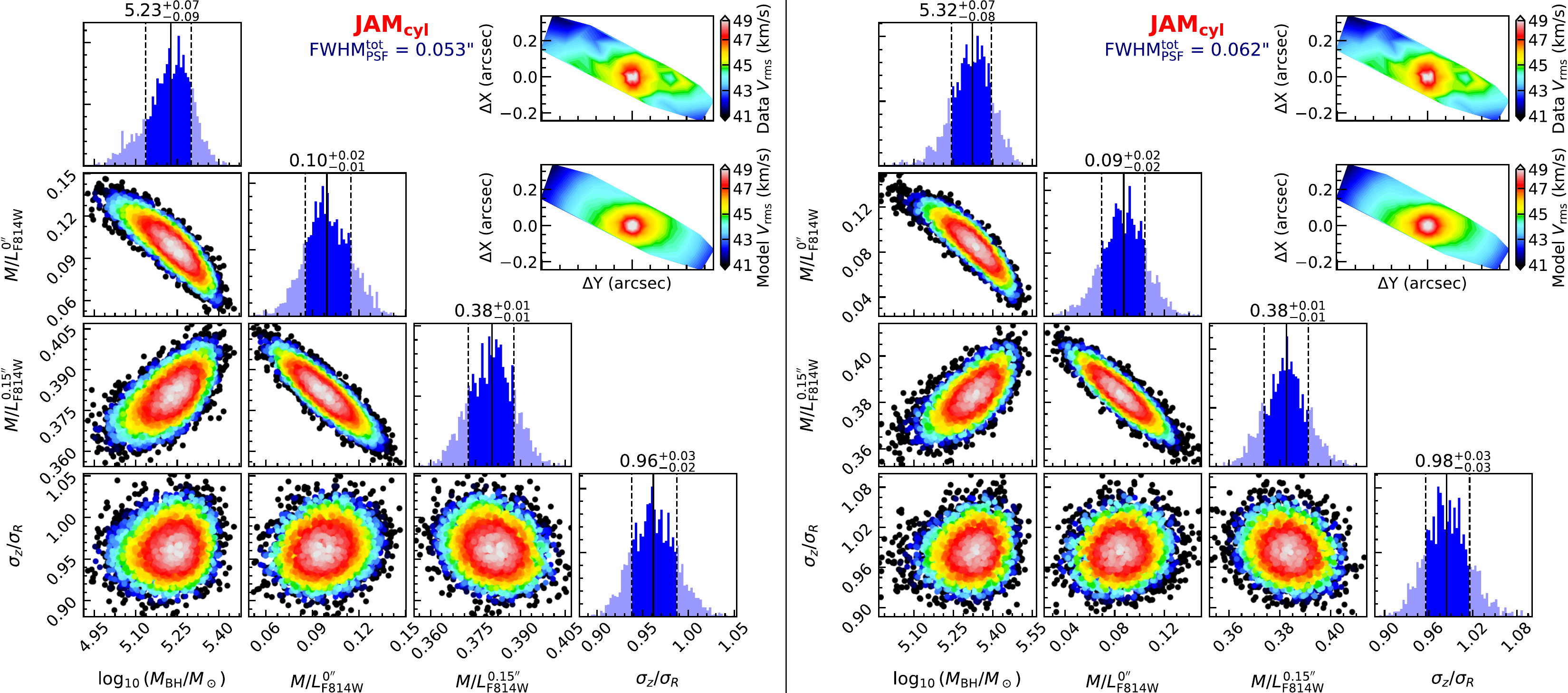}  
     \caption{Corner plot summarizing the \adamet\ MCMC posteriors for the \jamcyl\ fits to the \osiris\ nuclear kinematics of IC~342, with all chains shown post-burn-in. Each triangle plot shows the 2D projected probability distributions of the model parameters, with 1D marginalized histograms along the diagonal; solid and dashed vertical lines indicate the best-fit values and $1\sigma$ uncertainties, respectively.  \textit{Insets:} Direct comparison of the observed (top) and model (bottom) \vrms\ maps on a common velocity stretch, illustrating how closely each model reproduces the kinematics across the \osiris\ FOV.  These maps are symmetrized assuming axisymmetry and rotated so that the major and minor axes align with the horizontal and vertical directions using the {\tt symmetrize\_velfield} routine from the PLOTBIN package\footnote{v3.1.7: \url{https://pypi.org/project/plotbin/}}, and are displayed with {\tt plot\_velfield}.}
     \label{fig:adamet_cyl}
\end{figure*}

\subsubsection{Logistic Anisotropy for \jamsph\ Models}
\label{sec:logistic_anisotropy}

The radial anisotropy profile we adopt for the \jamsph\ models is motivated by a now well-established empirical pattern: stellar orbits across galaxies of very different masses and morphologies tend to settle into a tangentially biased configuration within the BH's SOI, while remaining closer to isotropic or mildly radial further out \citep[][their Figure~10]{Cappellari2026}. The leading physical explanation for this pattern in massive ellipticals invokes binary SMBH scouring during dry mergers, a process that preferentially depletes radial orbits near the center \citep[e.g.,][]{Milosavljevic2001, Rantala2024}; the fact that a similar tangential bias appears even in lower-mass, fast-rotating systems with no comparable merger history suggests the behavior may be a more general feature of galactic nuclei rather than a merger relic specifically. Rather than treating this as a qualitative expectation only, we encode it directly as a prior on the model, following the implementation of \citet{Simon2024}, in which the anisotropy $\beta_r(r) \equiv 1-\sigma_\theta^2/\sigma_r^2$ varies with $\log r$ according to a logistic function: 
\begin{equation}
\beta_r(r) = \beta_{r,0} + 
\frac{\beta_{r,\infty} - \beta_{r,0}}
{1 + (r_a/r)^\alpha},
\end{equation}
with $\beta_{r,0}$ and $\beta_{r,\infty}$ setting the inner and outer anisotropy plateaus and $r_a$ marking the radius of transition between them; for ease of interpretation, we report this throughout as the equivalent dispersion ratio $\sigma_r/\sigma_\theta = (1-\beta_r)^{-1/2}$.

The value of imposing this particular functional form rests on its track record: \citet{Simon2024}, \citet{Nguyen_2026_4258}, and \citet{Nguyen_2026_M81} applied the identical prescription to M87, NGC~4258, and NGC~3031, respectively, and recovered \Mbh\ values in each case consistent with independent literature measurements within their reported uncertainties. These three systems span a substantial range in morphology, from the giant elliptical M87 to the spirals NGC~4258 and NGC~3031, providing a reasonably broad test of the prescription's validity prior to its application here.  We adopt the same prior ranges here on that basis, fixing $\alpha=2$ to keep the parameter space manageable and bounding the inner and outer dispersion ratios to $0.5<(\sigma_r/\sigma_\theta)_0<1$ and $1<(\sigma_r/\sigma_\theta)_\infty<1.3$, consistent with the tangentially-biased-to-isotropic transition described above. The transition radius itself is held fixed at $r_a=0\farcs09$, approximately matching the wavelength-stable effective radius of the nuclear cluster reported by \citet{Carson15}. This value sits outside the SOI of the IC~342 BH ($R_\mathrm{SOI}\approx0\farcs05$; Section~\ref{sec:smbh}) and the OSIRIS PSF FWHM so that the logistic curve has room to complete its transition from the inner tangential regime to the outer, more isotropic one within the radial range actually constrained by the data.

\begin{figure*}[!t]
\centering
     \includegraphics[width=0.8\textwidth]{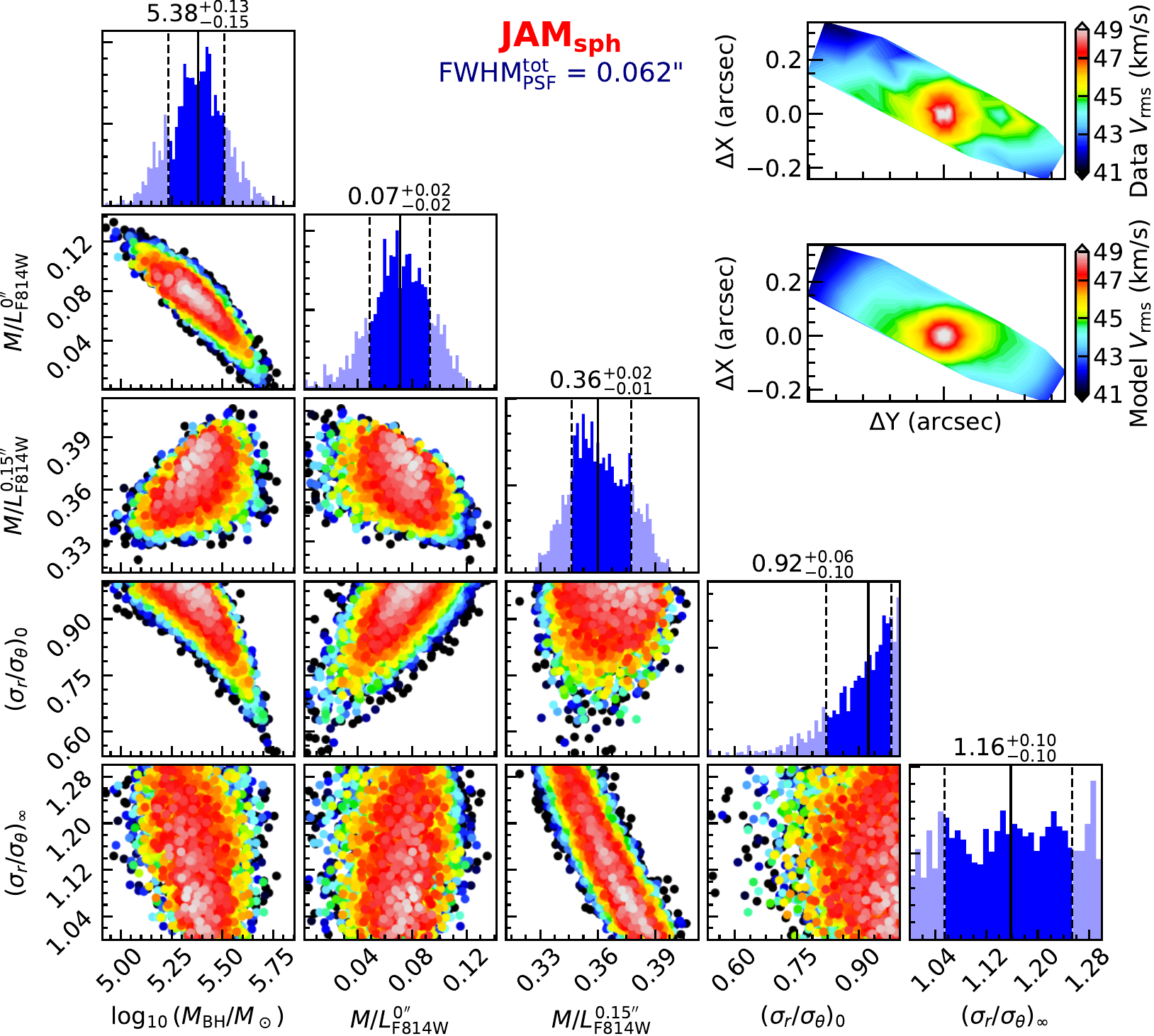} 
     \caption{Same as Figure~\ref{fig:adamet_cyl} but for the \jamsph\ fit with the empirical OSIRIS PSF (Table~\ref{tab:mge_psf}).}
     \label{fig:adamet_sph_psf1}
\end{figure*}

\begin{figure*}[!t]
\centering
     \includegraphics[width=0.8\textwidth]{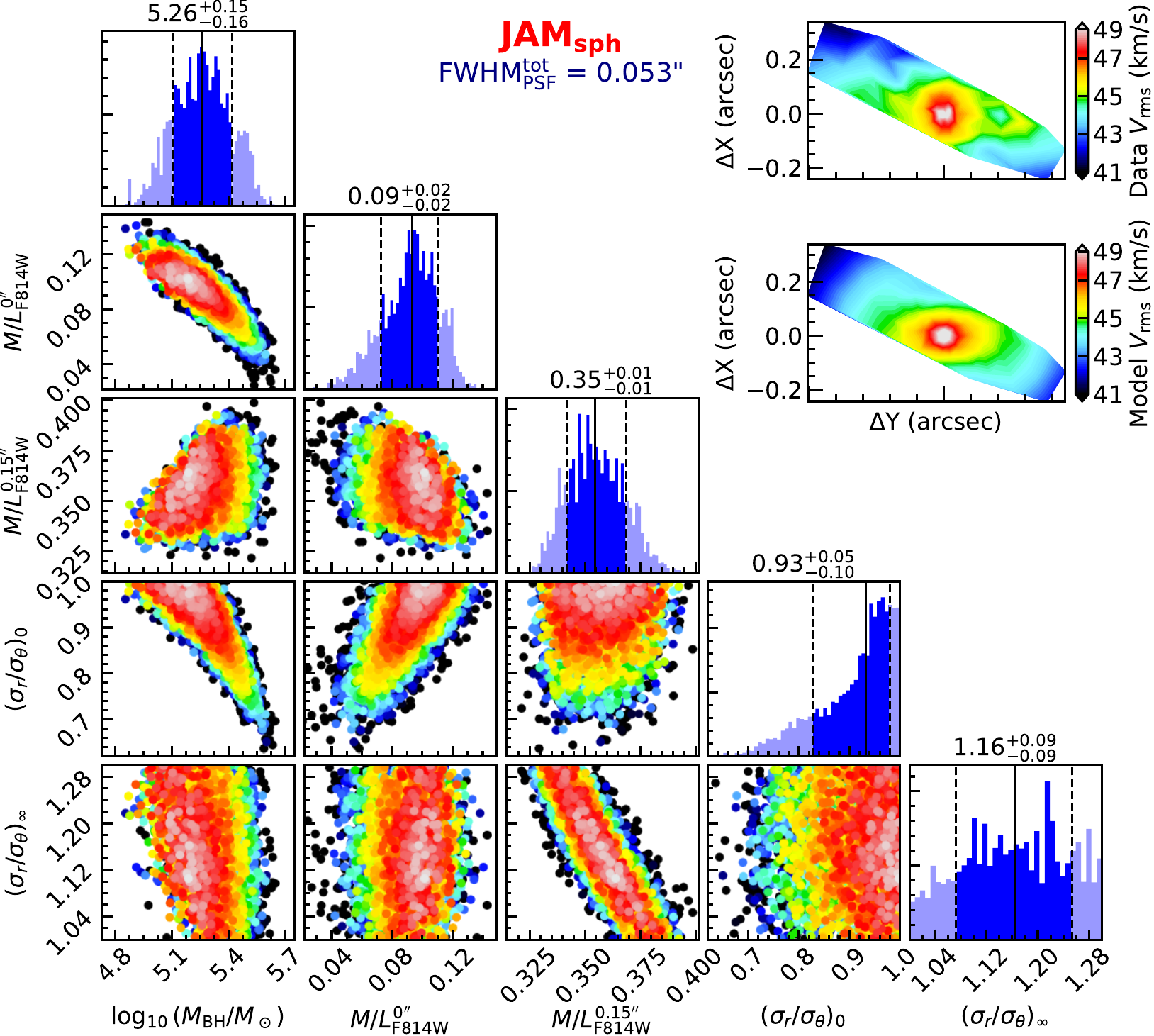} 
     \caption{Same as Figure~\ref{fig:adamet_cyl} but for the \jamsph\ fit with the diffraction-limited OSIRIS PSF (Table~\ref{tab:mge_psf}).}
     \label{fig:adamet_sph_psf2}
\end{figure*}

\begin{figure}[!h]
\centering
\includegraphics[width=0.47\textwidth]{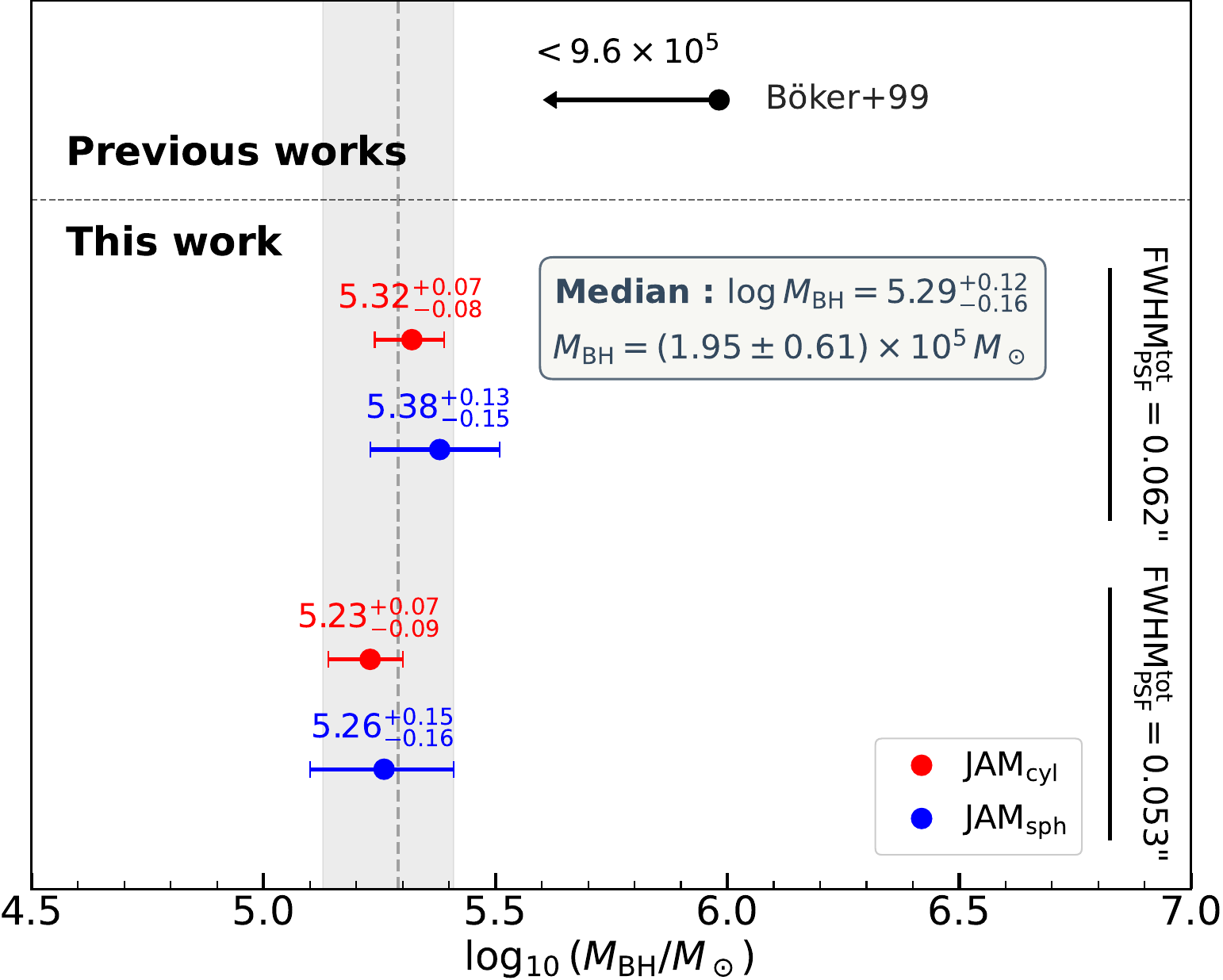}
\caption{Four independent \jam\ measurements of \Mbh\ for IC~342, obtained from the Keck/\osiris\ stellar kinematics by combining two velocity-ellipsoid alignments (\jamcyl, \jamsph) with two treatments of the \osiris\ PSF (empirical, diffraction-limited). Points with error bars give each measurement and its $1\sigma$ uncertainty; the dashed vertical line marks our adopted \Mbh, with the shaded band showing the total uncertainty after combining the statistical and systematic components in quadrature.}
\label{fig:MBHsummary}
\end{figure}

\subsection{MCMC Fitting and Uncertainty Estimation}\label{sec:mcmc_uncertainty}

Posterior distributions for the \jam\ parameters were generated by Markov chain Monte Carlo (MCMC) sampling, specifically the adaptive Metropolis algorithm of \citet{Haario2001} as implemented in \adamet\footnote{v2.0.9: \url{https://pypi.org/project/adamet/}} \citep{Cappellari2013a}, running each chain for $5\times10^4$ steps and discarding the first 20\% as burn-in. We take the maximum-likelihood solution as the best-fit value for each parameter, with $1\sigma$ and $3\sigma$ confidence levels (CLs) read off the marginalized posteriors at the 16th--84th and 0.14th--99.86th percentiles, respectively.

A subtlety of fitting such a large number of high-precision \osiris\ kinematic bins is that the resulting parameter uncertainties can end up unrealistically small if any systematics remain unaccounted for in the model, and that the $\chi^2$ can become dominated by the sheer number of bins at large radii rather than by the comparatively few bins inside the SOI where the BH actually dominates the gravitational force on the kinematics. In the conventional approach, both issues can be mitigated using the $\Delta\chi^{2}$ inflation scheme introduced by \citet{vandenBosch2009}, which in a Bayesian framework corresponds to rescaling the kinematic uncertainties by a factor of $(2N)^{1/4}$ \citep[e.g.,][]{Mitzkus2017, Nguyen20, Nguyen21, Ngo2025a}. To keep the central kinematics fully informative on \Mbh, the inflation is applied only to bins outside the BH SOI, while bins within the SOI retain their formal uncertainties. This selective-inflation strategy has become common practice in both stellar- and gas-dynamical \Mbh\ measurements \citep[e.g.,][]{Drehmer2015, Thater22, Nguyen_2026_4258, Nguyen_2026_M81, Simon2024}, and has been adopted in recent analyses (T H. T. Ho et al. submitted; H. N. Ngo et al. submitted).

For IC~342, however, we deliberately choose not to apply the selective-inflation scheme, for two reasons. First, the total number of kinematic bins used in the JAM fit is small ($N = 70$), so the nominal inflation factor $(2N)^{1/4} \approx 1.9$ would uniformly enlarge the outer-bin errors without a compelling physical justification. Second, the BH SOI inferred here ($r_{\rm SOI} \approx 0\farcs05$, Section~\ref{sec:smbh}) is only marginally resolved, being comparable to, and in fact slightly smaller than, the effective spatial resolution of the Keck/OSIRIS data. Under these conditions, the SOI/non-SOI partition on which selective inflation relies is not cleanly defined, and applying it would risk introducing rather than mitigating a systematic bias on \Mbh. We therefore retain the formal kinematic uncertainties for all bins and rely on the model-averaging procedure described above to capture the associated systematic budget.

\subsection{The IMBH Dynamical Mass}\label{sec:smbh}

Table~\ref{tab:adamet} summarizes the best-fit parameters and $1\sigma$ CLs from all four \jam\ configurations. The inferred \Mbh\ spans $\log_{10}(M_\mathrm{BH}/M_\odot) \approx 5.23$--$5.38$ across the ensemble, a factor of $\sim$1.4 range that reflects the combined sensitivity to velocity-ellipsoid geometry and PSF treatment. Both the inner and outer stellar mass-to-light ratios, $M/L^{0\arcsec}_{\rm F814W} \approx 0.07$--$0.10$~\Msun/\Lsun\ and $M/L^{0\farcs15}_{\rm F814W} \approx 0.35$--$0.38$~\Msun/\Lsun, are stable across all configurations, confirming that the photometric normalization at both small and large radii is insensitive to the assumed orbital geometry. The \jamsph\ models systematically return higher \Mbh\ than their \jamcyl\ counterparts by $\sim$0.03--0.06~dex, with the \jamsph\ ensemble spanning $\log_{10}(M_\mathrm{BH}/M_\odot) \approx 5.26$--$5.38$ against the \jamcyl\ range of $\log_{10}(M_\mathrm{BH}/M_\odot) \approx 5.23$--$5.32$.

Figure~\ref{fig:1D-profile} compares each best-fit JAM model to the observed \vrms\ profile along the kinematic major and minor axes. The \osiris\ data span $|\Delta r| \lesssim 0\farcs4$ and $\lesssim$$0\farcs2$ along those axes, respectively, sampling kinematics from within the IMBH SOI out to the surrounding stellar component. Both PSF variants reproduce the global shape of the two profiles well, including the central \vrms\ peak at $|\Delta r| \lesssim 0\farcs1$ that constitutes the primary kinematic signature of the central dark mass. Residuals between data and model are typically $\lesssim$$2\%$ ($\sim$1~\kms) across the full kinematic field. Crucially, the $M_\mathrm{BH}=0$ model (grey dashed curve) produces a central $V_\mathrm{rms}$ \emph{dip} in place of the observed peak and fails to reproduce the data in all four configurations. We nonetheless regard this as a marginal detection: because $R_\mathrm{SOI} \approx 0\farcs04$--$0\farcs06$ is comparable to, or slightly below, the OSIRIS PSF FWHM ($0\farcs053$--$0\farcs062$), the SOI is only marginally resolved, so the formal uncertainties likely understate the true systematic error on $M_\mathrm{BH}$.

\begin{table*}
\caption{\jam\ models best-fitting parameters and formal uncertainties} 
\vspace{-4mm}\centering
\begin{tabular}{lcccc|ccc} 
\hline\hline
Parameter (Unit) & Search range & Best-fitting & $1\sigma$ Error& $3\sigma$ Error& Best-fitting & $1\sigma$ Error& $3\sigma$ Error\\
   &  & value & (16--84)\% & (0.14--99.86)\% & value   & (16--84)\% & (0.14--99.86)\% \\  
\multicolumn{1}{c}{(1)} & (2) & (3) & (4) & (5) & (3) & (4) & (5) \\  
\hline 
Empirical PSF&\multicolumn{2}{c}{FWHM$^{\rm tot}_{\rm PSF}=0\farcs062$}&\multicolumn{2}{c}{\jamcyl}&\multicolumn{3}{c}{\jamsph}\\[1mm] 
\hline 
$\log_{10}$(\Mbh/\Msun)&  $(0\to7)$ &  $5.32$  &$-0.08$, $+0.07$ &$-0.42$, $+0.18$ & $5.38$   & $-0.15$, $+0.13$ & $-0.40$, $+0.37$\\
\ml$^{0\arcsec}_{\rm F814W}$ (\Msun/\Lsun)& $(0.1\to3)$ & $0.09$ &$-0.02$, $+0.02$ &$-0.05$, $+0.06$ & $0.07$ &$-0.02$, $+0.02$ &$-0.05$, $+0.06$\\
\ml$^{0\farcs15}_{\rm F814W}$ (\Msun/\Lsun)& $(0.1\to3)$ & $0.38$ &$-0.01$, $+0.01$ &$-0.03$, $+0.03$ & $0.36$ &$-0.01$, $+0.02$ &$-0.03$, $+0.04$\\ 
%$q_{\rm min}$         &$(200\to240)$& $218.30$ &$-0.61$, $+0.62$ &$-1.71$, $+1.72$ & $218.23$ & $-0.77$, $+0.75$ & $-2.14$, $+2.08$\\
($\sigma_z/\sigma_R$)$_{\rm cyl}$   & $(0.1\to2)$ & $0.98$  &$-0.03$, $+0.03$ &$-0.08$, $+0.10$ & --  & --  & --  \\
($\sigma_\theta/\sigma_r$)$_0$      & $(0.5\to1)$   &  --  &  --  &  --  & $0.92$ & $-0.10$, $+0.06$ & $-0.38$, $+0.08$\\
($\sigma_\theta/\sigma_r$)$_\infty$ & $(1\to1.3)$   &  --  &  --  &  --  & $1.16$ & $-0.10$, $+0.10$ & $-0.16$, $+0.14$\\
%$r_a$ (\arcsec)                     & $(0\to0.5)$   &  --  &  --  &  --  & $68.14$ & $-0.73$, $+0.68$ & $-2.12$, $+1.88$\\[1mm]
\hline
Diffraction-limited PSF&\multicolumn{2}{c}{FWHM$^{\rm tot}_{\rm PSF}=0\farcs053$}&\multicolumn{2}{c}{\jamcyl}&\multicolumn{3}{c}{\jamsph}\\[1mm] 
\hline 
$\log_{10}$(\Mbh/\Msun)&  $(0\to7)$ &  $5.23$  &$-0.09$, $+0.07$ &$-0.27$, $+0.20$ & $5.26$   & $-0.16$, $+0.15$ & $-0.39$, $+0.37$\\
\ml$^{0\arcsec}_{\rm F814W}$ (\Msun/\Lsun)&  $(0.1\to3)$  &  $0.10$  &$-0.01$, $+0.02$ &$-0.04$, $+0.04$ & $0.09$   &$-0.02$, $+0.02$  &$-0.06$, $+0.04$\\
\ml$^{0\farcs15}_{\rm F814W}$ (\Msun/\Lsun)&  $(0.1\to3)$  &  $0.38$  &$-0.01$, $+0.01$ &$-0.02$, $+0.02$ & $0.35$   &$-0.01$, $+0.01$  &$-0.03$, $+0.04$\\ 
%$q_{\rm min}$             &$(200\to240)$  & $218.30$ &$-0.61$, $+0.62$ &$-1.71$, $+1.72$ & $218.23$ & $-0.77$, $+0.75$ & $-2.14$, $+2.08$\\
($\sigma_z/\sigma_R$)$_{\rm cyl}$   & $(0.1\to2)$ & $0.96$  &$-0.02$, $+0.03$ &$-0.07$, $+0.08$ & --  & --  & --  \\
($\sigma_r/\sigma_\theta$)$_0$      & $(0.5\to1)$   &  --  &  --  &  --  & $0.93$ & $-0.10$, $+0.05$ & $-0.28$, $+0.07$\\
($\sigma_r/\sigma_\theta$)$_\infty$ & $(1\to1.3)$   &  --  &  --  &  --  & $1.16$ & $-0.09$, $+0.09$ & $-0.16$, $+0.13$\\
%$r_a$ (\arcsec)                     & $(0\to0.5)$   &  --  &  --  &  --  & $68.14$ & $-0.73$, $+0.68$ & $-2.12$, $+1.88$\\[1mm]
\hline
\end{tabular}
\label{tab:adamet}
\end{table*} 

\begin{figure*}
	\centering
    \includegraphics[width=0.85\textwidth]{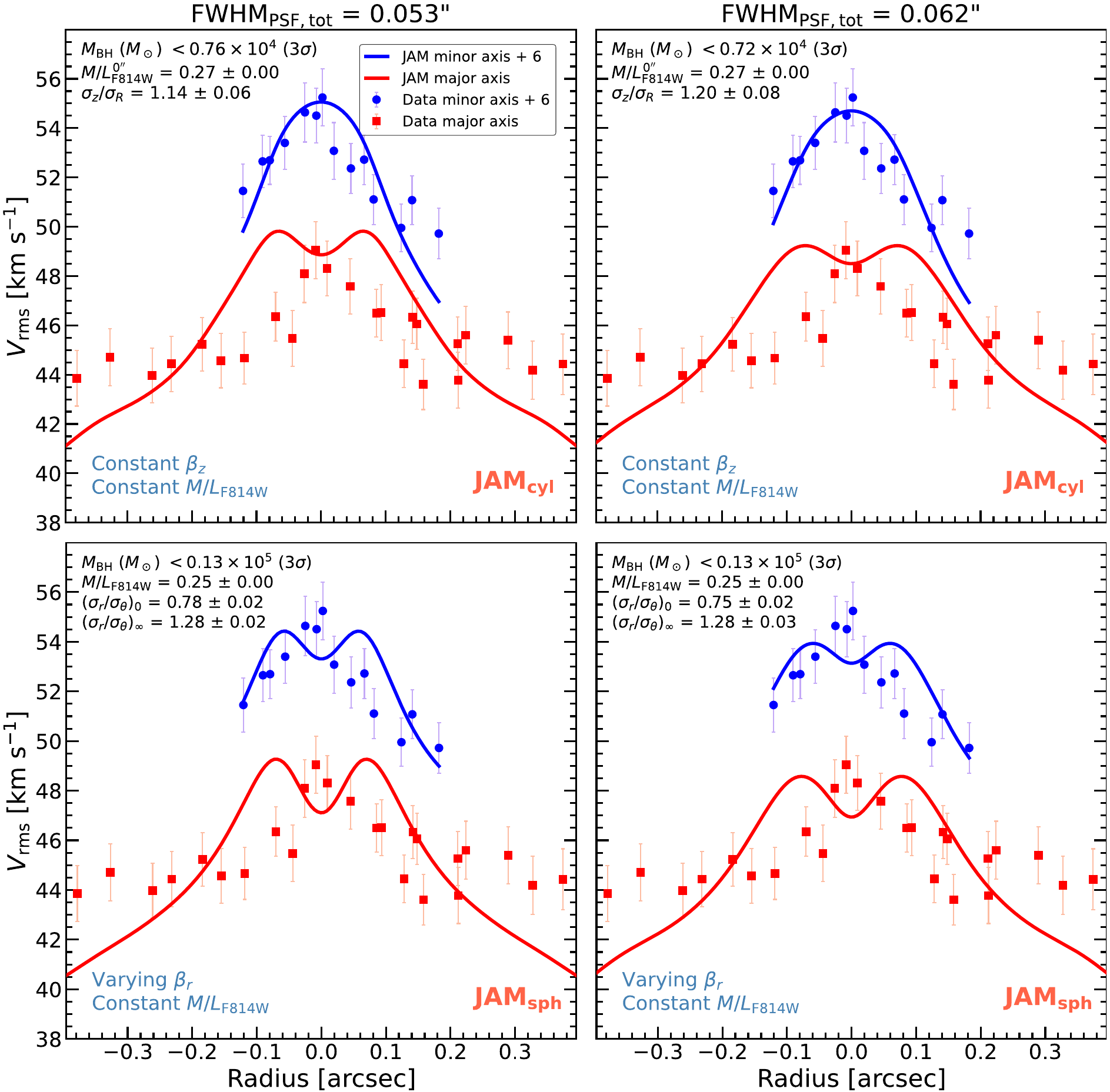} 
	 \caption{Same as Figure~\ref{fig:1D-profile} but for the constant-\mlhst\ \jam\ configurations. \textit{Top:} \jamcyl; \textit{Bottom:} \jamsph; diffraction-limited (\textit{left}) and empirical (\textit{right}) OSIRIS PSF in each case. All four models fail to reproduce simultaneously the central $V_\mathrm{rms}$ peak and the outer kinematic bins, leaving larger systematic residuals ($\sim$4~\kms) at both radial extremes compare to the the detection case as shown in Figure~\ref{fig:1D-profile}. The posteriors consequently include $M_\mathrm{BH} = 0$~\Msun\ within $1\sigma$ (see Figure~\ref{adamet_upperLimit}), motivating the adoption of a radially varying \mlhst\ profile (Section~\ref{sec:mlconst}).}
	 \label{capfit_upperLimit}
\end{figure*}

Figure~\ref{fig:adamet_cyl} shows the 2D posterior distributions from the \adamet\ MCMC optimization for the two \jamcyl\ configurations (empirical and diffraction-limited OSIRIS PSF). All four free parameters: $\log_{10}$(\Mbh/\Msun), $M/L^{0\arcsec}_{\rm F814W}$, $M/L^{0\farcs15}_{\rm F814W}$, and $\sigma_z/\sigma_R$, are recovered with well-defined, unimodal posteriors in both cases.  Two distinct covariance structures are visible in the joint distributions. First, \Mbh\ exhibits a \textit{negative} covariance with $M/L^{0\arcsec}_{\rm F814W}$: a higher central stellar $M/L$ assigns more mass to the stellar component at the scale of the BH SOI, reducing the BH contribution required to reproduce the observed central $V_{\rm rms}$ peak, and vice versa. This is the classical BH--$M/L$ degeneracy operating at the resolution scale of the kinematic data \citep{Cappellari08, Thater22}. Second, \Mbh\ shows a \textit{positive} covariance with $M/L^{0\farcs15}_{\rm F814W}$, the outer $M/L$ value: because $M/L^{0\farcs15}$ governs the stellar mass budget at radii beyond the SOI, increasing it raises the overall stellar mass normalization of the model, which in turn requires a more massive BH to maintain the observed $V_{\rm rms}$ profile at the center. This positive covariance is partially degenerate with the limited radial extent of the \osiris\ FOV relative to the spatial scale over which the $M/L$ gradient operates, meaning the two $M/L$ values cannot be fully decoupled from \Mbh\ by the data alone. Both covariances are nonetheless contained within the $3\sigma$ credible region, and the 1D marginalized histograms of \Mbh\ remains well-constrained in both PSF variants.

The joint distribution between $M/L^{0\arcsec}_{\rm F814W}$ and $M/L^{0\farcs15}_{\rm F814W}$ reveals a \textit{negative} covariance: when the central $M/L$ is high, the stellar mass near the BH SOI is already elevated, and the model compensates by preferring a lower outer $M/L$ to maintain a consistent fit to the global $V_{\rm rms}$ profile shape. This anti-correlation between the two endpoints of the linear $M/L$ gradient reflects the limited radial leverage of the \osiris\ FOV, which cannot fully decouple the inner and outer mass normalization without wider kinematic coverage.

The \Mbh--$\sigma_z/\sigma_R$ joint distribution shows a \textit{positive} covariance, a direct manifestation of the classical mass--anisotropy degeneracy \citep{Binney1982}: more radially biased orbits ($\sigma_z/\sigma_R > 1$) inflate the projected LOS $V_{\rm rms}$, which the model accommodates by simultaneously increasing \Mbh. However, the tight constraint on $\sigma_z/\sigma_R$ near isotropy ($\approx$$0.96$--$0.98$) significantly limits the range over which this degeneracy can operate, leaving the \Mbh\ posterior well-constrained despite the covariance.

Both \jamcyl\ configurations recover a well-constrained orbital anisotropy of $\sigma_z/\sigma_R = 0.96 \pm 0.03$ and $0.98 \pm 0.03$ for the diffraction-limited and empirical PSF models, respectively, with narrow, symmetric posteriors in both cases. These values are consistent with isotropic stellar orbits ($\beta_z \equiv 1 - (\sigma_z/\sigma_R)^2 \approx 0$) within $1\sigma$, indicating that the nuclear stellar kinematics of IC~342 require no significant vertical anisotropy to reproduce the observed \vrms\ field. The near-isotropic solution is stable across both PSF treatments, confirming that this result is not sensitive to the adopted PSF model. Near-isotropic orbits are a characteristic feature of fast-rotating, disk-dominated systems \citep[e.g.,][]{Cappellari08} and have been consistently recovered in stellar dynamical studies of similar late-type spirals \citep[e.g.,][]{Nguyen_2026_4258, Nguyen_2026_M81}, lending further confidence that the \jamcyl\ orbital structure recovered here is physically realistic.

Similarly, the 2D \jamsph\ posteriors share the same \Mbh--$M/L$ covariance structure as the \jamcyl\ models: \Mbh\ exhibits a negative covariance with $M/L^{0\arcsec}_{\rm F814W}$ and a positive covariance with $M/L^{0\farcs15}_{\rm F814W}$, while the two $M/L$ endpoints are anti-correlated with each other. However, the \jamsph\ posteriors reveal additional covariance structures arising from the logistic anisotropy parameterization that are absent in the \jamcyl\ case.

Most notably, \Mbh\ shows a strong negative covariance with the inner dispersion ratio $(\sigma_r/\sigma_\theta)_0$: a more tangentially biased central orbit distribution (lower $(\sigma_r/\sigma_\theta)_0$) suppresses the projected $V_\mathrm{rms}$ in the innermost bins, requiring a more massive BH to compensate and reproduce the observed central peak. This is a manifestation of the mass--anisotropy degeneracy operating specifically at the scale of the BH SOI, as similar as the \jamcyl, and represents the dominant source of the broader \Mbh\ uncertainties in \jamsph. Conversely, \Mbh\ exhibits a mild positive covariance with the outer dispersion ratio $(\sigma_r/\sigma_\theta)_\infty$: more radially biased orbits at large radii inflate the outer $V_\mathrm{rms}$, which is partially compensated by increasing \Mbh\ to steepen the central profile, though this covariance is weaker given that the outer anisotropy primarily affects bins well outside the SOI.

The two anisotropy parameters themselves are anti-correlated: a more tangentially biased center (low $(\sigma_r/\sigma_\theta)_0$) is preferentially paired with a more radially biased outer region (high $(\sigma_r/\sigma_\theta)_\infty$), reflecting the physical prior that the logistic function must produce a monotonically increasing anisotropy profile from center to large radii. 

The logistic anisotropy profile is well-behaved across both \jamsph\ configurations. The inner dispersion ratio $(\sigma_r/\sigma_\theta)_0$ is constrained to the tangentially biased regime at $\approx$$0.92$--$0.93$, consistent with the physically motivated prior $(\sigma_r/\sigma_\theta)_0 < 1$ and confirming that the \osiris\ kinematics genuinely prefer tangential orbits within the BH SOI rather than being driven to this solution solely by the prior boundary. The outer dispersion ratio $(\sigma_r/\sigma_\theta)_\infty \approx 1.16$ is recovered in the mildly radial regime, also consistent with its prior range, with a narrow posterior indicating that the kinematic data at larger radii provide meaningful constraints on the outer orbital structure. Together, these two parameters define a logistic transition from tangential to radial anisotropy with increasing radius, qualitatively consistent with the anisotropy profiles reported in stellar dynamical studies of other galactic nuclei hosting central dark masses \citep[e.g.,][]{Simon2024, Nguyen_2026_4258, Nguyen_2026_M81}.

Figure~\ref{fig:MBHsummary} presents all four \Mbh\ measurements with their $1\sigma$ uncertainties, spanning the two velocity-ellipsoid alignments and two PSF treatments. To extract a single fiducial estimate from this ensemble, we take the median across the four models and propagate the spread via $1\sigma$ bootstrap resampling, obtaining \Mbh~$= (1.95^{+0.12}_{-0.16})\times10^5$~\Msun.

The total uncertainty on this fiducial mass combines two components added in quadrature. The first is the statistical component, taken as the MCMC CL of each individual model. The second is a systematic component quantifying the model-to-model scatter across the four \jam\ runs, which captures the sensitivity of \Mbh\ to the choice of velocity-ellipsoid geometry and PSF model. Distance uncertainty is handled separately (Section~\ref{sec:dist}) rather than folded into this budget, since \Mbh\ scales linearly with $D$ and its contribution is most transparently reported in isolation. Combining all terms yields our adopted black hole mass of \Mbh~$= (1.95 \pm 0.61)\times10^5$~\Msun\ for IC~342.

As mentioned in Section~\ref{sec:target}, IC~342 has no prior direct, dynamical \Mbh\ measurement. The only existing constraint is an upper limit of $M_\mathrm{BH} < 9.6 \times 10^5$~\Msun, scaled to our adopted TRGB distance of $D = 3.45$~Mpc \citep{Wu2014}, was inferred by \citet{Boker1999} from 1D spherical Jeans models of the NSC using seeing-limited ground-based spectroscopy at $\sim$5\arcsec\ resolution, a scale more than two orders of magnitude larger than the BH SOI. At such resolution, the BH gravitational influence is entirely unresolved and the upper limit carries little constraining power on the actual \Mbh. Our Keck/\osiris\ LGS-AO data, with a PSF FWHM of $0\farcs062$, marginally resolve the BH SOI ($R_{\rm SOI} \approx 0\farcs05$ when adopted the NSC velocity dispersion of 33~\kms\ \citep{Boker1999} and represent the first dataset capable of delivering a direct, stellar-dynamical \Mbh\ measurement in IC~342. The result presented here is therefore the first direct dynamical \Mbh\ determination for this galaxy, filling a long-standing gap in the black hole census of nearby late-type spirals.

With $M_{\rm BH}=(1.95\pm0.61)\times10^{5}$~\Msun, IC~342 joins the small sample of IMBHs with direct dynamical mass measurements, becoming the eighth such measurement to date: NGC~205 \citep{Nguyen19}, NGC~5102 and NGC~5206 \citep{Nguyen18, Nguyen19}, NGC~404 \citep{Davis20}, NGC~3593 \citep{Nguyen22}, NGC~4395 \citep{denBrok15}, and IC~750 \citep{Zaw2020}. IC~342 is also now the second bulgeless spiral galaxy (i.e., it joins NGC~4395) with a direct dynamical \Mbh\ measurement, extending this rare dynamical census to a galaxy without a classical bulge.

\subsection{Additional Uncertainty Budgets}\label{sec:uncertainty}

\subsubsection{Distances}\label{sec:dist}

We adopt the TRGB distance $D = 3.45 \pm 0.13$~Mpc \citep{Wu2014} as our fiducial value throughout. The NASA/IPAC Extragalactic Database lists 17 independent distance determinations for IC~342, spanning 1.0--6.0~Mpc and yielding a formal mean of $\langle D \rangle = 3.51 \pm 1.42$~Mpc; the large scatter is driven by older, less reliable Tully-Fisher estimates rather than by genuine uncertainty in the true distance. Restricting to the six most reliable measurements: Cepheids \citep{Saha2002} and TRGB \citep{Wu2014}, yields a weighted mean of $3.39 \pm 0.07$~Mpc, in excellent agreement with our adopted value and confirming that $D \approx 3.45$~Mpc is well-supported by the best available distance indicators. Since $M_\mathrm{BH} \propto D$, the $\pm 0.13$~Mpc uncertainty on our adopted distance propagates directly into a systematic uncertainty of $\pm$3.8\% on \Mbh.

\begin{figure*}
	\centering
    \includegraphics[width=\textwidth]{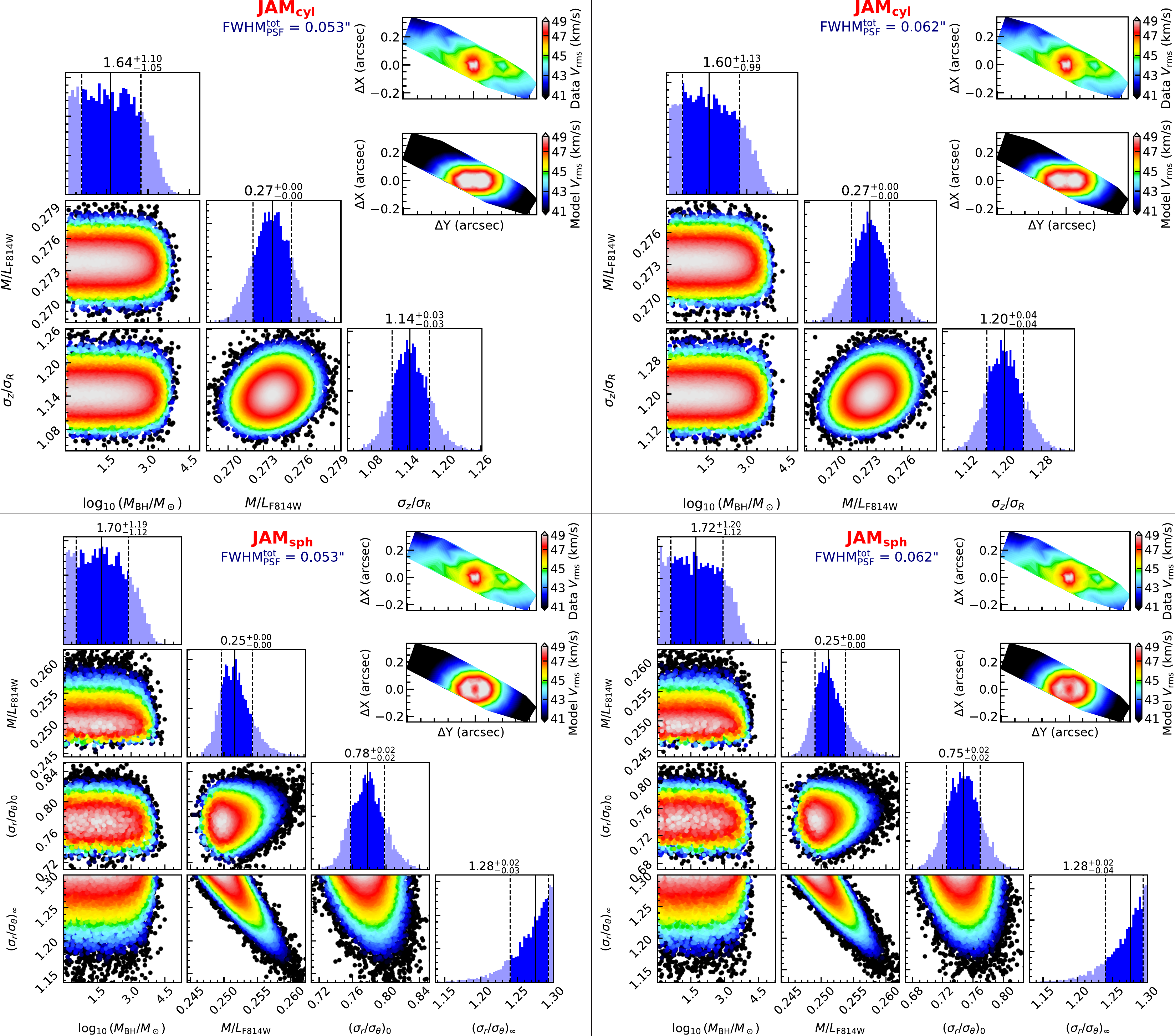} 
	 \caption{Same as Figures~\ref{fig:adamet_cyl} and \ref{fig:adamet_sph_psf1}--\ref{fig:adamet_sph_psf2}, but for JAM models assuming a spatially constant \mlhst\ across the \osiris\ FOV. \textit{Top panels:} \jamcyl\ models with the diffraction-limited (\textit{left}) and empirical (\textit{right}) PSF. \textit{Bottom panels:} \jamsph\ models with the same PSF ordering. In all four configurations, the posteriors include $M_\mathrm{BH} = 0$ within $1\sigma$, yielding only an upper limit on \Mbh\ rather than a detection. The \osiris\ stellar kinematics are inconsistent with a constant \mlhst\ parameterization and instead require a radially varying \mlhst\ profile (Section~\ref{sec:mlconst}).}
	 \label{adamet_upperLimit}
\end{figure*}

\subsubsection{Constant versus Varying \mlhst\ Parameterization}\label{sec:mlconst}

\begin{figure*}
	\centering
    	 \includegraphics[width=0.9\textwidth]{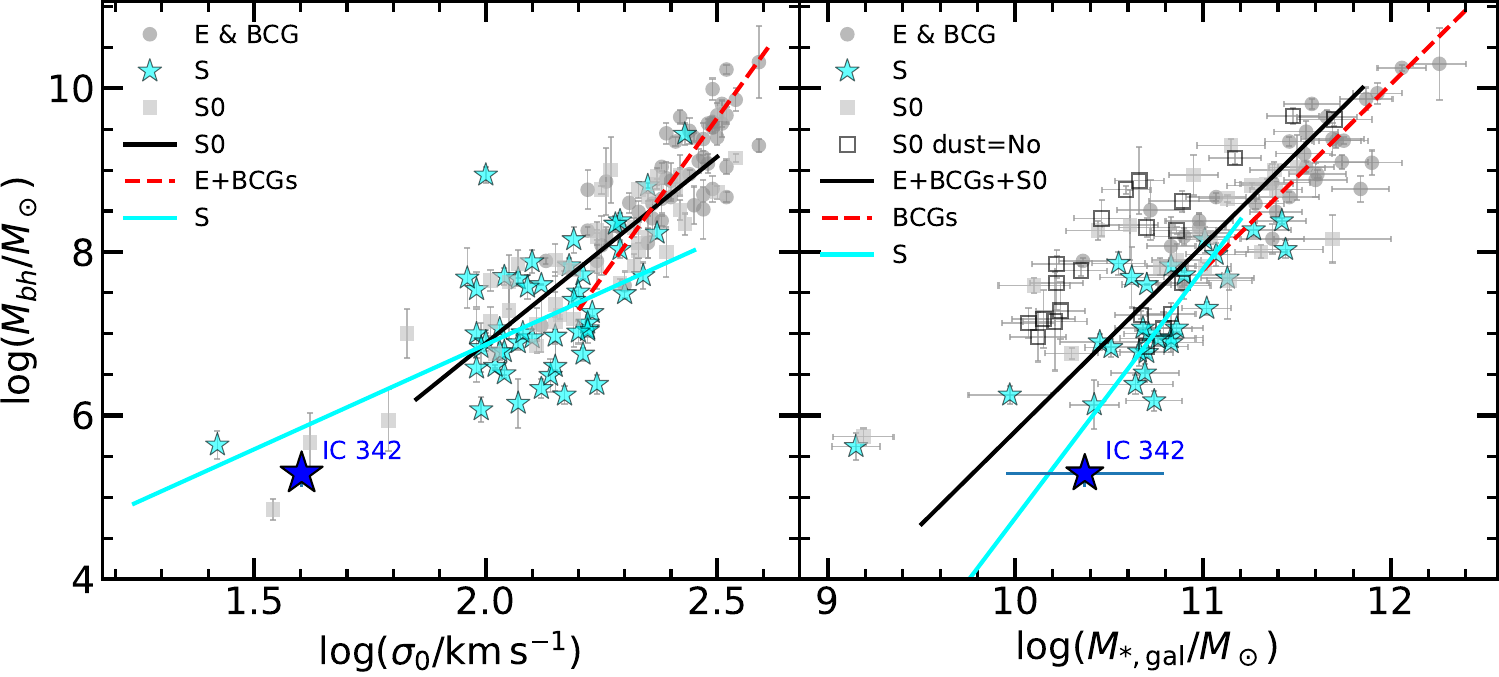} 
	 \caption{Location of IC~342 (blue star) on the $M_\mathrm{BH}$--$\sigma_0$ (\textit{left}) and $M_\mathrm{BH}$--$M_{\star,\mathrm{gal}}$ (\textit{right}) scaling diagrams. Symbol types and scaling relation lines follow \citet{Graham2023MNRAS.518.6293G, Graham2023MNRAS.518.2177G, Graham2026MNRAS.549ag782G} as indicated in the legend.}
	 \label{bhmass_sigma}
\end{figure*}

As a test of whether the radial variation in $M/L_{\rm F814W}$ is required by the stellar kinematics, we repeated all four JAM configurations using a spatially constant $M/L_{\rm F814W}$. The constant-$M/L$ models give systematically worse fits than the corresponding varying-$M/L_{\rm F814W}$ models, with $\chi^2_{{\rm constant}\,M/L} - \chi^2_{{\rm varying}\,M/L} = 206$--$241$ across the four JAM configurations. Since the varying model introduces only one additional free parameter, this difference quantifies the preference of the stellar kinematics for a radial $M/L_{\rm F814W}$ gradient.

In all four JAM configurations, the constant-$M/L_{\rm F814W}$ models allow $M_{\rm BH}=0$ within $1\sigma$, providing only an upper limit rather than a black-hole detection (Figure~\ref{capfit_upperLimit}). This is in contrast to the models with a varying $M/L_{\rm F814W}$, which consistently give a non-zero black-hole mass.

The difference between the two parameterizations is also apparent in the fitted \vrms\ profiles. The constant-$M/L_{\rm F814W}$ models produce systematically larger residuals, exceeding $\sim$4~\kms\ in the innermost bins and near the outer edge of the OSIRIS field, whereas the varying-$M/L_{\rm F814W}$ models generally reproduce the observed kinematics to within $\lesssim$2\% ($\sim$1~\kms; Section~\ref{sec:smbh}). Figures \ref{capfit_upperLimit} and \ref{adamet_upperLimit} show the corresponding 2D posterior distributions and 1D marginalized distributions. None of the four constant-$M/L_{\rm F814W}$ configurations can simultaneously reproduce the observed \vrms\ field and constrain $M_{\rm BH}$ above zero.

The preference for a varying $M/L_{\rm F814W}$ is also physically plausible. IC 342 hosts a prominent NSC with an estimated age of $\sim$60--630~Myr \citep{Boker1999}, younger than the surrounding stellar population. The dynamical models give a central $M/L^{0\arcsec}_{\rm F814W}\approx$~(0.07--0.10)~\Msun/\Lsun, increasing to $M/L^{0\arcsec}_{\rm F814W}\approx$~(0.35--0.38)~\Msun/\Lsun\ at larger radii. This lower central $M/L_{\rm F814W}$ is consistent with the presence of a younger stellar population in the nucleus. The central dynamical $M/L_{\rm F814W}$ represents a luminosity-weighted average of the nuclear cluster and surrounding stars within the innermost resolution elements, and therefore is expected to be higher than the $M/L_{\rm F814W}$ of the nuclear cluster alone. For comparison, stellar-population models for a young nuclear cluster with a Kroupa IMF and solar metallicity give $M/L_I\sim0.015$--$0.16$~\Msun/\Lsun\ \citep{Bruzual2003}, broadly consistent with the lower central dynamical values once the contribution from the surrounding older population is included.

The strong foreground extinction toward IC~342 ($A_{\mathrm{F814W}} \approx 0.74$~mag; \citealt{Schlafly2011}), however, prevents an independent optical stellar-population measurement of the central $M/L_{\rm F814W}$ gradient. We therefore rely on the stellar kinematics to determine whether a radial $M/L_{\rm F814W}$ variation is required, while the young nuclear population provides independent physical motivation for the inferred gradient. Taken together, the model comparison and the physical properties of the nuclear cluster support the use of a radially varying $M/L_{\rm F814W}$ in our dynamical modeling.

\section{Discussion}\label{sec:discussion}

\subsection{Isotropic Nuclear Orbits as a Signature of  Merger-Free Secular Growth}\label{sec:isotropy_secular}

Beyond providing a robust \Mbh, the orbital structure recovered by our JAM models carries direct information about the assembly history of the IC~342 nucleus. Both \jamcyl\ configurations return tightly constrained, near-isotropic orbits (e.g., $\sigma_z/\sigma_R \approx 0.96$--$0.98$; $\beta_z \approx 0$), and the \jamsph\ models find only a mild tangential bias confined to the innermost region  (e.g., ($\sigma_r/\sigma_\theta)_\infty \approx 0.92$--$0.93$; $\beta_r < 0$). This stands in contrast to the anisotropy signatures characteristic of merger-built nuclei: violent relaxation during dissipationless mergers imprints radially biased orbits in classical bulges \citep[e.g.,][]{Hernquist1992}, while binary SMBH scouring in gas-poor major mergers preferentially ejects stars on radial orbits, leaving strongly tangentially biased cores \citep{Milosavljevic2001, Rantala2024}. The absence of either signature in IC~342 indicates a phase space that has never been dynamically disturbed by a significant merger: a kinematic counterpart to the photometric evidence for a bulgeless disk (Section~\ref{sec:photomodel}). The nuclear structure of IC~342, comprising a young NSC and an IMBH embedded in molecular gas with ongoing inflow \citep{Schinnerer2003}, is therefore most naturally explained by purely secular, gas-driven growth, which is consistent with the bar-driven inflow channel discussed in Section~\ref{sec:intro}. In this sense, the near-isotropic orbits constitute a kinematic ``smoking gun'' for merger-free CMO assembly, and establish IC~342 as a clean laboratory for the secular-growth mode explored in the remainder of this section.

\subsection{IC~342 on the BH-Galaxy Scaling Relations}\label{sec:scaling_gal}

The secular, merger-free assembly history inferred above provides the natural context for interpreting the position of IC~342 on the BH--galaxy scaling relations. Figure~\ref{bhmass_sigma} shows the location of IC~342 on the \Mbh--$\sigma$ and \Mbh--$M_\star$ diagrams alongside galaxies with direct dynamical \Mbh\ measurements, velocity dispersions, and stellar masses 
% \citet[][and references therein]{Graham2023MNRAS.518.6293G, Graham2023MNRAS.518.2177G, Graham2026MNRAS.549ag782G}. 
\citep[][and references therein]{Sahu19b, Graham2026MNRAS.549ag782G}. In both panels, IC~342 lies below the spiral galaxy relations (cyan lines), indicating that its IMBH is somewhat undermassive for its host galaxy properties, even relative to other late-type galaxies. Although these offsets are modest ($\sim$0.45~dex in the \Mbh--$\sigma$ plane and $\sim$0.3~dex in the \Mbh--$M_\star$ plane), IC~342 remains well within the intrinsic 1$\sigma$ scatter of the spiral galaxy relations.

Nonetheless, its systematic location below the relations may be physically meaningful. IC~342 is a confirmed bulgeless SABcd galaxy whose nucleus is dominated by an NSC and has evolved primarily through secular processes rather than through the violent relaxation associated with the formation of a classical bulge. Bulgeless and pseudobulge galaxies have long been recognised as preferentially occupying the lower-right side of the classical-bulge sequences in both the \Mbh--$\sigma$ and \Mbh--$M_\star$ diagrams \citep{Graham2008, Hu2008, Greene2010, Kormendy13}, although spiral galaxies as a population are likewise systematically offset (Figure~\ref{bhmass_sigma}).

These offsets likely reflect differences in the co-evolution of black holes and their host galaxies. At fixed $\sigma$, spiral galaxies appear to host systematically lower-mass black holes than early-type galaxies. Moreover, their larger disk-to-bulge stellar mass ratios naturally displace them further from the $M_{\rm bh}$--$M_{\rm \star,bulge}$ relation than is the case for early-type galaxies in the $M_{\rm bh}$--$M_\star$ plane.
The position of IC~342 below the spiral-galaxy scaling relations is therefore consistent with a growth history in which the galaxy assembled largely through secular evolution, efficiently building its nuclear star cluster while, thus far, experiencing comparatively limited black-hole growth.

The scientific significance of this measurement extends beyond its position on the scaling diagrams. At the low-mass end ($M_\mathrm{BH} \lesssim 10^6$~\Msun), direct dynamical \Mbh\ measurements remain extremely scarce, and the shape, normalization, and scatter of the $M_\mathrm{BH}$--$\sigma$ relation are poorly constrained \citep{Greene2010, Nguyen18, Nguyen19}. This is precisely the regime where theoretical models of BH seeding diverge most strongly: light-seed models from the remnants of Population~III stars \citep{Madau2001}, predict a wide scatter and high BH occupation fraction in low-mass galaxies, while heavy-seed models via the direct collapse of gas clouds \citep{Lodato2006} predict a lower occupation fraction but a higher minimum BH mass \citep{Volonteri2008}. Populating the IMBH end of the scaling relations with direct, well-resolved dynamical measurements in well-characterized hosts is therefore essential for discriminating between competing seeding scenarios. As a nearby, bulgeless spiral with a well-studied NSC and now a directly measured IMBH, IC~342 provides one of the most robust anchors currently available for this low-mass regime, and demonstrates that the $M_\mathrm{BH}$--$\sigma$ relation for spiral galaxies extends continuously down to at least $M_\mathrm{BH} \sim 10^5$~\Msun.

\begin{figure}[!h]
    \centering
    \includegraphics[width=0.45\textwidth]{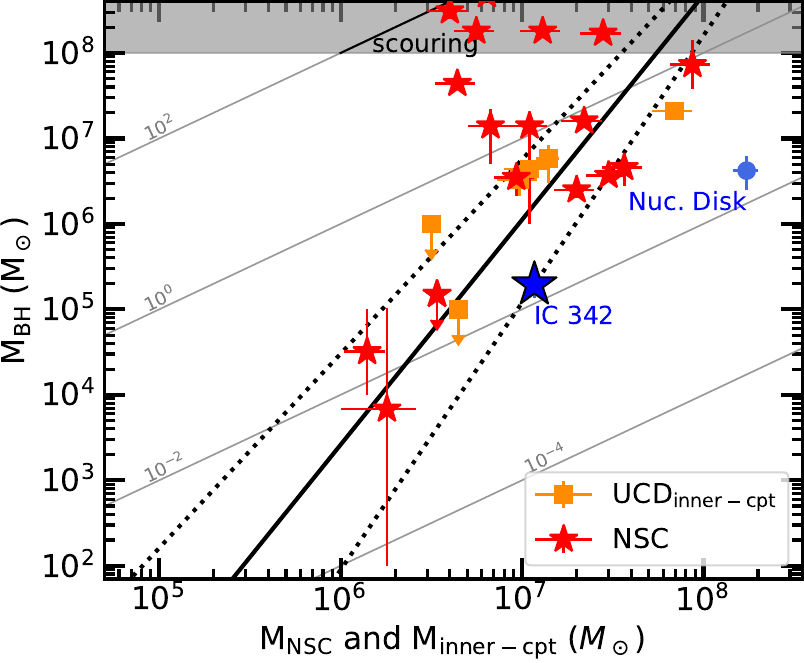}
    \caption{\Mbh--\Mnsc\ scaling diagram adapted from \citet{Graham2020} with updates from \citet{Graham2025}, showing IC~342 (blue star) at its dynamically measured \Mbh\ and NSC mass $M_\mathrm{NSC} \approx 1.17 \times 10^7$~\Msun\ \citep{Boker1999}. Red stars and orange squares denote systems hosting NSCs and UCDs, respectively; gray diagonal lines indicate constant $M_\mathrm{BH}/M_\mathrm{NSC}$ ratios as labeled. The solid line shows the best-fit relation with $1\sigma$ intrinsic scatter indicated by the dotted lines. Systems with $M_\mathrm{BH} > 10^8$~\Msun\ (gray shaded region) are omitted as their NSCs are expected to have been depleted by binary SMBH scouring during dry mergers \citep{Bekki2010}; flattened nuclear disks, which can extend to tens of parsecs in size \citep[e.g.,][]{Balcells2007}, are likewise excluded as they do not follow the NSC--BH relation. IC~342 falls within the $1\sigma$ scatter of the relation, confirming its membership in the NSC--BH coexistence regime at the IMBH mass scale.}
    \label{fig:bh-nsc}
\end{figure}

\subsection{IC~342 on the BH--NSC Scaling Relation}\label{sec:scaling_nsc}

Narrowing from galaxy-wide properties to the CMO themselves, Figure~\ref{fig:bh-nsc} shows the location of IC~342 on the \Mbh--\Mnsc\ diagram of \citet{Graham2020}, updated with recent measurements from \citet{Graham2025}. IC~342 (blue star) is plotted at its dynamically measured $M_\mathrm{BH} \approx (1.95 \pm 0.61) \times 10^5$~\Msun\ and NSC mass $M_\mathrm{NSC} \approx 1.17 \times 10^7$~\Msun\ \citep{Boker1999}, scaled to our adopted distance. IC~342 falls below the best-fit relation (solid line) but within the $1\sigma$ scatter region (dotted lines), with a mass ratio $M_\mathrm{BH}/M_\mathrm{NSC} \approx$0.017, placing it near the $10^{-2}$ iso-ratio line. This is consistent with the general picture in which lower-mass systems tend to be NSC-dominated, with the BH contributing only a small fraction of the total CMO mass \citep{Graham2009, Graham2020, Neumayer2020}. 

The gray shaded region at the top of Figure~\ref{fig:bh-nsc} marks the regime $M_\mathrm{BH} \gtrsim 10^8$~\Msun\ where massive merger-built ETGs dominate and NSCs start to be eroded by binary SMBH scouring \citep{Bekki2010}; IC~342 is far from this regime, firmly in the NSC--BH coexistence zone where both CMOs grow simultaneously through secular processes. The position of IC~342 on this diagram is particularly significant because it represents one of the few systems where $M_\mathrm{BH}$ at the IMBH mass scale ($M_\mathrm{BH} \sim 10^5$~\Msun) is directly and dynamically measured, rather than inferred from indirect indicators, making it a robust anchor for the low-mass end of the \Mbh--\Mnsc\ relation. 

Beyond galaxy evolution, the calibration of the \Mbh--\Mnsc\ relation at the IMBH end also bears directly on predictions for gravitational wave science, a point we return to in Section~\ref{sec:emri_lisa}.

\subsection{An Evolutionary Sequence of In-Situ CMO Growth: from M33 to IC~342}\label{sec:cmo_sequence}

The nuclear region of IC~342 may be showing us NSC formation in action. As illustrated in Figure~\ref{fig:IC342_nucleus-2}, the circumnuclear disk of gas and dust hosts several off-center knots of star formation. Such clumps are the natural outcome of a cold, gas-rich disk fragmenting and collapsing once local overdensities exceed the Jeans criterion; once formed, dynamical friction against the background disk drains their orbital energy, driving them to spiral inward and merge at the bottom of the potential well \citep[e.g.,][]{Tremaine1975, Bekki2010, Arca-Sedda2014, Neumayer2020}. The observed configuration: a young, massive NSC surrounded by off-center star-forming clumps \citep{Boker1997, Carson15} and ongoing molecular gas inflow \citep{Schinnerer2003}, suggests that both the NSC and the IMBH of IC~342 are still actively growing through this secular, in-situ channel.

This picture invites comparison with M33, a nearby bulgeless spiral that appears to sit \emph{earlier} along the same evolutionary pathway. M33 hosts a well-studied NSC \citep{Carson15}, yet carries one of the most stringent stellar-dynamical BH limits known, \Mbh~$\lesssim(1.5-3)\times10^3$~\Msun\ \citep{Gebhardt01, Merritt2001}: its nucleus has evidently not yet grown, or fed, a substantial central BH. If the in-situ channel operating in IC~342 is generic, the sequence M33~$\rightarrow$~IC~342 traces the gradual, NSC-mediated assembly of central BHs in secular disk galaxies. This constitutes an argument \emph{against} the universality of massive ($\gtrsim$$10^5$~\Msun) BH seeds, since it implies that at least some disk galaxies begin with little or no central BH and grow one only gradually within their nucleus \citep[e.g.,][]{NeumaerSeth2011, Volonteri2008, GrahamSoria2019}. We note that the Large Magellanic Cloud, which lacks a prominent NSC, may nonetheless host a $\simeq$$6\times10^5$~\Msun\ IMBH inferred from hypervelocity-star trajectories \citep{Han2025}; if confirmed, its inverted configuration, a BH without an NSC, would point to a distinct growth channel outside the in-situ sequence traced here. Direct dynamical measurements at the IMBH scale in further bulgeless disks will be required to populate, or refute, this proposed sequence.

\subsection{IC~342 as a Live Analog of Stripped Nuclei: $\omega$~Centauri and B023-G078}\label{sec:stripped_nuclei}

The NSC--IMBH configuration of IC~342 finds a striking counterpart in two nearby stellar systems believed to be the surviving nuclei of tidally disrupted galaxies. $\omega$~Centauri, the most massive globular cluster of the Milky Way, is now widely regarded as the stripped NSC of an accreted dwarf galaxy \citep{1988IAUS..126..603Z, Bekki2003, Massari2019, Ibata2019}, most plausibly the Gaia-Enceladus/Sequoia progenitor accreted $\sim$10~Gyr ago. \citet{Haberle2024} recently identified seven fast-moving stars within its central $0.08$~pc, establishing a firm lower limit of $M_\mathrm{BH} \gtrsim 8{,}200$~\Msun\ (and plausibly up to $\sim$$5\times10^4$~\Msun) for a central IMBH. An even closer analog in mass is B023-G078, the most massive globular cluster of M31 and likewise a stripped-nucleus candidate, for which \citet{Pechetti22} measured a dynamical BH mass of $\sim$$10^5$~\Msun, which is nearly identical to our measurement in IC~342.

These systems plausibly represent two evolutionary stages of the same phenomenon. IC~342 shows the \emph{in-situ} configuration: an IMBH of $\sim$$10^5$~\Msun\ embedded in a massive, actively growing NSC ($M_\mathrm{NSC} \approx 1.17\times10^7$~\Msun) at the center of an intact, gas-rich, bulgeless disk galaxy, still fed by molecular inflows \citep{Schinnerer2003}. Should a galaxy like IC~342 be accreted by a more massive host, tidal stripping would remove its disk while the dense nucleus, and its IMBH, survives, leaving behind precisely the kind of massive, chemically complex star cluster observed as $\omega$~Cen and B023-G078 \citep{Pfeffer2021, Neumayer2020}. IC~342 thus offers a rare view of the \emph{progenitor} state of stripped nuclei, with both CMO masses dynamically measured, anchoring the initial conditions that stripped-nucleus and globular-cluster IMBH studies must reproduce.

\subsection{IC~342 as a Local EMRI Laboratory and a Caution on NSC Structural Fits} \label{sec:emri_lisa}

The evolutionary framework above also positions IC~342 as a valuable target for gravitational wave science. Dense NSCs surrounding IMBHs are expected to be the dominant sites of extreme-mass-ratio inspirals (EMRIs), in which stellar-mass compact remnants spiral into the central BH through gravitational wave emission \citep{Amaro-Seoane2017}. At $D = 3.45$~Mpc, IC~342 is among the nearest systems in which both the NSC mass ($\approx$$1.17\times10^7$~\Msun) and the IMBH mass ($\approx$$10^5$~\Msun), squarely within the optimal LISA sensitivity band) are directly and dynamically characterized. The directly measured NSC--IMBH system of IC~342 therefore provides an ideal local test bed and target template for modeling EMRI capture rates and waveform templates for the upcoming LISA mission.

We close with an important caveat for such modeling efforts. EMRI rate calculations require the stellar density profile of the NSC, typically encoded in S\'{e}rsic, King, or Plummer fits to the observed light distribution. However, as the archival \hst\ imaging makes clear, the nuclear region of IC~342 is a \emph{lumpy conglomerate} of star-forming clumps and clusters rather than a single, smoothly declining stellar system akin to a globular cluster \citep{Boker1997, Carson15}. Single-component structural fits to distant, poorly resolved NSCs implicitly assume such smooth profiles, and the resulting sizes and masses, which propagate directly into published NSC scaling relations and EMRI rate predictions, may consequently carry substantial, unquantified systematic errors. The reported $M_\mathrm{NSC}$ of IC~342 itself, derived from aperture-integrated kinematics \citep{Boker1999}, should be understood as the mass of the nuclear conglomerate rather than of a single relaxed cluster. High-resolution imaging and spectroscopy of nearby, resolved systems such as IC~342 are therefore essential for calibrating the structural assumptions underpinning both NSC demographics and LISA-era EMRI forecasts.

\section{Conclusions}\label{sec:conclusion}

We have presented the first direct stellar-dynamical measurement of an IMBH in IC~342 using NIR LGS-AO IFS with Keck/\osiris. The unprecedented combination of high angular resolution (PSF FWHM $\approx 0\farcs053-0\farcs062$, corresponding to $\approx$0.8--1~pc at $D = 3.45$~Mpc) and $K$-band observations that penetrate the heavy foreground extinction ($A_{\mathrm{F814W}} \approx 0.74$~mag) delivers the first spatially resolved view of the nuclear stellar kinematics in this system. Our main results are as follows:

\begin{enumerate}

\item \textbf{High-resolution nuclear stellar kinematics.} The 2D LOSVDs extracted from the CO bandheads ($\sim$2.3~$\micron$) via \ppxf\ reveal a velocity dispersion rising from $\sim$42~\kms\ at $r \gtrsim 0\farcs2$ to a central peak of $\sim$49~\kms\ within the innermost resolution element. This central increase in both $\sigma$ and \Vrms\ provides unambiguous kinematic evidence for a central dark mass embedded within the IC~342 NSC.

\item \textbf{First direct IMBH mass measurement.} Modeling the observed \Vrms\ maps with an ensemble of four \jam\ models, spanning cylindrical (\jamcyl) and spherical (\jamsph) velocity-ellipsoid alignments and two \osiris\ PSF treatments (empirical and diffraction-limited), combined with a three-component \mge\ decomposition of archival \hst/WFC3 F814W imaging, yields an IMBH mass of $M_\mathrm{BH} = (1.95 \pm 0.61) \times 10^5$~\Msun. All four model variants are consistent within $1\sigma$, and the SOI ($R_\mathrm{SOI} \approx 0\farcs05$) is marginally resolved by the \osiris\ PSF ($\approx$0\farcs053--0\farcs062), making this the first direct dynamical \Mbh\ determination for IC~342.

\item \textbf{Stellar orbital structure.} Both \jamcyl\ and \jamsph\ consistently recover near-isotropic orbits in the nuclear region ($\sigma_z/\sigma_R \approx 0.96$--$0.98$ for \jamcyl; $\beta_z \approx 0$), consistent with the orbital structure typical of fast-rotating, disk-dominated systems \citep{Cappellari08, Nguyen_2026_4258, Nguyen_2026_M81}. The \jamsph\ logistic anisotropy profile confirms the physically expected transition from tangentially biased orbits in the center $(\sigma_r/\sigma_\theta)_0 \approx 0.92$--$0.93$, to mildly radial orbits at large radii $(\sigma_r/\sigma_\theta)_\infty \approx 1.16$, consistent with independent validation on M87 \citep{Simon2024}, NGC~4258 \citep{Nguyen_2026_4258}, NGC~3031 \citep{Nguyen_2026_M81}, NGC~5055 (T. H. T. Ho et al. submitted), and NGC~4736 (H. N. Ngo et al. submitted).

\item \textbf{Radially varying stellar $M/L_{\rm F814W}$.} The \osiris\ kinematics consistently disfavor a spatially constant \mlhst: constant-$M/L$ models include $M_\mathrm{BH} = 0$ within $1\sigma$ and produce systematically worse residuals exceeding $\sim$4~\kms\ in both the innermost and the edge bins. The fiducial linearly varying profile, with inner value $M/L^{0\arcsec}_\mathrm{F814W} \approx 0.07$--$0.10$~\Msun/\Lsun\ and outer value $M/L^{0\farcs15}_\mathrm{F814W} \approx 0.35$--$0.38$~\Msun/\Lsun, is qualitatively consistent with the young NSC stellar population ($\sim$60--630~Myr; \citealt{Boker1999}) dominating the innermost parsecs atop an older surrounding component. This result serves as a critical caveat for dynamical searches of IMBHs in LTGs: assuming a constant $M/L$ in systems with young, luminous NSCs systematically overestimates the central stellar mass. This ``phantom" stellar mass artificially accounts for the central rise in velocity dispersion, thereby absorbing the gravitational signature of any central compact object and yielding false-negative null results (i.e., mere upper limits). Correctly modeling stellar population and $M/L$ gradients is therefore a vital prerequisite for conducting a robust, unbiased IMBH census at the low-mass end of the scaling relations.

\item \textbf{Photometric and systematic robustness.} Replacing our four-component photometric decomposition with the single-S\'{e}rsic GALFIT model of \citet{Carson15} yields \Mbh\ values consistent within $1\sigma$ across all four \jam\ configurations, confirming insensitivity to the photometric model choice. Conversely, omitting dust masking from the decomposition shifts \Mbh\ by more than $3\sigma$, validating dust masking as an essential step for a reliable dynamical measurement in this heavily obscured system.

\item \textbf{NSC--BH coexistence.} With $M_\mathrm{BH} \approx 1.95 \times 10^5$~\Msun\ and $M_\mathrm{NSC} \approx 1.17 \times 10^7$~\Msun, IC~342 falls in the transition regime where NSCs and BHs are expected to coexist, with a mass ratio $M_\mathrm{BH}/M_\mathrm{NSC} \approx 0.017$. This measurement directly probes the NSC--BH coexistence regime and provides a new, well-constrained anchor for theoretical models of CMO co-evolution in the secular-evolution regime. This direct dynamical measurement at the IMBH mass scale anchors the low-mass end of the \Mbh--\Mnsc\ relation, which is a key input for forecasting EMRI event rates detectable by future gravitational wave observatories \citep[e.g.,][]{Amaro-Seoane2017}.

\item \textbf{An evolutionary context for in-situ CMO growth.} The nuclear configuration of IC~342: a young NSC surrounded by off-center star-forming clumps and fed by ongoing molecular gas inflow, captures the in-situ, secular growth of both CMOs in action. This places IC~342 within a broader evolutionary sequence of NSC-mediated BH growth in disk galaxies. At the early end sits M33, whose well-studied NSC hosts at most a $\lesssim$$(1.5$--$3)\times10^3$~\Msun\ BH \citep{Gebhardt01, Merritt2001}. IC~342 represents the active phase,with both CMOs growing concurrently. The end state is exemplified by $\omega$~Centauri \citep{Haberle2024} and B023-G078 \citep{Pechetti22}: massive clusters now classified as globular clusters, but widely regarded as the stripped nuclei of disrupted galaxies, each retaining an IMBH of mass comparable to that of IC~342. This sequence argues against the universality of massive ($\gtrsim$$10^5$~\Msun) BH seeds, suggesting instead that central BHs in at least some disk galaxies assemble gradually within their NSCs.

\item \textbf{Scaling relations.} Our \Mbh\ measurement places IC~342 below but within the scatter of the $M_\mathrm{BH}$--$\sigma$ and $M_\mathrm{BH}$--$M_{\star,\mathrm{gal}}$ relations for spiral galaxies, consistent with the known systematic offset of bulgeless hosts from relations calibrated on classical-bulge and merger-built systems \citep{Kormendy13, Graham2023MNRAS.518.2177G}. Each direct dynamical \Mbh\ at the IMBH mass scale carries disproportionate weight for constraining the low-mass end of these relations, where the scatter is largest and the shape remains poorly determined. This regime is of particular theoretical importance: light-seed models predict a wide mass distribution and high BH occupation fraction in low-mass galaxies, while heavy-seed models predict a higher minimum \Mbh\ and lower occupation fraction \citep{Volonteri2008, Madau2001, Lodato2006}. As the sample of direct dynamical measurements in bulgeless spirals grows, currently populated by only a handful of systems including NGC~4395 \citep{denBrok2015}, NGC~2748 \citep{Atkinson2005}, possibly NGC~5055 (T. H. T. Ho et al. submitted), and now IC~342, it will become possible to statistically discriminate between competing BH seeding scenarios directly from the observed distribution of \Mbh\ at fixed host-galaxy properties.

\item \textbf{Bulgeless spiral with a direct dynamical \Mbh.} The surface brightness decomposition of IC~342 shows no evidence for a classical bulge component, with all S\'{e}rsic indices $n < 1$, confirming its bulgeless nature. IC~342 thus joins the rare class of bulgeless spirals with a direct dynamical \Mbh\ determination: a regime currently populated by only a handful of galaxies including NGC~4395 \citep{denBrok2015} and NGC~2748 \citep{Atkinson2005}, and extends this census down to the IMBH mass scale ($M_\mathrm{BH} \sim 10^5$~\Msun).  This is particularly significant because bulgeless systems are thought to follow a more stochastic evolutionary path decoupled from the tight co-evolutionary channel governing classical spheroids \citep{Kormendy13}, yet direct \Mbh\ measurements to test this prediction at the low-mass end remain extremely scarce.

\end{enumerate}

\section*{Acknowledgements} 
DDN acknowledges support from the ELT postdoctoral research fellowship at the Department of Astronomy, University of Michigan. TNL would like to thank the ICISE staﬀ for their enthusiastic support. This work is partially supported by a grant from the Simons Foundation to IFIRSE, ICISE (916424, N.H.). RS acknowledges support from the INAF grant number 1.05.23.04.04. IC’s research is supported by the SAO Telescope Data Center. He also acknowledges support from the NASA ADAP-22-0102 and HST-GO-16739 grants. % 

Some of the data presented herein were obtained at Keck Observatory, which is a private 501(c)3 non-profit organization operated as a scientific partnership among the California Institute of Technology, the University of California, and the National Aeronautics and Space Administration. The Observatory was made possible by the generous financial support of the W. M. Keck Foundation. These observations are associated with the Keck Program ID: 2024B\_W294.

The authors wish to recognize and acknowledge the very significant cultural role and reverence that the summit of Maunakea has always had within the Native Hawaiian community. We are most fortunate to have the opportunity to conduct observations from this mountain. 
\facility{Keck/\osiris\ and HST/WFC3.}

\software{{\tt Python~3.12:} \citep{VanRossum2009}, 
{\tt Matplotlib~3.6:} \citep{Hunter2007}, 
{\tt NumPy~1.22:} \citep{Harris2020}, 
{\tt SciPy~1.3:} \citep{Virtanen2020},  
{\tt photutils~0.7:} \citep{bradley2024}, 
{\tt AstroPy~5.1} \citep{AstropyCollaboration2022}, 
{\tt AdaMet 2.0} \citep{Cappellari2013a}, 
{\tt JamPy~7.2.4} \citep{Cappellari2020}, 
{\tt MgeFit~6.0.4} \citep{Cappellari2002},
{\tt Profiler} \citep{Ciambur2015}, and 
{\tt Profiler} \citep{Ciambur2016}.
}

\bibliographystyle{aasjournalv7}
\bibliography{ic342}

\begin{thebibliography}{}
\expandafter\ifx\csname natexlab\endcsname\relax\def\natexlab#1{#1}\fi
\providecommand{\url}[1]{\href{#1}{#1}}
\providecommand{\dodoi}[1]{doi:~\href{http://doi.org/#1}{\nolinkurl{#1}}}
\providecommand{\doeprint}[1]{\href{http://ascl.net/#1}{\nolinkurl{http://ascl.net/#1}}}
\providecommand{\doarXiv}[1]{\href{https://arxiv.org/abs/#1}{\nolinkurl{https://arxiv.org/abs/#1}}}

\bibitem[{P. {Amaro-Seoane} {et~al.}(2017){Amaro-Seoane}, {Audley}, {Babak},
  {Baker}, {Barausse}, {Bender}, {Berti}, {Binetruy}, {Born}, {Bortoluzzi},
  {Camp}, {Caprini}, {Cardoso}, {Colpi}, {Conklin}, {Cornish}, {Cutler},
  {Danzmann}, {Dolesi}, {Ferraioli}, {Ferroni}, {Fitzsimons}, {Gair}, {Gesa
  Bote}, {Giardini}, {Gibert}, {Grimani}, {Halloin}, {Heinzel}, {Hertog},
  {Hewitson}, {Holley-Bockelmann}, {Hollington}, {Hueller}, {Inchauspe},
  {Jetzer}, {Karnesis}, {Killow}, {Klein}, {Klipstein}, {Korsakova}, {Larson},
  {Livas}, {Lloro}, {Man}, {Mance}, {Martino}, {Mateos}, {McKenzie},
  {McWilliams}, {Miller}, {Mueller}, {Nardini}, {Nelemans}, {Nofrarias},
  {Petiteau}, {Pivato}, {Plagnol}, {Porter}, {Reiche}, {Robertson},
  {Robertson}, {Rossi}, {Russano}, {Schutz}, {Sesana}, {Shoemaker}, {Slutsky},
  {Sopuerta}, {Sumner}, {Tamanini}, {Thorpe}, {Troebs}, {Vallisneri},
  {Vecchio}, {Vetrugno}, {Vitale}, {Volonteri}, {Wanner}, {Ward}, {Wass},
  {Weber}, {Ziemer}, \& {Zweifel}}]{Amaro-Seoane2017}
{Amaro-Seoane}, P., {Audley}, H., {Babak}, S., {et~al.} 2017,
  \bibinfo{title}{{Laser Interferometer Space Antenna},} arXiv e-prints,
  arXiv:1702.00786, \dodoi{10.48550/arXiv.1702.00786}

\bibitem[{F. {Antonini} {et~al.}(2015{\natexlab{a}}){Antonini}, {Barausse}, \&
  {Silk}}]{Antonini2015ApJ...812...72A}
{Antonini}, F., {Barausse}, E., \& {Silk}, J. 2015{\natexlab{a}},
  \bibinfo{title}{{The Coevolution of Nuclear Star Clusters, Massive Black
  Holes, and Their Host Galaxies},} \apj, 812, 72,
  \dodoi{10.1088/0004-637X/812/1/72}

\bibitem[{F. {Antonini} {et~al.}(2015{\natexlab{b}}){Antonini}, {Barausse}, \&
  {Silk}}]{Antonini2015ApJ...806L...8A}
{Antonini}, F., {Barausse}, E., \& {Silk}, J. 2015{\natexlab{b}},
  \bibinfo{title}{{The Imprint of Massive Black Hole Mergers on the Correlation
  between Nuclear Star Clusters and Their Host Galaxies},} \apjl, 806, L8,
  \dodoi{10.1088/2041-8205/806/1/L8}

\bibitem[{F. {Antonini} {et~al.}(2012){Antonini}, {Capuzzo-Dolcetta},
  {Mastrobuono-Battisti}, \& {Merritt}}]{Antonini2013}
{Antonini}, F., {Capuzzo-Dolcetta}, R., {Mastrobuono-Battisti}, A., \&
  {Merritt}, D. 2012, \bibinfo{title}{{Dissipationless Formation and Evolution
  of the Milky Way Nuclear Star Cluster},} \apj, 750, 111,
  \dodoi{10.1088/0004-637X/750/2/111}

\bibitem[{M. {Arca-Sedda} \& R. {Capuzzo-Dolcetta}(2014){Arca-Sedda} \&
  {Capuzzo-Dolcetta}}]{Arca-Sedda2014}
{Arca-Sedda}, M., \& {Capuzzo-Dolcetta}, R. 2014, \bibinfo{title}{{The globular
  cluster migratory origin of nuclear star clusters},} \mnras, 444, 3738,
  \dodoi{10.1093/mnras/stu1683}

\bibitem[{ {Astropy Collaboration} {et~al.}(2022){Astropy Collaboration},
  {Price-Whelan}, {Lim}, {Earl}, {Starkman}, {Bradley}, {Shupe}, {Patil},
  {Corrales}, {Brasseur}, {N{\"o}the}, {Donath}, {Tollerud}, {Morris},
  {Ginsburg}, {Vaher}, {Weaver}, {Tocknell}, {Jamieson}, {van Kerkwijk},
  {Robitaille}, {Merry}, {Bachetti}, {G{\"u}nther}, {Aldcroft},
  {Alvarado-Montes}, {Archibald}, {B{\'o}di}, {Bapat}, {Barentsen},
  {Baz{\'a}n}, {Biswas}, {Boquien}, {Burke}, {Cara}, {Cara}, {Conroy},
  {Conseil}, {Craig}, {Cross}, {Cruz}, {D'Eugenio}, {Dencheva}, {Devillepoix},
  {Dietrich}, {Eigenbrot}, {Erben}, {Ferreira}, {Foreman-Mackey}, {Fox},
  {Freij}, {Garg}, {Geda}, {Glattly}, {Gondhalekar}, {Gordon}, {Grant},
  {Greenfield}, {Groener}, {Guest}, {Gurovich}, {Handberg}, {Hart},
  {Hatfield-Dodds}, {Homeier}, {Hosseinzadeh}, {Jenness}, {Jones}, {Joseph},
  {Kalmbach}, {Karamehmetoglu}, {Ka{\l}uszy{\'n}ski}, {Kelley}, {Kern},
  {Kerzendorf}, {Koch}, {Kulumani}, {Lee}, {Ly}, {Ma}, {MacBride}, {Maljaars},
  {Muna}, {Murphy}, {Norman}, {O'Steen}, {Oman}, {Pacifici}, {Pascual},
  {Pascual-Granado}, {Patil}, {Perren}, {Pickering}, {Rastogi}, {Roulston},
  {Ryan}, {Rykoff}, {Sabater}, {Sakurikar}, {Salgado}, {Sanghi}, {Saunders},
  {Savchenko}, {Schwardt}, {Seifert-Eckert}, {Shih}, {Jain}, {Shukla}, {Sick},
  {Simpson}, {Singanamalla}, {Singer}, {Singhal}, {Sinha}, {Sip{\H{o}}cz},
  {Spitler}, {Stansby}, {Streicher}, {{\v{S}}umak}, {Swinbank}, {Taranu},
  {Tewary}, {Tremblay}, {de Val-Borro}, {Van Kooten}, {Vasovi{\'c}}, {Verma},
  {de Miranda Cardoso}, {Williams}, {Wilson}, {Winkel}, {Wood-Vasey}, {Xue},
  {Yoachim}, {Zhang}, {Zonca}, \& {Astropy Project
  Contributors}}]{AstropyCollaboration2022}
{Astropy Collaboration}, {Price-Whelan}, A.~M., {Lim}, P.~L., {et~al.} 2022,
  \bibinfo{title}{{The Astropy Project: Sustaining and Growing a
  Community-oriented Open-source Project and the Latest Major Release (v5.0) of
  the Core Package},} \apj, 935, 167, \dodoi{10.3847/1538-4357/ac7c74}

\bibitem[{J.~W. {Atkinson} {et~al.}(2005){Atkinson}, {Collett}, {Marconi},
  {Axon}, {Alonso-Herrero}, {Batcheldor}, {Binney}, {Capetti}, {Carollo},
  {Dressel}, {Ford}, {Gerssen}, {Hughes}, {Macchetto}, {Maciejewski},
  {Merrifield}, {Scarlata}, {Sparks}, {Stiavelli}, {Tsvetanov}, \& {van der
  Marel}}]{Atkinson2005}
{Atkinson}, J.~W., {Collett}, J.~L., {Marconi}, A., {et~al.} 2005,
  \bibinfo{title}{{Supermassive black hole mass measurements for NGC 1300 and
  2748 based on Hubble Space Telescope emission-line gas kinematics},} \mnras,
  359, 504, \dodoi{10.1111/j.1365-2966.2005.08904.x}

\bibitem[{M. {Balcells} {et~al.}(2003){Balcells}, {Graham},
  {Dom{\'\i}nguez-Palmero}, \& {Peletier}}]{Balcells2003}
{Balcells}, M., {Graham}, A.~W., {Dom{\'\i}nguez-Palmero}, L., \& {Peletier},
  R.~F. 2003, \bibinfo{title}{{Galactic Bulges from Hubble Space Telescope
  Near-Infrared Camera Multi-Object Spectrometer Observations: The Lack of
  r$^{1/4}$ Bulges},} \apjl, 582, L79, \dodoi{10.1086/367783}

\bibitem[{M. {Balcells} {et~al.}(2007){Balcells}, {Graham}, \&
  {Peletier}}]{Balcells2007}
{Balcells}, M., {Graham}, A.~W., \& {Peletier}, R.~F. 2007,
  \bibinfo{title}{{Galactic Bulges from Hubble Space Telescope NICMOS
  Observations: Central Galaxian Objects, and Nuclear Profile Slopes},} \apj,
  665, 1084, \dodoi{10.1086/519752}

\bibitem[{D.~S. {Balser} {et~al.}(2017){Balser}, {Wenger}, {Goss}, {Johnson},
  \& {Kepley}}]{Balser2017}
{Balser}, D.~S., {Wenger}, T.~V., {Goss}, W.~M., {Johnson}, K.~E., \& {Kepley},
  A.~A. 2017, \bibinfo{title}{{JVLA Observations of IC 342: Probing Star
  Formation in the Nucleus},} \apj, 844, 73, \dodoi{10.3847/1538-4357/aa7a01}

\bibitem[{F.~E. {Bauer} {et~al.}(2003){Bauer}, {Brandt}, \&
  {Lehmer}}]{Bauer2003}
{Bauer}, F.~E., {Brandt}, W.~N., \& {Lehmer}, B. 2003, \bibinfo{title}{{The
  X-Ray Properties of the Nearby Star-forming Galaxy IC 342: The XMM-Newton
  View},} \aj, 126, 2797, \dodoi{10.1086/379139}

\bibitem[{K. {Bekki} \& K.~C. {Freeman}(2003){Bekki} \& {Freeman}}]{Bekki2003}
{Bekki}, K., \& {Freeman}, K.~C. 2003, \bibinfo{title}{{Formation of
  {\ensuremath{\omega}} Centauri from an ancient nucleated dwarf galaxy in the
  young Galactic disc},} \mnras, 346, L11,
  \dodoi{10.1046/j.1365-2966.2003.07275.x}

\bibitem[{K. {Bekki} \& A.~W. {Graham}(2010){Bekki} \& {Graham}}]{Bekki2010}
{Bekki}, K., \& {Graham}, A.~W. 2010, \bibinfo{title}{{On the Transition from
  Nuclear-cluster- to Black-hole-dominated Galaxy Cores},} \apjl, 714, L313,
  \dodoi{10.1088/2041-8205/714/2/L313}

\bibitem[{J. {Binney} \& G.~A. {Mamon}(1982){Binney} \& {Mamon}}]{Binney1982}
{Binney}, J., \& {Mamon}, G.~A. 1982, \bibinfo{title}{{M/L and velocity
  anisotropy from observations of spherical galaxies, or must M87 have a
  massive black hole ?},} \mnras, 200, 361, \dodoi{10.1093/mnras/200.2.361}

\bibitem[{T. {B{\"o}ker} {et~al.}(1997){B{\"o}ker}, {Forster-Schreiber}, \&
  {Genzel}}]{Boker1997}
{B{\"o}ker}, T., {Forster-Schreiber}, N.~M., \& {Genzel}, R. 1997,
  \bibinfo{title}{{Near-Infrared Imaging Spectroscopy of IC~342: Evolution of a
  Bar-Driven Central Starburst},} \aj, 114, 1883, \dodoi{10.1086/118624}

\bibitem[{T. {B{\"o}ker} {et~al.}(2002){B{\"o}ker}, {Laine}, {van der Marel},
  {Sarzi}, {Rix}, {Ho}, \& {Shields}}]{Boker2002}
{B{\"o}ker}, T., {Laine}, S., {van der Marel}, R.~P., {et~al.} 2002,
  \bibinfo{title}{{A Hubble Space Telescope Census of Nuclear Star Clusters in
  Late-Type Spiral Galaxies. I. Observations and Image Analysis},} \aj, 123,
  1389, \dodoi{10.1086/339025}

\bibitem[{T. {B{\"o}ker} {et~al.}(2004){B{\"o}ker}, {Sarzi}, {McLaughlin}, {van
  der Marel}, {Rix}, {Ho}, \& {Shields}}]{Boker2004}
{B{\"o}ker}, T., {Sarzi}, M., {McLaughlin}, D.~E., {et~al.} 2004,
  \bibinfo{title}{{A Hubble Space Telescope Census of Nuclear Star Clusters in
  Late-Type Spiral Galaxies. II. Cluster Sizes and Structural Parameter
  Correlations},} \aj, 127, 105, \dodoi{10.1086/380911}

\bibitem[{T. {B{\"o}ker} {et~al.}(1999){B{\"o}ker}, {van der Marel}, \&
  {Vacca}}]{Boker1999}
{B{\"o}ker}, T., {van der Marel}, R.~P., \& {Vacca}, W.~D. 1999,
  \bibinfo{title}{{CO Band Head Spectroscopy of IC 342: Mass and Age of the
  Nuclear Star Cluster},} \aj, 118, 831, \dodoi{10.1086/300985}

\bibitem[{L. Bradley {et~al.}(2024)Bradley, Sip{\H o}cz, Robitaille, Tollerud,
  Vin{\'{\i}}cius, Deil, Barbary, Wilson, Busko, Donath, G{\"u}nther, Cara,
  Lim, Me{\ss}linger, Conseil, Burnett, Bostroem, Droettboom, Bray, Bratholm,
  Ginsburg, Jamieson, Barentsen, Craig, Morris, Perrin, Rathi, Pascual, \&
  Georgiev}]{bradley2024}
Bradley, L., Sip{\H o}cz, B., Robitaille, T., {et~al.} 2024,
  \bibinfo{title}{astropy/photutils: 2.0.2,}, 2.0.2 Zenodo,
  \dodoi{10.5281/zenodo.13989456}

\bibitem[{G. {Bruzual} \& S. {Charlot}(2003){Bruzual} \&
  {Charlot}}]{Bruzual2003}
{Bruzual}, G., \& {Charlot}, S. 2003, \bibinfo{title}{{Stellar population
  synthesis at the resolution of 2003},} \mnras, 344, 1000,
  \dodoi{10.1046/j.1365-8711.2003.06897.x}

\bibitem[{M. Cappellari(2002)Cappellari}]{Cappellari2002}
Cappellari, M. 2002, \bibinfo{title}{Efficient multi-Gaussian expansion of
  galaxies,} \mnras, 333, 400, \dodoi{10.1046/j.1365-8711.2002.05412.x}

\bibitem[{M. Cappellari(2008)Cappellari}]{Cappellari08}
Cappellari, M. 2008, \bibinfo{title}{Measuring the inclination and
  mass-to-light ratio of axisymmetric galaxies via anisotropic Jeans models of
  stellar kinematics,} \mnras, 390, 71,
  \dodoi{10.1111/j.1365-2966.2008.13754.x}

\bibitem[{M. {Cappellari}(2017){Cappellari}}]{Cappellari2017}
{Cappellari}, M. 2017, \bibinfo{title}{{Improving the full spectrum fitting
  method: accurate convolution with Gauss-Hermite functions},} \mnras, 466,
  798, \dodoi{10.1093/mnras/stw3020}

\bibitem[{M. Cappellari(2020)Cappellari}]{Cappellari2020}
Cappellari, M. 2020, \bibinfo{title}{{Efficient solution of the anisotropic
  spherically aligned axisymmetric Jeans equations of stellar hydrodynamics for
  galactic dynamics},} \mnras, 494, 4819, \dodoi{10.1093/mnras/staa959}

\bibitem[{M. {Cappellari}(2023){Cappellari}}]{Cappellari23}
{Cappellari}, M. 2023, \bibinfo{title}{{Full spectrum fitting with photometry
  in PPXF: stellar population versus dynamical masses, non-parametric star
  formation history and metallicity for 3200 LEGA-C galaxies at redshift z
  {\ensuremath{\approx}} 0.8},} \mnras, 526, 3273,
  \dodoi{10.1093/mnras/stad2597}

\bibitem[{M. {Cappellari}(2025){Cappellari}}]{Cappellari2025}
{Cappellari}, M. 2025, \bibinfo{title}{{PowerBin: fast adaptive data binning
  with Centroidal Power Diagrams},} \mnras, 544, 1432,
  \dodoi{10.1093/mnras/staf1726}

\bibitem[{M. Cappellari(2026)Cappellari}]{Cappellari2026}
Cappellari, M. 2026, in Encyclopedia of Astrophysics, Volume 4, ed. I.~Mandel,
  Vol.~4 (Amsterdam, The Netherlands: Elsevier), 122--152,
  \dodoi{10.1016/B978-0-443-21439-4.00109-7}

\bibitem[{M. {Cappellari} \& E. {Emsellem}(2004){Cappellari} \&
  {Emsellem}}]{Cappellari2004}
{Cappellari}, M., \& {Emsellem}, E. 2004, \bibinfo{title}{{Parametric Recovery
  of Line-of-Sight Velocity Distributions from Absorption-Line Spectra of
  Galaxies via Penalized Likelihood},} \pasp, 116, 138, \dodoi{10.1086/381875}

\bibitem[{M. {Cappellari} {et~al.}(2009){Cappellari}, {Neumayer}, {Reunanen},
  {van der Werf}, {de Zeeuw}, \& {Rix}}]{Cappellari2009}
{Cappellari}, M., {Neumayer}, N., {Reunanen}, J., {et~al.} 2009,
  \bibinfo{title}{{The mass of the black hole in Centaurus A from SINFONI
  AO-assisted integral-field observations of stellar kinematics},} \mnras, 394,
  660, \dodoi{10.1111/j.1365-2966.2008.14377.x}

\bibitem[{M. {Cappellari} {et~al.}(2013){Cappellari}, {Scott}, {Alatalo},
  {Blitz}, {Bois}, {Bournaud}, {Bureau}, {Crocker}, {Davies}, {Davis}, {de
  Zeeuw}, {Duc}, {Emsellem}, {Khochfar}, {Krajnovi{\'c}}, {Kuntschner},
  {McDermid}, {Morganti}, {Naab}, {Oosterloo}, {Sarzi}, {Serra}, {Weijmans}, \&
  {Young}}]{Cappellari2013a}
{Cappellari}, M., {Scott}, N., {Alatalo}, K., {et~al.} 2013,
  \bibinfo{title}{{The ATLAS$^{3D}$ project - XV. Benchmark for early-type
  galaxies scaling relations from 260 dynamical models: mass-to-light ratio,
  dark matter, Fundamental Plane and Mass Plane},} \mnras, 432, 1709,
  \dodoi{10.1093/mnras/stt562}

\bibitem[{D.~J. {Carson} {et~al.}(2015){Carson}, {Barth}, {Seth}, {den Brok},
  {Cappellari}, {Greene}, {Ho}, \& {Neumayer}}]{Carson15}
{Carson}, D.~J., {Barth}, A.~J., {Seth}, A.~C., {et~al.} 2015,
  \bibinfo{title}{{The Structure of Nuclear Star Clusters in Nearby Late-type
  Spiral Galaxies from Hubble Space Telescope Wide Field Camera 3 Imaging},}
  \aj, 149, 170, \dodoi{10.1088/0004-6256/149/5/170}

\bibitem[{D. {Chattopadhyay} {et~al.}(2023){Chattopadhyay}, {Stegmann},
  {Antonini}, {Barber}, \& {Romero-Shaw}}]{2023MNRAS.526.4908C}
{Chattopadhyay}, D., {Stegmann}, J., {Antonini}, F., {Barber}, J., \&
  {Romero-Shaw}, I.~M. 2023, \bibinfo{title}{{Double black hole mergers in
  nuclear star clusters: eccentricities, spins, masses, and the growth of
  massive seeds},} \mnras, 526, 4908, \dodoi{10.1093/mnras/stad3048}

\bibitem[{B.~C. {Ciambur}(2015){Ciambur}}]{Ciambur2015}
{Ciambur}, B.~C. 2015, \bibinfo{title}{{Beyond Ellipse(s): Accurately Modelling
  the Isophotal Structure of Galaxies with ISOFIT and CMODEL},} \apj, 810, 120,
  \dodoi{10.1088/0004-637X/810/2/120}

\bibitem[{B.~C. {Ciambur}(2016){Ciambur}}]{Ciambur2016}
{Ciambur}, B.~C. 2016, \bibinfo{title}{{Profiler - A Fast and Versatile New
  Program for Decomposing Galaxy Light Profiles},} \pasa, 33, e062,
  \dodoi{10.1017/pasa.2016.60}

\bibitem[{P. {C{\^o}t{\'e}} {et~al.}(2006){C{\^o}t{\'e}}, {Piatek},
  {Ferrarese}, {Jord{\'a}n}, {Merritt}, {Peng}, {Ha{\textcommabelow s}egan},
  {Blakeslee}, {Mei}, {West}, {Milosavljevi{\'c}}, \& {Tonry}}]{Cote2006}
{C{\^o}t{\'e}}, P., {Piatek}, S., {Ferrarese}, L., {et~al.} 2006,
  \bibinfo{title}{{The ACS Virgo Cluster Survey. VIII. The Nuclei of Early-Type
  Galaxies},} \apjs, 165, 57, \dodoi{10.1086/504042}

\bibitem[{L.~P. {Crosthwaite} {et~al.}(2000){Crosthwaite}, {Turner}, \&
  {Ho}}]{Crosthwaite2000}
{Crosthwaite}, L.~P., {Turner}, J.~L., \& {Ho}, P.~T.~P. 2000,
  \bibinfo{title}{{The Neutral Hydrogen Distribution in IC~342},} \aj, 119,
  1720, \dodoi{10.1086/301302}

\bibitem[{L.~P. Crosthwaite {et~al.}(2000)Crosthwaite, Turner, \&
  Ho}]{Crosthwaite_2000}
Crosthwaite, L.~P., Turner, J.~L., \& Ho, P. T.~P. 2000,
  \bibinfo{title}{Structure in the Neutral Hydrogen Disk of the Spiral Galaxy
  IC 342,} The Astronomical Journal, 119, 1720, \dodoi{10.1086/301302}

\bibitem[{B.~L. {Davis} {et~al.}(2018){Davis}, {Graham}, \&
  {Cameron}}]{2018ApJ...869..113D}
{Davis}, B.~L., {Graham}, A.~W., \& {Cameron}, E. 2018, \bibinfo{title}{{Black
  Hole Mass Scaling Relations for Spiral Galaxies. II. M $_{BH}$-M $_{*,tot}$
  and M $_{BH}$-M $_{*,disk}$},} \apj, 869, 113,
  \dodoi{10.3847/1538-4357/aae820}

\bibitem[{T.~A. {Davis} {et~al.}(2020){Davis}, {Nguyen}, {Seth}, {Greene},
  {Nyland}, {Barth}, {Bureau}, {Cappellari}, {den Brok}, {Iguchi}, {Lelli},
  {Liu}, {Neumayer}, {North}, {Onishi}, {Sarzi}, {Smith}, \&
  {Williams}}]{Davis20}
{Davis}, T.~A., {Nguyen}, D.~D., {Seth}, A.~C., {et~al.} 2020,
  \bibinfo{title}{{Revealing the intermediate-mass black hole at the heart of
  the dwarf galaxy NGC 404 with sub-parsec resolution ALMA observations},}
  \mnras, 496, 4061, \dodoi{10.1093/mnras/staa1567}

\bibitem[{V.~P. {Debattista} {et~al.}(2006){Debattista}, {Mayer}, {Carollo},
  {Moore}, {Wadsley}, \& {Quinn}}]{Debattista2006}
{Debattista}, V.~P., {Mayer}, L., {Carollo}, C.~M., {et~al.} 2006,
  \bibinfo{title}{{The Secular Evolution of Disk Structural Parameters},} \apj,
  645, 209, \dodoi{10.1086/504147}

\bibitem[{M. {den Brok} {et~al.}(2015{\natexlab{a}}){den Brok}, {Seth},
  {Barth}, {Carson}, {Neumayer}, {Cappellari}, {Debattista}, {Ho}, {Hood}, \&
  {McDermid}}]{denBrok15}
{den Brok}, M., {Seth}, A.~C., {Barth}, A.~J., {et~al.} 2015{\natexlab{a}},
  \bibinfo{title}{{Measuring the Mass of the Central Black Hole in the
  Bulgeless Galaxy NGC 4395 from Gas Dynamical Modeling},} \apj, 809, 101,
  \dodoi{10.1088/0004-637X/809/1/101}

\bibitem[{M. {den Brok} {et~al.}(2015{\natexlab{b}}){den Brok}, {Seth},
  {Barth}, {Carson}, {Neumayer}, {Cappellari}, {Debattista}, {Ho}, {Hood}, \&
  {McDermid}}]{denBrok2015}
{den Brok}, M., {Seth}, A.~C., {Barth}, A.~J., {et~al.} 2015{\natexlab{b}},
  \bibinfo{title}{{Measuring the Mass of the Central Black Hole in the
  Bulgeless Galaxy NGC 4395 from Gas Dynamical Modeling},} \apj, 809, 101,
  \dodoi{10.1088/0004-637X/809/1/101}

\bibitem[{P. {Dominiak} {et~al.}(2026){Dominiak}, {Bureau}, {Liang},
  {Cappellari}, {Davis}, {Lelli}, {Ruffa}, {Williams}, \&
  {Zhang}}]{Dominiak2026MNRAS.548ag546D}
{Dominiak}, P., {Bureau}, M., {Liang}, F.-H., {et~al.} 2026,
  \bibinfo{title}{{WISDOM project ─ XXVIII. Molecular gas measurement of the
  supermassive black hole mass of the galaxy NGC 1387},} \mnras, 548, stag546,
  \dodoi{10.1093/mnras/stag546}

\bibitem[{D.~A. {Drehmer} {et~al.}(2015){Drehmer}, {Storchi-Bergmann},
  {Ferrari}, {Cappellari}, \& {Riffel}}]{Drehmer2015}
{Drehmer}, D.~A., {Storchi-Bergmann}, T., {Ferrari}, F., {Cappellari}, M., \&
  {Riffel}, R.~A. 2015, \bibinfo{title}{{The benchmark black hole in NGC 4258:
  dynamical models from high-resolution two-dimensional stellar kinematics},}
  \mnras, 450, 128, \dodoi{10.1093/mnras/stv536}

\bibitem[{E. {Emsellem} {et~al.}(1994){Emsellem}, {Monnet}, \&
  {Bacon}}]{Emsellem94}
{Emsellem}, E., {Monnet}, G., \& {Bacon}, R. 1994, \bibinfo{title}{{The
  multi-gaussian expansion method: a tool for building realistic photometric
  and kinematical models of stellar systems I. The formalism},} \aap, 285, 723

\bibitem[{L. {Ferrarese} \& H. {Ford}(2005){Ferrarese} \&
  {Ford}}]{Ferrarese2005}
{Ferrarese}, L., \& {Ford}, H. 2005, \bibinfo{title}{{Supermassive Black Holes
  in Galactic Nuclei: Past, Present and Future Research},} \ssr, 116, 523,
  \dodoi{10.1007/s11214-005-3947-6}

\bibitem[{L. {Ferrarese} \& D. {Merritt}(2000){Ferrarese} \&
  {Merritt}}]{Ferrarese00}
{Ferrarese}, L., \& {Merritt}, D. 2000, \bibinfo{title}{{A Fundamental Relation
  between Supermassive Black Holes and Their Host Galaxies},} \apjl, 539, L9,
  \dodoi{10.1086/312838}

\bibitem[{K. {Gebhardt} {et~al.}(2000){Gebhardt}, {Bender}, {Bower},
  {Dressler}, {Faber}, {Filippenko}, {Green}, {Grillmair}, {Ho}, {Kormendy},
  {Lauer}, {Magorrian}, {Pinkney}, {Richstone}, \& {Tremaine}}]{Gebhardt00}
{Gebhardt}, K., {Bender}, R., {Bower}, G., {et~al.} 2000, \bibinfo{title}{{A
  Relationship between Nuclear Black Hole Mass and Galaxy Velocity
  Dispersion},} \apjl, 539, L13, \dodoi{10.1086/312840}

\bibitem[{K. {Gebhardt} {et~al.}(2001){Gebhardt}, {Lauer}, {Kormendy},
  {Pinkney}, {Bower}, {Green}, {Gull}, {Hutchings}, {Kaiser}, {Nelson},
  {Richstone}, \& {Weistrop}}]{Gebhardt01}
{Gebhardt}, K., {Lauer}, T.~R., {Kormendy}, J., {et~al.} 2001,
  \bibinfo{title}{{M33: A Galaxy with No Supermassive Black Hole},} \aj, 122,
  2469, \dodoi{10.1086/323481}

\bibitem[{R. {Genzel} \& C.~H. {Townes}(1987){Genzel} \&
  {Townes}}]{1987ARA&A..25..377G}
{Genzel}, R., \& {Townes}, C.~H. 1987, \bibinfo{title}{{Physical conditions,
  dynamics, and mass distribution in the center of the galaxy.},} \araa, 25,
  377, \dodoi{10.1146/annurev.aa.25.090187.002113}

\bibitem[{R.~M. {Gonz{\'a}lez Delgado} {et~al.}(2008){Gonz{\'a}lez Delgado},
  {P{\'e}rez}, {Cid Fernandes}, \& {Schmitt}}]{2008AJ....135..747G}
{Gonz{\'a}lez Delgado}, R.~M., {P{\'e}rez}, E., {Cid Fernandes}, R., \&
  {Schmitt}, H. 2008, \bibinfo{title}{{HST/WFPC2 Imaging of the Circumnuclear
  Structure of Low-Luminosity Active Galactic Nuclei. I. Data and Nuclear
  Morphology},} \aj, 135, 747, \dodoi{10.1088/0004-6256/135/3/747}

\bibitem[{A.~W. {Graham}(2008){Graham}}]{Graham2008}
{Graham}, A.~W. 2008, \bibinfo{title}{{Fundamental Planes and the Barless
  M$_{BH}$-{\ensuremath{\sigma}} Relation for Supermassive Black Holes},} \apj,
  680, 143, \dodoi{10.1086/587473}

\bibitem[{A.~W. {Graham}(2016{\natexlab{a}}){Graham}}]{Graham2016}
{Graham}, A.~W. 2016{\natexlab{a}}, in IAU Symposium, Vol. 312, Star Clusters
  and Black Holes in Galaxies across Cosmic Time, ed. Y.~{Meiron}, S.~{Li},
  F.-K. {Liu}, \& R.~{Spurzem}, 269--273, \dodoi{10.1017/S1743921315008017}

\bibitem[{A.~W. {Graham}(2016{\natexlab{b}}){Graham}}]{2016ASSL..418..263G}
{Graham}, A.~W. 2016{\natexlab{b}}, in Astrophysics and Space Science Library,
  Vol. 418, Galactic Bulges, ed. E.~{Laurikainen}, R.~{Peletier}, \&
  D.~{Gadotti}, 263, \dodoi{10.1007/978-3-319-19378-6_11}

\bibitem[{A.~W. {Graham}(2020){Graham}}]{Graham2020}
{Graham}, A.~W. 2020, \bibinfo{title}{{A consistency test for determining
  whether ultracompact dwarf galaxies could be the remnant nuclei of threshed
  galaxies},} \mnras, 492, 3263, \dodoi{10.1093/mnras/stz3547}

\bibitem[{A.~W. {Graham}(2023{\natexlab{a}}){Graham}}]{2023MNRAS.522.3588G}
{Graham}, A.~W. 2023{\natexlab{a}}, \bibinfo{title}{{Resequencing the Hubble
  sequence and the quadratic (black hole mass)-(spheroid stellar mass) relation
  for elliptical galaxies},} \mnras, 522, 3588, \dodoi{10.1093/mnras/stad1124}

\bibitem[{A.~W.
  {Graham}(2023{\natexlab{b}}){Graham}}]{Graham2023MNRAS.518.6293G}
{Graham}, A.~W. 2023{\natexlab{b}}, \bibinfo{title}{{Invoking the virial
  theorem to understand the impact of (dry) mergers on the
  M$_{bh}$-{\ensuremath{\sigma}} relation},} \mnras, 518, 6293,
  \dodoi{10.1093/mnras/stac3173}

\bibitem[{A.~W. {Graham}(2026){Graham}}]{Graham2026MNRAS.549ag782G}
{Graham}, A.~W. 2026, \bibinfo{title}{{Clash of the Trident and tuning fork:
  insights from bar and spiral strength in the (massive black hole)─stellar
  mass diagrams, and the 'Triangal' galaxy evolution schema},} \mnras, 549,
  stag782, \dodoi{10.1093/mnras/stag782}

\bibitem[{A.~W. {Graham} {et~al.}(2025){Graham}, {Chilingarian}, {Nguyen},
  {Soria}, {Durr{\'e}}, \& {Forbes}}]{Graham2025}
{Graham}, A.~W., {Chilingarian}, I., {Nguyen}, D.~D., {et~al.} 2025,
  \bibinfo{title}{{Dot to dot: High-z little red dots in {\textendash} diagrams
  with galaxy-morphology-specific scaling relations},} \pasa, 42, e068,
  \dodoi{10.1017/pasa.2025.10035}

\bibitem[{A.~W. {Graham} {et~al.}(2001){Graham}, {Erwin}, {Caon}, \&
  {Trujillo}}]{Graham2001}
{Graham}, A.~W., {Erwin}, P., {Caon}, N., \& {Trujillo}, I. 2001,
  \bibinfo{title}{{A Correlation between Galaxy Light Concentration and
  Supermassive Black Hole Mass},} \apjl, 563, L11, \dodoi{10.1086/338500}

\bibitem[{A.~W. {Graham} \& R. {Guzm{\'a}n}(2003){Graham} \&
  {Guzm{\'a}n}}]{Graham2003}
{Graham}, A.~W., \& {Guzm{\'a}n}, R. 2003, \bibinfo{title}{{HST Photometry of
  Dwarf Elliptical Galaxies in Coma, and an Explanation for the Alleged
  Structural Dichotomy between Dwarf and Bright Elliptical Galaxies},} \aj,
  125, 2936, \dodoi{10.1086/374992}

\bibitem[{A.~W. {Graham} \& N. {Sahu}(2023){Graham} \&
  {Sahu}}]{Graham2023MNRAS.518.2177G}
{Graham}, A.~W., \& {Sahu}, N. 2023, \bibinfo{title}{{Appreciating mergers for
  understanding the non-linear M$_{bh}$-M$_{*,spheroid}$ and M$_{bh}$-M$_{*,
  galaxy}$ relations, updated herein, and the implications for the (reduced)
  role of AGN feedback},} \mnras, 518, 2177, \dodoi{10.1093/mnras/stac2019}

\bibitem[{A.~W. {Graham} \& R. {Soria}(2019){Graham} \&
  {Soria}}]{GrahamSoria2019}
{Graham}, A.~W., \& {Soria}, R. 2019, \bibinfo{title}{{Expected
  Intermediate-Mass Black Holes in the Virgo Cluster. I. Early-Type Galaxies},}
  \mnras, 484, 794, \dodoi{10.1093/mnras/sty3068}

\bibitem[{A.~W. {Graham} \& L.~R. {Spitler}(2009){Graham} \&
  {Spitler}}]{Graham2009}
{Graham}, A.~W., \& {Spitler}, L.~R. 2009, \bibinfo{title}{{Quantifying the
  coexistence of massive black holes and dense nuclear star clusters},} \mnras,
  397, 2148, \dodoi{10.1111/j.1365-2966.2009.15118.x}

\bibitem[{J.~E. {Greene} {et~al.}(2010){Greene}, {Peng}, {Kim}, {Kuo},
  {Braatz}, {Impellizzeri}, {Condon}, {Lo}, {Henkel}, \& {Reid}}]{Greene2010}
{Greene}, J.~E., {Peng}, C.~Y., {Kim}, M., {et~al.} 2010,
  \bibinfo{title}{{Precise Black Hole Masses from Megamaser Disks: Black
  Hole-Bulge Relations at Low Mass},} \apj, 721, 26,
  \dodoi{10.1088/0004-637X/721/1/26}

\bibitem[{H. Haario {et~al.}(2001)Haario, Saksman, \& Tamminen}]{Haario2001}
Haario, H., Saksman, E., \& Tamminen, J. 2001, \bibinfo{title}{{An adaptive
  Metropolis algorithm},} Bernoulli, 7, 223

\bibitem[{M. {H{\"a}berle} {et~al.}(2024){H{\"a}berle}, {Neumayer}, {Seth},
  {Bellini}, {Libralato}, {Baumgardt}, {Whitaker}, {Dumont}, {Alfaro-Cuello},
  {Anderson}, {Clontz}, {Kacharov}, {Kamann}, {Feldmeier-Krause}, {Milone},
  {Nitschai}, {Pechetti}, \& {van de Ven}}]{Haberle2024}
{H{\"a}berle}, M., {Neumayer}, N., {Seth}, A., {et~al.} 2024,
  \bibinfo{title}{{Fast-moving stars around an intermediate-mass black hole in
  {\ensuremath{\omega}} Centauri},} \nat, 631, 285,
  \dodoi{10.1038/s41586-024-07511-z}

\bibitem[{J.~J. {Han} {et~al.}(2025){Han}, {El-Badry}, {Lucchini}, {Hernquist},
  {Brown}, {Garavito-Camargo}, {Conroy}, \& {Sari}}]{Han2025}
{Han}, J.~J., {El-Badry}, K., {Lucchini}, S., {et~al.} 2025,
  \bibinfo{title}{{Hypervelocity Stars Trace a Supermassive Black Hole in the
  Large Magellanic Cloud},} \apj, 982, 188, \dodoi{10.3847/1538-4357/adb967}

\bibitem[{N. {H{\"a}ring} \& H.-W. {Rix}(2004){H{\"a}ring} \& {Rix}}]{Haring04}
{H{\"a}ring}, N., \& {Rix}, H.-W. 2004, \bibinfo{title}{{On the Black Hole
  Mass-Bulge Mass Relation},} \apjl, 604, L89, \dodoi{10.1086/383567}

\bibitem[{C.~R. Harris {et~al.}(2020)Harris, Millman, van~der Walt, Gommers,
  Virtanen, Cournapeau, Wieser, Taylor, Berg, Smith, Kern, Picus, Hoyer, van
  Kerkwijk, Brett, Haldane, del R{\'{\i}}o, Wiebe, Peterson,
  G{\'{e}}rard-Marchant, Sheppard, Reddy, Weckesser, Abbasi, Gohlke, \&
  Oliphant}]{Harris2020}
Harris, C.~R., Millman, K.~J., van~der Walt, S.~J., {et~al.} 2020,
  \bibinfo{title}{Array programming with {NumPy},} Nature, 585, 357,
  \dodoi{10.1038/s41586-020-2649-2}

\bibitem[{L. {Hernquist}(1992){Hernquist}}]{Hernquist1992}
{Hernquist}, L. 1992, \bibinfo{title}{{Structure of Merger Remnants. I.
  Bulgeless Progenitors},} \apj, 400, 460, \dodoi{10.1086/172009}

\bibitem[{J. {Hu}(2008){Hu}}]{Hu2008}
{Hu}, J. 2008, \bibinfo{title}{{The black hole mass-stellar velocity dispersion
  correlation: bulges versus pseudo-bulges},} \mnras, 386, 2242,
  \dodoi{10.1111/j.1365-2966.2008.13195.x}

\bibitem[{J.~D. Hunter(2007)Hunter}]{Hunter2007}
Hunter, J.~D. 2007, \bibinfo{title}{Matplotlib: A 2D graphics environment,}
  Computing In Science \& Engineering, 9, 90, \dodoi{10.1109/MCSE.2007.55}

\bibitem[{T.-O. Husser {et~al.}(2013)Husser, Wende-von Berg, Dreizler, Homeier,
  Reiners, Barman, \& Hauschildt}]{Husser2013}
Husser, T.-O., Wende-von Berg, S., Dreizler, S., {et~al.} 2013,
  \bibinfo{title}{A new extensive library of PHOENIX stellar atmospheres and
  synthetic spectra,} \aap, 553, A6, \dodoi{10.1051/0004-6361/201219058}

\bibitem[{R.~A. {Ibata} {et~al.}(2019){Ibata}, {Bellazzini}, {Malhan},
  {Martin}, \& {Bianchini}}]{Ibata2019}
{Ibata}, R.~A., {Bellazzini}, M., {Malhan}, K., {Martin}, N., \& {Bianchini},
  P. 2019, \bibinfo{title}{{Identification of the long stellar stream of the
  prototypical massive globular cluster {\ensuremath{\omega}} Centauri},}
  Nature Astronomy, 3, 667, \dodoi{10.1038/s41550-019-0751-x}

\bibitem[{I.~D. {Karachentsev} {et~al.}(2003){Karachentsev}, {Sharina},
  {Dolphin}, {Grebel}, {Geisler}, {Guhathakurta}, {Hodge}, {Karachentseva},
  {Sarajedini}, \& {Seitzer}}]{Karachentsev2003}
{Karachentsev}, I.~D., {Sharina}, M.~E., {Dolphin}, A.~E., {et~al.} 2003,
  \bibinfo{title}{{Distances to Nearby Galaxies Around IC~342},} \aap, 404, 93,
  \dodoi{10.1051/0004-6361:20030170}

\bibitem[{J. Kormendy \& L.~C. Ho(2013)Kormendy \& Ho}]{Kormendy13}
Kormendy, J., \& Ho, L.~C. 2013, \bibinfo{title}{{Coevolution (Or Not) of
  Supermassive Black Holes and Host Galaxies},} \araa, 51, 511,
  \dodoi{10.1146/annurev-astro-082708-101811}

\bibitem[{D. {Krajnovi{\'c}} {et~al.}(2006){Krajnovi{\'c}}, {Cappellari}, {de
  Zeeuw}, \& {Copin}}]{Krajnovic2006}
{Krajnovi{\'c}}, D., {Cappellari}, M., {de Zeeuw}, P.~T., \& {Copin}, Y. 2006,
  \bibinfo{title}{{Kinemetry: a generalization of photometry to the higher
  moments of the line-of-sight velocity distribution},} \mnras, 366, 787,
  \dodoi{10.1111/j.1365-2966.2005.09902.x}

\bibitem[{P.-Y. {Lablanche} {et~al.}(2012){Lablanche}, {Cappellari},
  {Emsellem}, {et~al.}}]{Lablanche2012}
{Lablanche}, P.-Y., {Cappellari}, M., {Emsellem}, E., {et~al.} 2012,
  \bibinfo{title}{{The ATLAS$^{\rm 3D}$ project -- XII. Recovery of the
  mass-to-light ratio of simulated early-type barred galaxies with axisymmetric
  dynamical models},} \mnras, 424, 1495,
  \dodoi{10.1111/j.1365-2966.2012.21346.x}

\bibitem[{J. Larkin {et~al.}(2006)Larkin, Barczys, Krabbe, Adkins, Aliado,
  Amico, Brims, Campbell, Canfield, Gasaway, Honey, Iserlohe, Johnson, Kress,
  Lafreniere, Magnone, Magnone, McElwain, Moon, Quirrenbach, Skulason, Song,
  Spencer, Weiss, \& Wright}]{Larkin2006}
Larkin, J., Barczys, M., Krabbe, A., {et~al.} 2006, \bibinfo{title}{{OSIRIS: A
  diffraction limited integral field spectrograph for Keck},} New Astron Rev,
  50, 362, \dodoi{10.1016/j.newar.2006.02.005}

\bibitem[{T.~Q.~T. Le {et~al.}(2026)Le, Nguyen, Ngo, Ho, Le, \&
  Nguyen}]{LeTinh2026}
Le, T. Q.~T., Nguyen, D.~D., Ngo, H.~N., {et~al.} 2026, \bibinfo{title}{Probing
  the Variation of the Inner Surface-Brightness Profile of Nuclear Star
  Clusters on the Intermediate-Mass Black Hole Mass Measurements Using Mock
  Observations of ELT/MICADO and HARMONI,} Universe, 12, 160,
  \dodoi{10.3390/universe12060160}

\bibitem[{K.~E. {Lockhart} {et~al.}(2019){Lockhart}, {Do}, {Larkin}, {Boehle},
  {Campbell}, {Chappell}, {Chu}, {Ciurlo}, {Cosens}, {Fitzgerald}, {Ghez},
  {Lu}, {Lyke}, {Mieda}, {Rudy}, {Vayner}, {Walth}, \& {Wright}}]{Lockhart2019}
{Lockhart}, K.~E., {Do}, T., {Larkin}, J.~E., {et~al.} 2019,
  \bibinfo{title}{{Characterizing and Improving the Data Reduction Pipeline for
  the Keck OSIRIS Integral Field Spectrograph},} \aj, 157, 75,
  \dodoi{10.3847/1538-3881/aaf64e}

\bibitem[{G. {Lodato} \& P. {Natarajan}(2006){Lodato} \&
  {Natarajan}}]{Lodato2006}
{Lodato}, G., \& {Natarajan}, P. 2006, \bibinfo{title}{{Supermassive black hole
  formation during the assembly of pre-galactic discs},} \mnras, 371, 1813,
  \dodoi{10.1111/j.1365-2966.2006.10801.x}

\bibitem[{P. {Madau} \& M.~J. {Rees}(2001){Madau} \& {Rees}}]{Madau2001}
{Madau}, P., \& {Rees}, M.~J. 2001, \bibinfo{title}{{Massive Black Holes as
  Population III Remnants},} \apjl, 551, L27, \dodoi{10.1086/319848}

\bibitem[{J. {Magorrian}(1999){Magorrian}}]{Magorrian1999}
{Magorrian}, J. 1999, \bibinfo{title}{{Kinematical signatures of hidden stellar
  discs},} \mnras, 302, 530, \dodoi{10.1046/j.1365-8711.1999.02135.x}

\bibitem[{J. {Magorrian} {et~al.}(1998){Magorrian}, {Tremaine}, {Richstone},
  {Bender}, {Bower}, {Dressler}, {Faber}, {Gebhardt}, {Green}, {Grillmair},
  {Kormendy}, \& {Lauer}}]{1998AJ....115.2285M}
{Magorrian}, J., {Tremaine}, S., {Richstone}, D., {et~al.} 1998,
  \bibinfo{title}{{The Demography of Massive Dark Objects in Galaxy Centers},}
  \aj, 115, 2285, \dodoi{10.1086/300353}

\bibitem[{D.~S.~S. {Mak} {et~al.}(2008){Mak}, {Pun}, \& {Kong}}]{Mak2008}
{Mak}, D.~S.~S., {Pun}, C.~S.~J., \& {Kong}, A.~K.~H. 2008,
  \bibinfo{title}{{High-Resolution X-Ray Imaging of the Center of IC~342},}
  \apj, 686, 995, \dodoi{10.1086/591482}

\bibitem[{A. {Marconi} {et~al.}(2000){Marconi}, {Schreier}, {Koekemoer},
  {Capetti}, {Axon}, {Macchetto}, \& {Caon}}]{Marconi2000}
{Marconi}, A., {Schreier}, E.~J., {Koekemoer}, A., {et~al.} 2000,
  \bibinfo{title}{{Unveiling the Active Nucleus of Centaurus A},} \apj, 528,
  276, \dodoi{10.1086/308168}

\bibitem[{D. {Massari} {et~al.}(2019){Massari}, {Koppelman}, \&
  {Helmi}}]{Massari2019}
{Massari}, D., {Koppelman}, H.~H., \& {Helmi}, A. 2019, \bibinfo{title}{{Origin
  of the system of globular clusters in the Milky Way},} \aap, 630, L4,
  \dodoi{10.1051/0004-6361/201936135}

\bibitem[{D. {Merritt} {et~al.}(2001){Merritt}, {Ferrarese}, \&
  {Joseph}}]{Merritt2001}
{Merritt}, D., {Ferrarese}, L., \& {Joseph}, C.~L. 2001, \bibinfo{title}{{No
  Supermassive Black Hole in M33?},} Science, 293, 1116,
  \dodoi{10.1126/science.1063896}

\bibitem[{M. Milosavljevi{\'c} \& D. Merritt(2001)Milosavljevi{\'c} \&
  Merritt}]{Milosavljevic2001}
Milosavljevi{\'c}, M., \& Merritt, D. 2001, \bibinfo{title}{Formation of
  Galactic Nuclei,} \apj, 563, 34, \dodoi{10.1086/323830}

\bibitem[{M. {Mitzkus} {et~al.}(2017){Mitzkus}, {Cappellari}, \&
  {Walcher}}]{Mitzkus2017}
{Mitzkus}, M., {Cappellari}, M., \& {Walcher}, C.~J. 2017,
  \bibinfo{title}{{Dominant dark matter and a counter-rotating disc: MUSE view
  of the low-luminosity S0 galaxy NGC 5102},} \mnras, 464, 4789,
  \dodoi{10.1093/mnras/stw2677}

\bibitem[{N. {Neumayer} {et~al.}(2020){Neumayer}, {Seth}, \&
  {B{\"o}ker}}]{Neumayer2020}
{Neumayer}, N., {Seth}, A., \& {B{\"o}ker}, T. 2020, \bibinfo{title}{{Nuclear
  Star Clusters},} \aapr, 28, 4, \dodoi{10.1007/s00159-020-00125-0}

\bibitem[{N. {Neumayer} \& A.~C. {Seth}(2011){Neumayer} \&
  {Seth}}]{NeumaerSeth2011}
{Neumayer}, N., \& {Seth}, A.~C. 2011, \bibinfo{title}{{Robustly Detected
  Nuclear Star Clusters in Galaxies with and Without Active Galactic Nuclei},}
  \apj, 743, 120, \dodoi{10.1088/0004-637X/743/2/120}

\bibitem[{H.~N. {Ngo} {et~al.}(2025{\natexlab{a}}){Ngo}, {Nguyen}, {Le}, {Ho},
  {Nguyen}, \& {Dang}}]{Ngo2025c}
{Ngo}, H.~N., {Nguyen}, D.~D., {Le}, T. T.~Q., {et~al.} 2025{\natexlab{a}},
  \bibinfo{title}{{Detecting Intermediate-Mass Black Holes out to 20 Mpc with
  ELT/HARMONI: The Case of FCC 119},} arXiv e-prints, arXiv:2510.22959.
\newblock \doarXiv{2510.22959}

\bibitem[{H.~N. {Ngo} {et~al.}(2025{\natexlab{b}}){Ngo}, {Nguyen}, {Nguyen},
  {Dang}, \& {Ho}}]{Ngo2025b}
{Ngo}, H.~N., {Nguyen}, D.~D., {Nguyen}, T.~N., {Dang}, T.~H., \& {Ho}, T.
  H.~T. 2025{\natexlab{b}}, \bibinfo{title}{{Extending the simulations of
  intermediate-mass black hole mass measurements to Virgo Cluster using
  ELT/HARMONI high resolution integral-field stellar kinematics},} arXiv
  e-prints, arXiv:2509.03364.
\newblock \doarXiv{2509.03364}

\bibitem[{H.~N. {Ngo} {et~al.}(2025{\natexlab{c}}){Ngo}, {Nguyen}, {Le}, {Ho},
  {Ho}, {Gallo}, {Nyland}, {Imanishi}, {Nakanishi}, {Le}, {Pacucci}, \&
  {Girma}}]{Ngo2025a}
{Ngo}, H.~N., {Nguyen}, D.~D., {Le}, T. Q.~T., {et~al.} 2025{\natexlab{c}},
  \bibinfo{title}{{Revisiting the Supermassive Black Hole Mass of NGC 7052
  Using High Spatial Resolution Molecular Gas Observed with ALMA},} \apj, 992,
  211, \dodoi{10.3847/1538-4357/ae0455}

\bibitem[{D.~D. {Nguyen} {et~al.}(2023){Nguyen}, {Cappellari}, \&
  {Pereira-Santaella}}]{Nguyen23}
{Nguyen}, D.~D., {Cappellari}, M., \& {Pereira-Santaella}, M. 2023,
  \bibinfo{title}{{Simulating supermassive black hole mass measurements for a
  sample of ultramassive galaxies using ELT/HARMONI high-spatial-resolution
  integral-field stellar kinematics},} \mnras, 526, 3548,
  \dodoi{10.1093/mnras/stad2860}

\bibitem[{D.~D. {Nguyen} {et~al.}(2014){Nguyen}, {Seth}, {Reines}, {den Brok},
  {Sand}, \& {McLeod}}]{Nguyen14}
{Nguyen}, D.~D., {Seth}, A.~C., {Reines}, A.~E., {et~al.} 2014,
  \bibinfo{title}{{Extended Structure and Fate of the Nucleus in Henize 2-10},}
  \apj, 794, 34, \dodoi{10.1088/0004-637X/794/1/34}

\bibitem[{D.~D. {Nguyen} {et~al.}(2017){Nguyen}, {Seth}, {den Brok},
  {Neumayer}, {Cappellari}, {Barth}, {Caldwell}, {Williams}, \&
  {Binder}}]{Nguyen17}
{Nguyen}, D.~D., {Seth}, A.~C., {den Brok}, M., {et~al.} 2017,
  \bibinfo{title}{{Improved Dynamical Constraints on the Mass of the Central
  Black Hole in NGC 404},} \apj, 836, 237, \dodoi{10.3847/1538-4357/aa5cb4}

\bibitem[{D.~D. {Nguyen} {et~al.}(2018){Nguyen}, {Seth}, {Neumayer}, {Kamann},
  {Voggel}, {Cappellari}, {Picotti}, {Nguyen}, {B{\"o}ker}, {Debattista},
  {Caldwell}, {McDermid}, {Bastian}, {Ahn}, \& {Pechetti}}]{Nguyen18}
{Nguyen}, D.~D., {Seth}, A.~C., {Neumayer}, N., {et~al.} 2018,
  \bibinfo{title}{{Nearby Early-type Galactic Nuclei at High Resolution:
  Dynamical Black Hole and Nuclear Star Cluster Mass Measurements},} \apj, 858,
  118, \dodoi{10.3847/1538-4357/aabe28}

\bibitem[{D.~D. {Nguyen} {et~al.}(2019){Nguyen}, {Seth}, {Neumayer}, {Iguchi},
  {Cappellari}, {Strader}, {Chomiuk}, {Tremou}, {Pacucci}, {Nakanishi},
  {Bahramian}, {Nguyen}, {den Brok}, {Ahn}, {Voggel}, {Kacharov}, {Tsukui},
  {Ly}, {Dumont}, \& {Pechetti}}]{Nguyen19}
{Nguyen}, D.~D., {Seth}, A.~C., {Neumayer}, N., {et~al.} 2019,
  \bibinfo{title}{{Improved Dynamical Constraints on the Masses of the Central
  Black Holes in Nearby Low-mass Early-type Galactic Nuclei and the First Black
  Hole Determination for NGC 205},} \apj, 872, 104,
  \dodoi{10.3847/1538-4357/aafe7a}

\bibitem[{D.~D. {Nguyen} {et~al.}(2020){Nguyen}, {den Brok}, {Seth}, {Davis},
  {Greene}, {Cappellari}, {Jensen}, {Thater}, {Iguchi}, {Imanishi}, {Izumi},
  {Nyland}, {Neumayer}, {Nakanishi}, {Nguyen}, {Tsukui}, {Bureau}, {Onishi},
  {Nguyen}, \& {Le}}]{Nguyen20}
{Nguyen}, D.~D., {den Brok}, M., {Seth}, A.~C., {et~al.} 2020,
  \bibinfo{title}{{The MBHBM$_{{\ensuremath{\star}}}$ Project. I. Measurement
  of the Central Black Hole Mass in Spiral Galaxy NGC 3504 Using Molecular Gas
  Kinematics},} \apj, 892, 68, \dodoi{10.3847/1538-4357/ab77aa}

\bibitem[{D.~D. {Nguyen} {et~al.}(2021){Nguyen}, {Izumi}, {Thater}, {Imanishi},
  {Kawamuro}, {Baba}, {Nakano}, {Turner}, {Kohno}, {Matsushita}, {Mart{\'\i}n},
  {Meier}, {Nguyen}, \& {Nguyen}}]{Nguyen21}
{Nguyen}, D.~D., {Izumi}, T., {Thater}, S., {et~al.} 2021,
  \bibinfo{title}{{Black hole mass measurement using ALMA observations of [CI]
  and CO emissions in the Seyfert 1 galaxy NGC 7469},} \mnras, 504, 4123,
  \dodoi{10.1093/mnras/stab1002}

\bibitem[{D.~D. {Nguyen} {et~al.}(2022){Nguyen}, {Bureau}, {Thater}, {Nyland},
  {den Brok}, {Cappellari}, {Davis}, {Greene}, {Neumayer}, {Imanishi}, {Izumi},
  {Kawamuro}, {Baba}, {Nguyen}, {Iguchi}, {Tsukui}, {Lam}, \& {Ho}}]{Nguyen22}
{Nguyen}, D.~D., {Bureau}, M., {Thater}, S., {et~al.} 2022,
  \bibinfo{title}{{The MBHBM$^{{\ensuremath{\star}}}$ Project - II. Molecular
  gas kinematics in the lenticular galaxy NGC 3593 reveal a supermassive black
  hole},} \mnras, 509, 2920, \dodoi{10.1093/mnras/stab3016}

\bibitem[{D.~D. {Nguyen} {et~al.}(2025{\natexlab{a}}){Nguyen}, {Cappellari},
  {Ngo}, {Le}, {Le}, {Ho}, {Nguyen}, {On}, {Tong}, {Thatte}, \&
  {Pereira-Santaella}}]{Nguyen2025b}
{Nguyen}, D.~D., {Cappellari}, M., {Ngo}, H.~N., {et~al.} 2025{\natexlab{a}},
  \bibinfo{title}{{Simulating Intermediate Black Hole Mass Measurements for a
  Sample of Galaxies with Nuclear Star Clusters Using ELT/HARMONI High Spatial
  Resolution Integral-field Stellar Kinematics},} \aj, 170, 124,
  \dodoi{10.3847/1538-3881/ade9ba}

\bibitem[{D.~D. {Nguyen} {et~al.}(2025{\natexlab{b}}){Nguyen}, {Ngo}, {Le},
  {Graham}, {Soria}, {Chilingarian}, {Thatte}, {Phuong}, {Hoang},
  {Pereira-Santaella}, {Durre}, {Pham}, {Ngoc Tram}, {Ngoc}, \&
  {L{\^e}}}]{Nguyen2025a}
{Nguyen}, D.~D., {Ngo}, H.~N., {Le}, T. Q.~T., {et~al.} 2025{\natexlab{b}},
  \bibinfo{title}{{Supermassive black hole mass measurement in the spiral
  galaxy NGC 4736 using JWST/NIRSpec stellar kinematics},} \aap, 698, L9,
  \dodoi{10.1051/0004-6361/202554672}

\bibitem[{D.~D. {Nguyen} {et~al.}(2025{\natexlab{c}}){Nguyen}, {Cappellari},
  {Ngo}, {Le}, {Le}, {Ho}, {Nguyen}, {On}, {Tong}, {Thatte}, \&
  {Pereira-Santaella}}]{Nguyen_2025_NSC}
{Nguyen}, D.~D., {Cappellari}, M., {Ngo}, H.~N., {et~al.} 2025{\natexlab{c}},
  \bibinfo{title}{{Simulating Intermediate Black Hole Mass Measurements for a
  Sample of Galaxies with Nuclear Star Clusters Using ELT/HARMONI High Spatial
  Resolution Integral-field Stellar Kinematics},} \aj, 170, 124,
  \dodoi{10.3847/1538-3881/ade9ba}

\bibitem[{D.~D. {Nguyen} {et~al.}(2026{\natexlab{a}}){Nguyen}, {Cappellari},
  {Le}, {Ngo}, {Gallo}, {Thatte}, {Zou}, {Ho}, {Le}, {Tong}, \&
  {Pereira-Santaella}}]{Nguyen_2026_highz}
{Nguyen}, D.~D., {Cappellari}, M., {Le}, T. Q.~T., {et~al.} 2026{\natexlab{a}},
  \bibinfo{title}{{Extending the frontier of spatially resolved supermassive
  black hole mass measurements to at 1 {\ensuremath{\lesssim}} z
  {\ensuremath{\lesssim}} 2: simulations with ELT/MICADO high-resolution mass
  models and HARMONI integral-field stellar kinematics},} \mnras, 546, stag238,
  \dodoi{10.1093/mnras/stag238}

\bibitem[{D.~D. {Nguyen} {et~al.}(2026{\natexlab{b}}){Nguyen}, {Ngo},
  {Cappellari}, {Le}, {Ho}, {Le}, {Gallo}, {Thatte}, {Zou}, {Perna}, \&
  {Pereira-Santaella}}]{Nguyen_2026_4258}
{Nguyen}, D.~D., {Ngo}, H.~N., {Cappellari}, M., {et~al.} 2026{\natexlab{b}},
  \bibinfo{title}{{Measuring the Central Dark Mass in NGC 4258 with
  JWST/NIRSpec Stellar Kinematics},} \apj, 999, 97,
  \dodoi{10.3847/1538-4357/ae32f2}

\bibitem[{D.~D. {Nguyen} {et~al.}(2026{\natexlab{c}}){Nguyen}, {Le},
  {Cappellari}, {Ngo}, {Le}, {Ho}, {Nguyen}, {Gallo}, {Zou}, {Perna}, {Thatte},
  \& {Pereira-Santaella}}]{Nguyen_2026_M81}
{Nguyen}, D.~D., {Le}, T.~N., {Cappellari}, M., {et~al.} 2026{\natexlab{c}},
  \bibinfo{title}{{The Supermassive Black Hole in the Nearby Spiral Galaxy M81:
  A Robust Mass from JWST/NIRSpec Stellar Dynamics},} \apj, 1003, 98,
  \dodoi{10.3847/1538-4357/ae64f0}

\bibitem[{R. {Pechetti} {et~al.}(2022){Pechetti}, {Seth}, {Kamann}, {Caldwell},
  {Strader}, {den Brok}, {Luetzgendorf}, {Neumayer}, \& {Voggel}}]{Pechetti22}
{Pechetti}, R., {Seth}, A., {Kamann}, S., {et~al.} 2022,
  \bibinfo{title}{{Detection of a 100,000 M $_{{\ensuremath{\odot}}}$ black
  hole in M31's Most Massive Globular Cluster: A Tidally Stripped Nucleus},}
  \apj, 924, 48, \dodoi{10.3847/1538-4357/ac339f}

\bibitem[{J. {Pfeffer} {et~al.}(2021){Pfeffer}, {Lardo}, {Bastian}, {Saracino},
  \& {Kamann}}]{Pfeffer2021}
{Pfeffer}, J., {Lardo}, C., {Bastian}, N., {Saracino}, S., \& {Kamann}, S.
  2021, \bibinfo{title}{{The accreted nuclear clusters of the Milky Way},}
  \mnras, 500, 2514, \dodoi{10.1093/mnras/staa3407}

\bibitem[{L.~S. {Pilyugin} {et~al.}(2023){Pilyugin},
  {Tautvai{\v{s}}ien{\.{e}}}, \& {Lara-L{\'o}pez}}]{Pilyugin2023}
{Pilyugin}, L.~S., {Tautvai{\v{s}}ien{\.{e}}}, G., \& {Lara-L{\'o}pez}, M.~A.
  2023, \bibinfo{title}{{Searching for Milky Way twins: Radial abundance
  distribution as a strict criterion},} \aap, 676, A57,
  \dodoi{10.1051/0004-6361/202346503}

\bibitem[{A. Rantala {et~al.}(2024)Rantala, Rawlings, Naab, Thomas, \&
  Johansson}]{Rantala2024}
Rantala, A., Rawlings, A., Naab, T., Thomas, J., \& Johansson, P.~H. 2024,
  \bibinfo{title}{The supermassive black hole merger-driven evolution of
  high-redshift red nuggets into present-day cored early-type galaxies,}
  \mnras, 535, 1202, \dodoi{10.1093/mnras/stae2424}

\bibitem[{G.~B. {Rybicki}(1987){Rybicki}}]{Rybicki1987}
{Rybicki}, G.~B. 1987, \bibinfo{title}{{Deprojection of Galaxies --- How Much
  Can Be Learned?},} in Structure and Dynamics of Elliptical Galaxies, IAU
  Symp. 127, 397

\bibitem[{R.~P. {Saglia} {et~al.}(2016){Saglia}, {Opitsch}, {Erwin}, {Thomas},
  {Beifiori}, {Fabricius}, {Mazzalay}, {Nowak}, {Rusli}, \&
  {Bender}}]{Saglia16}
{Saglia}, R.~P., {Opitsch}, M., {Erwin}, P., {et~al.} 2016,
  \bibinfo{title}{{The SINFONI Black Hole Survey: The Black Hole Fundamental
  Plane Revisited and the Paths of (Co)evolution of Supermassive Black Holes
  and Bulges},} \apj, 818, 47, \dodoi{10.3847/0004-637X/818/1/47}

\bibitem[{A. {Saha} {et~al.}(2002){Saha}, {Claver}, \& {Hoessel}}]{Saha2002}
{Saha}, A., {Claver}, J., \& {Hoessel}, J.~G. 2002, \bibinfo{title}{{Cepheid
  Distances to IC~342 and NGC~6946},} \aj, 124, 839, \dodoi{10.1086/341650}

\bibitem[{N. {Sahu} {et~al.}(2019{\natexlab{a}}){Sahu}, {Graham}, \&
  {Davis}}]{Sahu19a}
{Sahu}, N., {Graham}, A.~W., \& {Davis}, B.~L. 2019{\natexlab{a}},
  \bibinfo{title}{{Black Hole Mass Scaling Relations for Early-type Galaxies.
  I. M $_{BH}$-M $_{*,}$ $_{sph}$ and M $_{BH}$-M $_{*,gal}$},} \apj, 876, 155,
  \dodoi{10.3847/1538-4357/ab0f32}

\bibitem[{N. {Sahu} {et~al.}(2019{\natexlab{b}}){Sahu}, {Graham}, \&
  {Davis}}]{Sahu19b}
{Sahu}, N., {Graham}, A.~W., \& {Davis}, B.~L. 2019{\natexlab{b}},
  \bibinfo{title}{{Revealing Hidden Substructures in the M
  $_{BH}$-{\ensuremath{\sigma}} Diagram, and Refining the Bend in the
  L-{\ensuremath{\sigma}} Relation},} \apj, 887, 10,
  \dodoi{10.3847/1538-4357/ab50b7}

\bibitem[{A. {Sandage} {et~al.}(1985){Sandage}, {Binggeli}, \&
  {Tammann}}]{1985AJ.....90.1759S}
{Sandage}, A., {Binggeli}, B., \& {Tammann}, G.~A. 1985,
  \bibinfo{title}{{Studies of the Virgo cluster. V. Luminosity functions of
  Virgo cluster galaxies.},} \aj, 90, 1759, \dodoi{10.1086/113875}

\bibitem[{E. {Schinnerer} {et~al.}(2003){Schinnerer}, {B{\"o}ker}, \&
  {Meier}}]{Schinnerer2003}
{Schinnerer}, E., {B{\"o}ker}, T., \& {Meier}, D.~S. 2003,
  \bibinfo{title}{{Molecular Gas and the Nuclear Star Cluster in IC 342:
  Sufficient Inflow for Recurring Star Formation Events?},} \apjl, 591, L115,
  \dodoi{10.1086/377118}

\bibitem[{E.~F. {Schlafly} \& D.~P. {Finkbeiner}(2011){Schlafly} \&
  {Finkbeiner}}]{Schlafly2011}
{Schlafly}, E.~F., \& {Finkbeiner}, D.~P. 2011, \bibinfo{title}{{Measuring
  Reddening with Sloan Digital Sky Survey Stellar Spectra and Recalibrating
  SFD},} \apj, 737, 103, \dodoi{10.1088/0004-637X/737/2/103}

\bibitem[{R. {Sch{\"o}del} {et~al.}(2014){Sch{\"o}del}, {Feldmeier},
  {Kunneriath}, {Stolovy}, {Neumayer}, {Amaro-Seoane}, \&
  {Nishiyama}}]{Schodel2014}
{Sch{\"o}del}, R., {Feldmeier}, A., {Kunneriath}, D., {et~al.} 2014,
  \bibinfo{title}{{Surface brightness profile of the Milky Way's nuclear star
  cluster},} \aap, 566, A47, \dodoi{10.1051/0004-6361/201423481}

\bibitem[{N. {Scott} \& A.~W. {Graham}(2013){Scott} \& {Graham}}]{Scott2013}
{Scott}, N., \& {Graham}, A.~W. 2013, \bibinfo{title}{{Updated Mass Scaling
  Relations for Nuclear Star Clusters and a Comparison to Supermassive Black
  Holes},} \apj, 763, 76, \dodoi{10.1088/0004-637X/763/2/76}

\bibitem[{E. {Serabyn} \& J.~H. {Lacy}(1985){Serabyn} \&
  {Lacy}}]{1985ApJ...293..445S}
{Serabyn}, E., \& {Lacy}, J.~H. 1985, \bibinfo{title}{{NE II observations of
  the galactic center : evidence for a massive black hole.},} \apj, 293, 445,
  \dodoi{10.1086/163250}

\bibitem[{A.~C. {Seth} {et~al.}(2008){Seth}, {Blum}, {Bastian}, {Caldwell}, \&
  {Debattista}}]{Seth2008}
{Seth}, A.~C., {Blum}, R.~D., {Bastian}, N., {Caldwell}, N., \& {Debattista},
  V.~P. 2008, \bibinfo{title}{{The Rotating Nuclear Star Cluster in NGC~4244},}
  \apj, 678, 116, \dodoi{10.1086/533955}

\bibitem[{A.~C. {Seth} {et~al.}(2006){Seth}, {Dalcanton}, {Hodge}, \&
  {Debattista}}]{Seth2006}
{Seth}, A.~C., {Dalcanton}, J.~J., {Hodge}, P.~W., \& {Debattista}, V.~P. 2006,
  \bibinfo{title}{{Clues to Nuclear Star Cluster Formation from Edge-On
  Spirals},} \aj, 132, 2539, \dodoi{10.1086/508994}

\bibitem[{A.~C. {Seth} {et~al.}(2010){Seth}, {van den Bosch}, {Mieske},
  {Baumgardt}, {den Brok}, {Strader}, {Neumayer}, {Chilingarian}, {Hilker},
  {McDermid}, {Spitler}, {Brodie}, {Frank}, \& {Hau}}]{Seth2010}
{Seth}, A.~C., {van den Bosch}, R., {Mieske}, S., {et~al.} 2010,
  \bibinfo{title}{{A Supermassive Black Hole in an Ultra-Compact Dwarf
  Galaxy},} \nat, 466, 1077, \dodoi{10.1038/nature09255}

\bibitem[{J.~D. {Silge} \& K. {Gebhardt}(2003){Silge} \&
  {Gebhardt}}]{Silge2003}
{Silge}, J.~D., \& {Gebhardt}, K. 2003, \bibinfo{title}{{Dust and the Infrared
  Kinematic Properties of Early-Type Galaxies},} \aj, 125, 2809,
  \dodoi{10.1086/375324}

\bibitem[{D.~A. {Simon} {et~al.}(2024){Simon}, {Cappellari}, \&
  {Hartke}}]{Simon2024}
{Simon}, D.~A., {Cappellari}, M., \& {Hartke}, J. 2024,
  \bibinfo{title}{{Supermassive black hole mass in the massive elliptical
  galaxy M87 from integral-field stellar dynamics using OASIS and MUSE with
  adaptive optics: assessing systematic uncertainties},} \mnras, 527, 2341,
  \dodoi{10.1093/mnras/stad3309}

\bibitem[{M.~D. {Smith} {et~al.}(2019){Smith}, {Bureau}, {Davis}, {Cappellari},
  {Liu}, {North}, {Onishi}, {Iguchi}, \& {Sarzi}}]{Smith2019}
{Smith}, M.~D., {Bureau}, M., {Davis}, T.~A., {et~al.} 2019,
  \bibinfo{title}{{WISDOM project - IV. A molecular gas dynamical measurement
  of the supermassive black hole mass in NGC 524},} \mnras, 485, 4359,
  \dodoi{10.1093/mnras/stz625}

\bibitem[{S. {Thater} {et~al.}(2023){Thater}, {Lyubenova}, {Fahrion},
  {Mart{\'\i}n-Navarro}, {Jethwa}, {Nguyen}, \& {van de Ven}}]{Thater23}
{Thater}, S., {Lyubenova}, M., {Fahrion}, K., {et~al.} 2023,
  \bibinfo{title}{{Effect of the initial mass function on the dynamical SMBH
  mass estimate in the nucleated early-type galaxy FCC 47},} \aap, 675, A18,
  \dodoi{10.1051/0004-6361/202245362}

\bibitem[{S. {Thater} {et~al.}(2022){Thater}, {Krajnovi{\'c}}, {Weilbacher},
  {Nguyen}, {Bureau}, {Cappellari}, {Davis}, {Iguchi}, {McDermid}, {Onishi},
  {Sarzi}, \& {van de Ven}}]{Thater22}
{Thater}, S., {Krajnovi{\'c}}, D., {Weilbacher}, P.~M., {et~al.} 2022,
  \bibinfo{title}{{Cross-checking SMBH mass estimates in NGC 6958 - I. Stellar
  dynamics from adaptive optics-assisted MUSE observations},} \mnras, 509,
  5416, \dodoi{10.1093/mnras/stab3210}

\bibitem[{S.~D. {Tremaine} {et~al.}(1975){Tremaine}, {Ostriker}, \&
  {Spitzer}}]{Tremaine1975}
{Tremaine}, S.~D., {Ostriker}, J.~P., \& {Spitzer}, Jr., L. 1975,
  \bibinfo{title}{{The formation of the nuclei of galaxies. I. M31.},} \apj,
  196, 407, \dodoi{10.1086/153422}

\bibitem[{M.~A. {van Dam} {et~al.}(2006){van Dam}, {Bouchez}, {Le Mignant},
  {Johansson}, {Wizinowich}, {Campbell}, {Chin}, {Hartman}, {Lafon}, {Stomski},
  \& {Summers}}]{vanDam2006}
{van Dam}, M.~A., {Bouchez}, A.~H., {Le Mignant}, D., {et~al.} 2006,
  \bibinfo{title}{{The W. M. Keck Observatory Laser Guide Star Adaptive Optics
  System: Performance Characterization},} \pasp, 118, 310,
  \dodoi{10.1086/499498}

\bibitem[{R.~C.~E. {van den Bosch} \& G. {van de Ven}(2009){van den Bosch} \&
  {van de Ven}}]{vandenBosch2009}
{van den Bosch}, R. C.~E., \& {van de Ven}, G. 2009,
  \bibinfo{title}{{Recovering the intrinsic shape of early-type galaxies},}
  \mnras, 398, 1117, \dodoi{10.1111/j.1365-2966.2009.15177.x}

\bibitem[{R.~P. {van der Marel} \& M. {Franx}(1993){van der Marel} \&
  {Franx}}]{vanderMarel1993}
{van der Marel}, R.~P., \& {Franx}, M. 1993, \bibinfo{title}{{A New Method for
  the Identification of Non-Gaussian Line Profiles in Elliptical Galaxies},}
  \apj, 407, 525, \dodoi{10.1086/172534}

\bibitem[{G. Van~Rossum \& F.~L. Drake(2009)Van~Rossum \&
  Drake}]{VanRossum2009}
Van~Rossum, G., \& Drake, F.~L. 2009, Python 3 Reference Manual (Scotts Valley,
  CA: CreateSpace)

\bibitem[{K. {Verro} {et~al.}(2022){Verro}, {Trager}, {Peletier},
  {Lan{\c{c}}on}, {Gonneau}, {Vazdekis}, {Prugniel}, {Chen}, {Coelho},
  {S{\'a}nchez-Bl{\'a}zquez}, {Martins}, {Arentsen}, {Lyubenova},
  {Falc{\'o}n-Barroso}, \& {Dries}}]{Verro2022}
{Verro}, K., {Trager}, S.~C., {Peletier}, R.~F., {et~al.} 2022,
  \bibinfo{title}{{The X-shooter Spectral Library (XSL): Data Release 3},}
  \aap, 660, A34, \dodoi{10.1051/0004-6361/202142388}

\bibitem[{P. Virtanen {et~al.}(2020)Virtanen, Gommers, Oliphant, Haberland,
  Reddy, Cournapeau, Burovski, Peterson, Weckesser, Bright, van~der Walt,
  Brett, Wilson, Millman, Mayorov, Nelson, Jones, Kern, Larson, Carey, Polat,
  Feng, Moore, VanderPlas, Laxalde, Perktold, Cimrman, Henriksen, Quintero,
  Harris, Archibald, Ribeiro, Pedregosa, van Mulbregt, Vijaykumar, Bardelli,
  Rothberg, Hilboll, Kloeckner, Scopatz, Lee, Rokem, Woods, Fulton, Masson,
  Häggström, Fitzgerald, Nicholson, Hagen, Pasechnik, Olivetti, Martin,
  Wieser, Silva, Lenders, Wilhelm, Young, Price, Ingold, Allen, Lee, Audren,
  Probst, Dietrich, Silterra, Webber, Slavi{\v{c}}, Nothman, Buchner, Kulick,
  Schönberger, de~Miranda~Cardoso, Reimer, Harrington, Rodr{\'{\i}}guez,
  Nunez-Iglesias, Kuczynski, Tritz, Thoma, Newville, Kümmerer, Bolingbroke,
  Tartre, Pak, Smith, Nowaczyk, Shebanov, Pavlyk, Brodtkorb, Lee, McGibbon,
  Feldbauer, Lewis, Tygier, Sievert, Vigna, Peterson, More, Pudlik, Oshima,
  Pingel, Robitaille, Spura, Jones, Cera, Leslie, Zito, Krauss, Upadhyay,
  Halchenko, \& V{\'{a}}zquez-Baeza}]{Virtanen2020}
Virtanen, P., Gommers, R., Oliphant, T.~E., {et~al.} 2020,
  \bibinfo{title}{{SciPy} 1.0: fundamental algorithms for scientific computing
  in Python,} Nature Methods, 17, 261, \dodoi{10.1038/s41592-019-0686-2}

\bibitem[{M. {Volonteri} {et~al.}(2008){Volonteri}, {Lodato}, \&
  {Natarajan}}]{Volonteri2008}
{Volonteri}, M., {Lodato}, G., \& {Natarajan}, P. 2008, \bibinfo{title}{{The
  evolution of massive black hole seeds},} \mnras, 383, 1079,
  \dodoi{10.1111/j.1365-2966.2007.12589.x}

\bibitem[{P.~L. Wizinowich {et~al.}(2006)Wizinowich, {Le Mignant}, Bouchez,
  Campbell, Chin, Contos, van Dam, Hartman, Johansson, Lafon, Lewis, Stomski,
  Summers, Brown, Danforth, Max, \& Pennington}]{Wizinowich2006}
Wizinowich, P.~L., {Le Mignant}, D., Bouchez, A.~H., {et~al.} 2006,
  \bibinfo{title}{{The W. M. Keck Observatory Laser Guide Star Adaptive Optics
  System: Overview},} PASA, 118, 297, \dodoi{10.1086/499290}

\bibitem[{P.-F. {Wu} {et~al.}(2014){Wu}, {Tully}, {Rizzi}, {Dolphin}, {Jacobs},
  \& {Karachentsev}}]{Wu2014}
{Wu}, P.-F., {Tully}, R.~B., {Rizzi}, L., {et~al.} 2014,
  \bibinfo{title}{{Infrared Tip of the Red Giant Branch and Distances to the
  Maffei/IC~342 Group},} \aj, 148, 7, \dodoi{10.1088/0004-6256/148/1/7}

\bibitem[{I. {Zaw} {et~al.}(2020){Zaw}, {Rosenthal}, {Katkov}, {Gelfand},
  {Chen}, {Greenhill}, {Brisken}, \& {Noori}}]{Zaw2020}
{Zaw}, I., {Rosenthal}, M.~J., {Katkov}, I.~Y., {et~al.} 2020,
  \bibinfo{title}{{An Accreting, Anomalously Low-mass Black Hole at the Center
  of Low-mass Galaxy IC 750},} \apj, 897, 111, \dodoi{10.3847/1538-4357/ab9944}

\bibitem[{C. {Zhang} {et~al.}(2024){Zhang}, {Wang}, {Tan}, {Gao}, {Lin}, \&
  {Xu}}]{Zhang2024}
{Zhang}, C., {Wang}, J., {Tan}, Q.-H., {et~al.} 2024,
  \bibinfo{title}{{Subkiloparsec Scaling Relations between Hot Gas, Dense Gas,
  and Star Formation Rate in Five Nearby Star-forming Galaxies},} \apjl, 967,
  L25, \dodoi{10.3847/2041-8213/ad4a69}

\bibitem[{H. {Zinnecker} {et~al.}(1988){Zinnecker}, {Keable}, {Dunlop},
  {Cannon}, \& {Griffiths}}]{1988IAUS..126..603Z}
{Zinnecker}, H., {Keable}, C.~J., {Dunlop}, J.~S., {Cannon}, R.~D., \&
  {Griffiths}, W.~K. 1988, in IAU Symposium, Vol. 126, The Harlow-Shapley
  Symposium on Globular Cluster Systems in Galaxies, ed. J.~E. {Grindlay} \&
  A.~G.~D. {Philip}, 603

\end{thebibliography}
%\appendix

\label{lastpage}
\end{document}